\documentclass[11pt,a4paper]{article}
\pdfoutput=1

\usepackage[utf8]{inputenc} 
\usepackage[english]{babel}
\usepackage{caption}
\usepackage{subcaption} % subfigures
\usepackage{multirow}

\usepackage[dvipsnames]{xcolor}
\usepackage{import}

\usepackage{amsfonts}
\usepackage{amsthm}
\usepackage{mathtools}
\usepackage{cool} % Trace, hypergeometric function, Pochhammer symbol, ...

\newcommand{\zs}{z_\star}
\newcommand{\zh}{z_h}

\newcommand{\AdS}{\text{AdS}}
\newcommand{\CFT}{\text{CFT}}

\newcommand{\zsh}{\left(\frac{\zs}{\zh} \right)}

\newcommand{\zshs}{\left(\tfrac{\zs}{\zh} \right)} % smaller version of \zsh if tight

\newcommand{\tpt}{^{\text{\tiny 2pt}}}
\newcommand{\EE}{^{\text{\textsc{ee}}}}
\usepackage{setspace}

\newcommand{\Area}[1]{\mathcal{A}}
\newcommand{\scl}[1]{\ell}
\newcommand{\wl}{^{\text{\textsc{wl}}}}
\renewcommand{\log}{\ln}
\newcommand{\EN}{\varepsilon}

\newcommand{\PGFcommands}{
  \renewcommand{\sffamily}{}
  \renewcommand{\frac}{\tfrac}
}

\def\dotsspace{0pt}
\def\dotsspaceend{1pt}
\def\mydots{{ \scriptstyle \hspace{\dotsspace}.\hspace{\dotsspace}.\hspace{\dotsspace}.\hspace{\dotsspaceend}}}

\usepackage{ifthen}
\newcommand{\Hgf}[4]{\left. {}_{\scriptscriptstyle #1}F_{\scriptscriptstyle #2} \left( #3 \right)  \ifthenelse{\equal{#4}{}}{\right.}{\right|_{\scriptscriptstyle #4}}
}

\usepackage[bottom]{footmisc}

\usepackage{pgf}

\usepackage{jheppub}
\graphicspath{{Inkscape/}{Plots/}}
\usepackage{hyperref}

  \title{Non-local observables at finite temperature in AdS/CFT}
\author[a,b]{Johanna Erdmenger,}
\author[a,b]{Nina Miekley\,}
\affiliation[a]{Lehrstuhl für Theoretische Physik III, Institut
	f\"ur Theoretische Physik und Astrophysik, \\ Julius-Maximilians-Universit\"at W\"urzburg,  
	Am Hubland, D-97074 W\"urzburg, Germany}
\affiliation[b]{Max-Planck-Institut f\"ur Physik (Werner-Heisenberg-Institut),\\
	F\"ohringer Ring 6,  D-80805 M\"unchen, Germany
}

\emailAdd{jke@mpp.mpg.de}
\emailAdd{nina.miekley@physik.uni-wuerzburg.de}
\abstract{Within gauge/gravity duality, we consider the AdS-Schwarzschild metric in arbitrary dimensions. We obtain
analytical closed-form results for the two-point function, Wilson
loop and entanglement entropy for strip geometries in the
finite-temperature field-theory dual. According to the duality,
these are given by the area of minimal surfaces of different
dimension in the gravity background. Our analytical results
involve generalised hypergeometric functions. We show that they
reproduce known numerical results to great accuracy.  Our results allow to
identify new physical behaviour: For instance, we consider the
entanglement density, i.e.\ the difference of entanglement
entropies at finite and vanishing temperature divided by the
volume of the entangling region. For field theories of dimension
seven or higher,  we find that the entanglement density displays
non-monotonic behaviour as function of $\ell \cdot T$, with $\ell$
the strip width and $T$ the temperature. This implies that the
area theorem, proven for RG flows in general dimensions, does not
apply here. This may signal the emergence of new degrees of
freedom for AdS Schwarzschild black holes in eight or more dimensions.
}
\keywords{AdS-CFT Correspondence, Gauge-gravity correspondence
}
\arxivnumber{1709.07016}
\begin{document}

  \hfill MPP-2017-236

  \maketitle
  
  \section{Introduction}

Non-local observables play an important role within the AdS/CFT correspondence. Examples include the entanglement entropy \cite{Ryu:2006bv,Ryu:2006ef} and the Wilson loop \cite{Maldacena:1998im}. The AdS/CFT correspondence \cite{Maldacena:1997re,Witten:1998qj,Gubser:1998bc} maps these observables to extremal surfaces in the bulk. Similarly, holography maps the two-point function for large scaling dimension to a geodesic in the bulk, i.e.\ to an extremal one-dimensional surface \cite{Balasubramanian:1999zv}. In this context, the two-point function is often treated together with the entanglement entropy and the Wilson loop as a non-local observable. These three field theory observables encode the geometry of the gravity side along their support. It is therefore interesting to look for characteristic signatures of features such as horizons in the dual gauge theory observables.\footnote{While we consider a time-independent case, it is also interesting to consider non-local observables in time-dependent situations such as thermalisation, as e.g.\ considered in \cite{Bellantuono:2016tkh}.} 

\begin{figure}[b]
	\vspace{1em}
	\hspace{0.24\linewidth}
	\def\svgwidth{0.6\linewidth}
	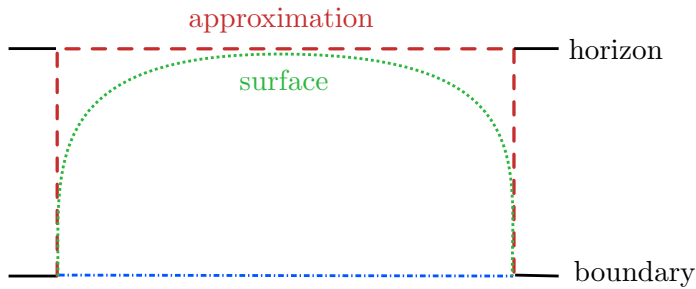
	\caption{Large-width behaviour.}
	\label{Fig:HighTBehaviour}
\end{figure}

It is challenging to obtain the result for the area of minimal surfaces at finite temperature in closed form. We consider the simplest example: finite-temperature field theories whose gravity dual is described by a planar AdS-Schwarzschild black hole \cite{Birmingham:1998nr,Witten:1998zw}. We study two-point function, Wilson loop and entanglement entropy associated to spatial surfaces anchored on a strip on the boundary. For a small strip, the results approach the well-known zero-temperature result \cite{Maldacena:1998im,Ryu:2006bv,Ryu:2006ef}. At high temperatures, the minimal area scales as the size of the strip. Figure \ref{Fig:HighTBehaviour} shows this limit: extremal surfaces (dotted green line) approach the horizon and wrap a large part of it.\footnote{This applies to  general metrics with a horizon, as examined in \cite{Hubeny:2012ry,Liu:2013una}.} This determines the leading contribution to the observables, yielding an exponential decay for the two-point function, an area term for the spatial Wilson loop and a volume term for the entanglement entropy. For the entanglement entropy, this volume term can be identified with the thermal entropy of the considered region \cite{Liu:2013una}. This leading contribution only depends on the geometry at the horizon and is correctly captured by the piecewise-smooth approximation (dashed red line) shown in Figure \ref{Fig:HighTBehaviour}. 

The first step in obtaining analytical expressions for the mentioned observables was made by Fischler and Kundu in \cite{Fischler:2012ca}. These authors obtained the minimal area of an $n$-dimensional surface in terms of power series in the value of the radial coordinate $\zs$ at the turning point,
\begin{align*}
\text{min. Area} &= \frac{A}{\zs^{n-1}}\sum_m c_m\cdot \zsh^{\gamma\cdot m}+\text{const.},
\end{align*}
where $\gamma$ is a positive, $n$-dependent constant and $\zh \propto T^{-1}$ is the position of the horizon.\footnote{The same technique was used in \cite{Kundu:2016dyk} to examine non-local observables in AdS-Reissner-Nordstr\"om.} This is an expansion in the characteristic length scale $\ell$, i.e.\ in the distance between the two points for the two-point function or in the width of the strip for the Wilson loop and the entanglement entropy. $A$ is the area of the $(n-1)$-dimensional boundary of the strip.\footnote{The strip is assumed to be infinitely long. For regularisation, we take the length $\tilde \ell \gg \ell$ such that the boundary area is $A=\tilde \ell^{n-1}$.} This power series diverges for $\zs \rightarrow \zh$. Additionally, these authors reorganised this series to obtain the large-width limit 
 \begin{align*}
 \text{min. Area}&\approx \# \frac{V}{\zh^{n}}+ \frac{A}{\zh^{n-1}} \sum_m \tilde c_m+\text{const.},
 \end{align*}
 where V is the volume of the strip $\ell\cdot A$. The series with coefficients $\tilde c$ converges. Hence, they obtained the next-to-leading order, width-independent contribution in terms of an infinite series. These contributions are of particular interest, since they are not captured by the approximation shown in Figure \ref{Fig:HighTBehaviour}.

In our paper, we simplify the power series results and bring them into a closed form involving generalised hypergeometric functions. The expressions obtained simplify further using Meijer $G$-functions, in terms of which they may be written as one term. These functions have many well-known properties, which can help us to understand the behaviour of the observables considered. In particular, we use these properties to consider the large-width limit, where we obtain closed forms of the next-to-leading order contributions derived in \cite{Fischler:2012ca}. These depend on the entire bulk metric, which makes it more difficult to calculate them. However, these subleading terms  contain further information about the dual field theory. We study examples demonstrating this below. 

As a first example, let us consider the small- and large-width limit of the entanglement entropy. For a small entangling region, entanglement thermodynamics applies. The first law of entanglement thermodynamics \cite{Blanco:2013joa} states that for two quantum states infinitely close in the Hilbert space of a QFT, their difference in entanglement entropy is equivalent to the difference of the expectation value  of the modular Hamiltonian. This follows from the positivity of their relative entropy. In this paper, we consider a strip geometry of width $\ell$  for the entangling region, for which the volume is $V =A\cdot \ell$.  According to \cite{Bhattacharya:2012mi}, for this configuration the difference of entanglement entropy is proportional to the change of energy inside the strip. For small strip widths $\ell$ and states with constant energy density $\langle T_{tt} \rangle $, this implies 
\begin{align}
S_{EE}-S_{EE}|_{T=0} \propto A\langle T_{tt} \rangle \cdot\ell^2 \label{eqn:IntroET}
\end{align}
for the entanglement entropy. We note that the left-hand side of \eqref{eqn:IntroET} grows with $\ell^2$ for small width $\ell$. For a CFT, the entanglement entropy at zero-temperature is of the form
\begin{align}
  S_{EE}|_{T=0} \propto \# \frac{A}{\epsilon^{d-2}} - ~ \mathcal{C}_d \frac{A}{\ell^{d-2}}, \label{eqn:IntroEE0}
\end{align}
where $\epsilon$ is a UV-cutoff, $A$ is the area of the entangling region and $\mathcal{C}_d$ is the central charge of the CFT. We omitted the dimension-dependent numerical constants in front of the UV-divergent term. For a large entangling region, the leading contribution is proportional to the volume of the entangling region and the subleading contribution is proportional to the area of the entangling surface,
\begin{align}
  S_{EE} = s\cdot V\, +\, \alpha \cdot A\,+\cdots, \label{eqn:IntroS}
\end{align}
where $s$ is the thermal entropy density and $\alpha$ is a constant \cite{Fischler:2012ca}. The further subleading terms of $\mathcal{O}(\ell)^{-1}$ are contained in the dots. In general, the entanglement entropy can also contain terms of the form $A \ln A$. As discussed in \cite{Swingle:2011np}, these correspond to an \textit{area law violation}.

There are several ansätze for obtaining a $c$-theorem related to the entanglement entropy. Motivated by the CFT result \eqref{eqn:IntroEE0}, the authors of \cite{Myers:2012ed} define a $c$-function 
\begin{align}
c_d \propto \frac{\ell^{d-1}}{A} \frac{\partial S_{EE} }{\partial \ell} \label{eqn:IntroCFct}
\end{align} 
and prove its monotonicity using the null energy condition. An example of this involving a torus was studied in \cite{Bueno:2016rma}. For a conformal field theory in even dimensions, $c_d$ is related to the central charge given by the topological contribution to the conformal anomaly \cite{Myers:2012vs,Myers:2010tj,Ryu:2006bv}. Considering the large-width limit $\ell \rightarrow \infty$ in \eqref{eqn:IntroS}, we note that the term proportional to the boundary area $A$ drops out and the holographic $c$-function \eqref{eqn:IntroCFct} depends on the $\mathcal{O}(\ell)^{-1}$ contributions contained in the dots in \eqref{eqn:IntroS}. In this work however, we focus our attention on the area term, i.e.\ the second term in \eqref{eqn:IntroS}. This term satisfies a variant of the $c$-theorem, the {\it area theorem}, which states that for a RG flow with a UV and an IR fixed point, the coefficient of the area law term contributing to the entanglement entropy must be larger in the UV than in the IR, i.e.\ $\alpha_\mathrm{UV} \geq \alpha_\mathrm{IR}$. Field-theory proofs exist for spherical entangling regions, for $d=3$ using strong subadditivity \cite{Casini:2012ei}, and for $d \geq 3$ using the positivity of relative entropy \cite{Casini:2016udt}. 

To examine this area term, it is useful to look at the entanglement density $\sigma$  introduced in \cite{Gushterov:2017vnr}.\footnote{This is not the entanglement density defined as variation of the entanglement entropy, as defined in \cite{Nozaki:2013wia,Bhattacharya:2014vja}.} This quantity is defined as
\begin{equation} \label{eq:sigma}
\sigma = \frac{ S_{EE} - S_{EE}|_{T=0}}{V} \, .
\end{equation}
where the zero-temperature result is subtracted for UV regularisation and the difference is divided by the volume of the entangling region. This yields a finite  cut-off independent expression. Entanglement thermodynamics determines the behaviour for a small width \eqref{eqn:IntroET}: the entanglement density vanishes for zero width $\ell$ and grows linearly for small values of $\ell$. For large $\ell \rightarrow \infty$, the entanglement density is (c.f.\ \eqref{eqn:IntroS})
\begin{equation}
\sigma = s - \Delta \alpha \frac{A}{V} + \cdots \, , \label{eqn:IntroED}
\end{equation}
where the subleading term is proportional to $\frac{A}{V}=\ell^{-1}$ and the constant $\Delta \alpha$ is defined as
\begin{align}
\Delta \alpha = \alpha|_{T=0}-\alpha. \label{eqn:IntroAlpha}
\end{align}
The sign of $\Delta \alpha$ determines whether the entanglement density approaches the thermal entropy density from below or above for $\ell \rightarrow \infty$. In the former case we have $\Delta \alpha>0$. The simplest behaviour for the entanglement density is to increase monotonically from zero to the thermal entropy density. This is not the case if the entanglement density approaches the thermal entropy density from above: In this case, we have $\Delta \alpha<0$ and the area theorem does not apply. The simplest behaviour for the entanglement density in this case is that it increases monotonically, reaches its maximum at a finite value for $\ell \cdot T$, after which it decreases and approaches the thermal entropy density asymptotically. Therefore, the sign of $\Delta \alpha$ can be easily determined from the qualitative behaviour of the entanglement density as function of $\ell$. For a RG flow, the difference $\Delta \alpha$ between the coefficient of the area law term in the UV and IR is of the form 
\begin{align}
\Delta \alpha = k\cdot  m^{d-2},
\end{align}
where $m$ is the mass scale of the RG flow and $k$ is a numerical constant, which is positive if the area theorem applies. At finite temperature, the mass scale is given by the temperature $T$. In contrast, it is not that straightforward to examine the $c$-function as considered by \cite{Myers:2012ed}; The characteristic $c_d/\ell^{d-2}$-term is contained in the subleading contributions of the large-width expansion \eqref{eqn:IntroED}. 

Using the analytical expressions for the entanglement entropy obtained in this paper for the strip geometry at finite temperature, we find analytical expressions for $\Delta \alpha$ in general dimensions. For this we expand our analytical expression to next-to-leading order at large widths in order to extract the area term. Strikingly, we find a critical dimension as follows: For field theories of spacetime dimension $d=6$ or smaller, the area theorem is always satisfied, whereas it does not apply for field theories of dimension $d=7$ or larger. A similar result and was found independently by studying the entanglement density numerically in \cite{Gushterov:2017vnr}. Our analytical calculation confirms these findings. We will discuss possible origins for this behaviour in section \ref{sec:ResultsEE} and suggest further studies in the conclusion. According to \cite{Gushterov:2017vnr}, a violation of the area theorem may be traced back to different scaling of time and the spatial coordinates. This is known to occur in the limit of infinite dimensions $d \rightarrow \infty$ \cite{Emparan:2013moa,Emparan:2013xia}.  It is remarkable that this happens here already in a large but finite range of dimensions.

For a large entangling region, the leading contribution to the entanglement entropy is the thermal entropy of the entangling region. This extensive term shows that the entanglement entropy for a mixed state contains classical contributions and is no longer a measure for entanglement. Instead, a proper measure is the entanglement negativity $\EN$ \cite{Calabrese:2012ew,Calabrese:2012nk,Calabrese:2014yza,2002PhRvA..65c2314V}. In holographic theories, it is proportional to the difference between entanglement and thermal entropy \cite{Chaturvedi:2016rcn,Chaturvedi:2016rft},
\begin{align}
\EN \propto S_{EE}-s\cdot V.
\end{align}
For a small entangling region, the entanglement negativity at finite temperature is smaller than at zero-temperature, as the subtracted thermal entropy is the leading-order contribution (c.f.\ \eqref{eqn:IntroET}). However, this is not in true for a general width. In the large-width limit, the extensive term is removed and only the area term remains,
\begin{align}
\EN \propto \alpha \cdot A + \cdots.
\end{align}
 The large-region behaviour is characterised by the sign of $\Delta \alpha$ as introduced in \eqref{eqn:IntroAlpha}.  In theories with $\Delta\alpha>0$, i.e.\ theories which obey the area theorem, turning on a temperature decreases entanglement on all scales. The situation for theories with $\Delta \alpha<0$ is more complex: while the temperature decreases short-range entanglement, it increases long-range entanglement. This may indicate that $\Delta \alpha$ is related to the change of the number of degrees of freedom and our results for thermal field theories in $d \geq 7$ may hint to the appearance of new degrees of freedom in the IR.

In addition to these considerations on entanglement entropy, we also obtain analytical results for the Wilson loop in the AdS soliton geometry, for arbitrary dimensions $d>2$.  The AdS soliton is a confining geometry which is obtained from the AdS-Schwarzschild solution by a double Wick rotation. We consider the quark-antiquark potential obtained from the Wilson loop and expand our analytical result for low energies, i.e.~for large quark separation. The first term in this expansion gives the expected linear confining contribution to the potential. The second term gives a finite energy independent of the quark separation, which corresponds to a finite mass renormalisation of the quarks bound in a pair. 

We begin the main part of our paper by briefly describing our gravity set up in section \ref{sec:Setup}. We then proceed by presenting our results for the observables the two-point function, the spatial Wilson loop and the entanglement entropy in sections \ref{sec:2pt}, \ref{sec:WL} and \ref{sec:HEE} respectively. Each of these sections has the same structure: we first shortly review the field-theory observables in field theories and their gravity dual. Then, we present our results in terms of generalised hypergeometric functions and, if applicable, our results in Meijer $G$-functions. In particular, we take the large-width limit of our expressions. Where applicable, we discuss the physical properties of the subleading term in this expansion.
For the entanglement entropy, we also consider two related quantities: the entanglement density \ref{sec:HED} and the entanglement negativity \ref{sec:HEN}. Moreover, we compare our results with the known expressions for the entanglement entropy in $d=2$ in  section \ref{sec:ResultsEE} and with numerical results of \cite{Balasubramanian:2011ur} in section \ref{sec:Numerics}. We conclude  with a short summary and outlook. 
Appendix \ref{sec:AppFunctions} contains a summary of the definitions and properties of generalised hypergeometric and Meijer $G$-functions.
Since the details of our analytical calculations are similar in all cases considered, we summarise the essential features of these calculations in  appendix  \ref{sec:MinimalSurface}.

  \section{Setup and conventions}
\label{sec:Setup}
We consider the duality between a QFT in $d$ dimensions at finite temperature and a gravity theory with a planar AdS-Schwarzschild black hole in $d+1$ dimensions \cite{Maldacena:1997re,Witten:1998qj,Aharony:1999ti,Birmingham:1998nr,Witten:1998zw}.
The metric is
\begin{subequations}
\begin{align}
 ds^2 &= \frac{L^2}{z^2} \left( -b(z) dt^2 + \frac{dz^2}{b(z)} + d \vec{x}^2 \right), \\
 b(z) &= 1-\frac{z^d}{z_h^d}, 
\end{align}\label{eqn:metric}%
\end{subequations}
where $L$ is the AdS-radius and $\zh$ is the horizon. The temperature of the dual field theory is the Hawking temperature of the black hole,
\begin{align}
T=\frac{d}{4 \pi \zh}. \label{eqn:Temperature}
\end{align}
The thermal entropy of the field theory is the black hole
entropy, whose density is
\begin{align}
 s &= \frac{1}{4 G_N} \left(\frac{L}{z_h}\right)^{d-1}\nonumber \\
 &= \frac{L^{d-1}}{4 G_N} \left(\frac{4 \pi}{d}\right)^{d-1} \cdot T^{d-1}. \label{eqn:ThermalEntropy}
\end{align}
The energy density is related to the asymptotic fall off of the metric \cite{deHaro:2000vlm}, which yields
\begin{align}
\langle T_{tt} \rangle = \frac{(d-1)L^{d-1}}{16\pi G \zh^d}. \label{eqn:EnergyDensity}
\end{align}

\begin{figure}[b]
	\center
	\def\svgwidth{0.45\linewidth}
	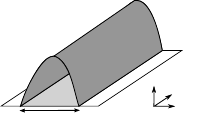
	\caption{Boundary region and associated bulk surface.\\
		The strip has the width $\ell$ in direction $x^1$ and length $\tilde \ell \gg \ell$ in the directions $x^i$ with  $i=2,\cdots,n$. The remaining directions (i.e.\ $x^j$ with $j=n+1,\cdots,d-1$) are not shown.
	}
\label{fig:StraightBeltIntro}
\end{figure}

Non-local observables correspond to minimal surfaces in the bulk. In
this paper, we consider the minimal area of spatial surfaces attached
to an $n$-dimensional strip as shown in Figure
\ref{fig:StraightBeltIntro}. This implies we consider a constant time slice.
As we explain in appendix \ref{sec:MinimalSurface}, we express the
width of the strip $\ell$ and the minimal area ${\cal A}$ of the
attached surface in terms of the turning point  $\zs$ of the surface. Both of these can be expressed as integrals (c.f.\ \eqref{eqn:SolutionAsIntegral}),
\begin{subequations}
	\begin{align}
	\ell %&= 2\int_0^\ell dz ~x'(z), \nonumber \\
	&=  2 \int\limits_{0}^{\zs} dz~ \left(\frac{z}{\zs}\right)^{\mathclap{n}} \frac{1}{\sqrt{b(z)}} \frac{1}{\sqrt{1-\left(z/\zs\right)^{2n}}},\\
	\Area{} &= 2L^n \tilde \ell^{n-1} \int\limits_{\epsilon}^{\zs}
                  dz~ z^{-n} \frac{1}{\sqrt{b(z)}}
                  \frac{1}{\sqrt{1-\left(z/\zs\right)^{2n}}} \, .
	\end{align}
	\label{eqn:SolutionAsIntegralIntro}%
\end{subequations}
In \cite{Fischler:2012ca}, the authors expanded the integrands in
these expressions as a power series, which we review in section \ref{sec:MinimalSurfacePS}. In this paper, we write this result as a finite sum containing generalised hypergeometric functions. Furthermore, we simplify this result further in terms of Meijer $G$-functions. For a review of these functions, see sections \ref{sec:AppHGF} and \ref{sec:AppMGF}. The detailed calculation can be found in sections \ref{sec:MinimalSurfaceHGF} and \ref{sec:MinimalSurfaceMGF}.

Two-point function, Wilson loop and entanglement entropy are related to $n=1$, $n=2$ and $n=(d-1)$-dimensional minimal surfaces, respectively. The following three sections are devoted to each of these cases.

  \section{Two-point function}
\label{sec:2pt}

For large scale dimensions $\Delta \geq 1$,
 the two-point function may be given in terms of the length of the geodesic 
between the two boundary points  \cite{Balasubramanian:1999zv}. This amounts to a semiclassical approximation.
In this case, the two-point function on the field theory side is written as \cite{Banks:1998dd}
\begin{align}
 \langle \mathcal{O}(t,\vec{x})  \mathcal{O}(t',\vec{y}) \rangle &=\lim \limits_{\epsilon \rightarrow 0}  \epsilon^{- 2 \Delta } \langle \varphi(b_x(\epsilon)) \varphi(b_y(\epsilon)) \rangle,
\end{align}
where $\langle \varphi\varphi \rangle $ is the two-point function of the dual field on the gravity side. The bulk positions $b(\epsilon)$ approach the corresponding boundary points,
\begin{subequations}
\begin{align}
  \lim\limits_{\epsilon \rightarrow 0} b_x(\epsilon) = \left(0,t,\vec{x}\right),\\
  \lim\limits_{\epsilon \rightarrow 0} b_y(\epsilon) = \left(0,t',\vec{y}\right).
\end{align}\label{eqn:2ptBulkPositions}
\end{subequations}
According to \cite{Balasubramanian:1999zv}, the two-point function on the gravity side is then given by
\begin{align}
 \langle \mathcal{\varphi}(\mathbf{x})  \mathcal{\varphi}(\mathbf{y}) \rangle &= \int \mathcal{DP}~\exp\left(i\Delta\cdot \frac{ \mathcal{L}(\mathcal{P})}{L}\right), \label{eq:Bala}
\end{align}
where $\mathbf{x}$ and $\mathbf{y}$ are points in the bulk connected by paths $\mathcal{P}$.\footnote{It would be interesting to find the explicit map between the standard calculation of the two-point function in terms of bulk-to-boundary propagators \cite{Witten:1998qj,Freedman:1998tz} and the geodesic approach used here. This is beyond the scope of the present paper.
We note that it was argued in \cite{Louko:2000tp} that \eqref{eq:Bala}  is a Green's function for the wave operator.} $\mathcal{L}(\mathcal{P})$ is the proper length of the path. The path-integral measure $\mathcal{DP}$  is not specified as we take the saddle-point approximation when considering $\Delta \gg 1$. The conventions are in such a way that spacelike geodesics have a positive imaginary length. The result for large conformal dimension $\Delta$ is then
\begin{align}
 \langle\mathcal{\varphi}(\mathbf{x})  \mathcal{\varphi}(\mathbf{y})  \rangle &= \exp\left(-\Delta \cdot\frac{\Area{1}}{L}\right),
\end{align}
where $\Area{1}$ is the length of the geodesic between $\left(0,t,\vec{x}\right)$ and $\left(0,t,\vec{y}\right)$. Applying this to \eqref{eqn:CFT2pt}, the two-point function in the field theory can be calculated as
\begin{align}
 \langle \mathcal{O}(t,\vec{x})  \mathcal{O}(t,\vec{y}) \rangle &=\lim \limits_{\epsilon \rightarrow 0}  \epsilon^{- 2 \Delta } \exp\left(-\Delta \cdot\frac{\Area{1}(\epsilon)}{L}\right),\label{eqn:CFT2pt}
\end{align}
where the bulk points approach the boundary as specified in \eqref{eqn:2ptBulkPositions}. The prefactor ensures that the two-point function is finite in the $\epsilon \rightarrow 0$ limit. Due to translational invariance, the result for the equal-time two-point function depends only on the distance $\scl{1}$
\begin{align}
  \scl{1} = |\vec{x}-\vec{y}|.
\end{align}
\begin{figure}
  \center
   \def\svgwidth{0.6\linewidth}
   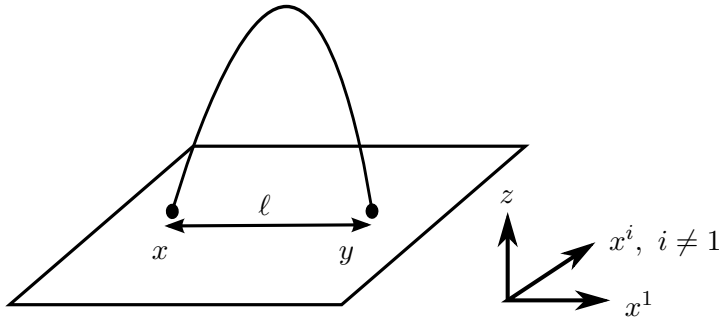
  \caption{Calculation of the two-point function from a geodesic in AdS space}
\label{fig:StraightLine}
\end{figure}

As the name already says, the two-point function depends on two points and is therefore not a non-local observable. However, holographically it is also associated with an extremal surface. Therefore, the calculation is similar and the two-point function is often considered together with non-local observables.

\subsection[Analytical result]{Analytical result for the two-point function}

The associated surface for the two-point function is a geodesic and hence one-dimensional (i.e.\ the two-point function corresponds to $n=1$, following our previous notation). The detailed calculation for the minimal area for a minimal surface anchored on a strip on the boundary can be found in appendix \ref{sec:MinimalSurface}. The starting point of our approach is the 
 expansion of the expressions \eqref{eqn:SolutionAsIntegralIntro} in power series in $\zs/\zh$ given in \cite{Fischler:2012ca} and reviewed in section \ref{sec:MinimalSurface}, where $\zs$ is the turning point of the minimal surface and $\zh$ the position of the horizon. We rearrange the sum in a particular form that allows us to write it as a finite number of generalised hypergeometric functions \footnote{See appendix \ref{sec:AppHGF} for a review of generalized hypergeometric functions.}.

According to \eqref{eqn:Parameters}, the parameters appearing in these hypergeometric functions are
\begin{align}
 a\tpt_i =& \frac{1}{2d}\left( \Delta m d +1 + 2 i\right),
\end{align}
which depend implicitly on the index of summation $\Delta m$. The distance between the two points is the one-dimensional analogue of the width of a strip appearing for instance in entanglement entropy calculations. Using  appendix \ref{sec:MinimalSurfaceHGF} and in particular \eqref{eqn:SolutionAsHGFWidth}, we have
\begin{subequations}
\begin{align}
 \scl{1} & =\frac{\sqrt{\pi}\zs}{2} \frac{\GammaFunc{\frac{d+2}{2}}}{\GammaFunc{\frac{d+3}{2}}}\zsh^{\mathclap{d}} \Hgf{d+2}{d+1}{a\tpt_{\frac{1}{2}},\mydots, a\tpt_{d-\frac{1}{2}}, \frac{3}{4},\frac{5}{4};  a\tpt_1,\mydots, a\tpt_{d}, \frac{3}{2};\zsh^{2d}}{\Delta m=1}\hspace{-2em} \nonumber \\
 &~~~+ 2\zs \Hgf{d+2}{d+1}{a\tpt_{\frac{1}{2}},\mydots, a\tpt_{d-\frac{1}{2}}, \frac{1}{4}, \frac{3}{4};  a\tpt_1,\mydots, a\tpt_{d}, \frac{1}{2};\zsh^{2d}}{\Delta m=0} . \label{eqn:Width2ptGeneral}
\end{align}
For even (boundary)  spacetime dimensions, this simplifies to
\begin{align}
 \scl{1} &= 2\zs \Hgf{\frac{d}{2}+1}{\frac{d}{2}}{2a\tpt_{\frac{1}{2}},\mydots, 2a\tpt_{\frac{d-1}{2}}, \frac{1}{2}; 2a\tpt_1,\mydots, 2a\tpt_{\frac{d}{2}};\zsh^{d}}{\Delta m=0}  \, .\label{eqn:Width2ptEven}
\end{align}\label{eqn:Width2pt}%
\end{subequations}
Here, $\zs$ is the turning point of the geodesic and $\zh$ is the position of the horizon, which is proportional to the inverse temperature $T$ (c.f.\ \eqref{eqn:Temperature}).

For the geodesic length, there are some subtleties as compared to the Wilson loop or entanglement entropy calculations since it has a logarithmic and not a power-law divergence in the UV limit. The corresponding calculation may be found in section \ref{sec:MinimalSurfacen1} and results in \eqref{eqn:SolutionAsHGFArean1}, yielding
\begin{subequations}
\begin{align}
 \Area{1}=
 &2L\ln\left(\frac{2\zs}{\epsilon}\right)+ \frac{3L}{8}\frac{\sqrt{\pi}\GammaFunc{d}}{\GammaFunc{\frac{2d+1}{2}}}\zsh^{\mathclap{~2d}} \nonumber \\
 &\times \Hgf{d+3}{d+2}{1,\frac{5}{4},\frac{7}{4},a\tpt_{-\frac{1}{2}},\mydots,a\tpt_{d-\frac{3}{2}};\frac{3}{2},2,a\tpt_0,\mydots,a\tpt_{d-1};\zsh^{2d}}{\Delta m=2}\hspace{-2em} \label{eqn:Area2ptGeneral} \\[0.5em]
 &+\frac{L}{2}\frac{\sqrt{\pi}\GammaFunc{\frac{d}{2}}}{\GammaFunc{\frac{d+1}{2}}} \zsh^{\mathclap{d}} \Hgf{d+2}{d+1}{\frac{3}{4},\frac{5}{4},a\tpt_{-\frac{1}{2}},\mydots,a\tpt_{d-\frac{3}{2}};\frac{3}{2},a\tpt_{0},\mydots,a\tpt_{d-1};\zsh^{2d} }{ \Delta m=1} \, ,\hspace{-0.5em}  \nonumber
\end{align}
which in even dimensions simplifies to
\begin{align}
 \Area{1} &= %2\ln (\zs) +\nonumber \\
  2L\ln\left(\frac{2\zs}{\epsilon}\right) +\frac{L}{2}  \frac{\sqrt{\pi}\GammaFunc{\tfrac{d}{2}}}{\GammaFunc{\tfrac{d+1}{2}
}}\zsh^{\mathclap{d}}  \nonumber \\[0.5em] &~~~\times \Hgf{\frac{d}{2}+2}{\frac{d}{2}+1}{\frac{3}{2},1,2a\tpt_{-\frac{1}{2}},\mydots,2a\tpt_{\frac{d}{2}-\frac{3}{2}};2,2a\tpt_{0},\mydots,a\tpt_{\frac{d}{2}-1};\zsh^{d} }{\Delta m=1}.\hspace{-2em}
\label{eqn:Area2ptEven}
\end{align}\label{eqn:Area2pt}%
\end{subequations}
This quantity is divergent when taking the bulk cut-off $\epsilon$ to zero.

A regular expression for the  field-theory two-point function is obtained from  \eqref{eqn:CFT2pt},
\begin{align}
\langle \mathcal{O}(t,\vec{x})  \mathcal{O}(t,\vec{y}) \rangle &=\left. \lim \limits_{\epsilon \rightarrow 0}  \epsilon^{- 2 \Delta } \exp\left(-\Delta \cdot\frac{\Area{1}}{L}\right)\right|_{\ell=|\vec{x}-\vec{y}|}.
\end{align}
The $\epsilon^{-2\Delta}$ factor eliminates the UV-divergent term of the area and ensures a finite result for $\epsilon \rightarrow 0$. Below we give explicit expressions for the two-point function and for the distance between the two  points in terms of the turning point $\zs$. Of course, this quantity does not have a physical meaning on the field-theory side and simply parametrizes the results. In contrast, the position of the horizon $\zh$ is related to the field-theory temperature (see \eqref{eqn:Temperature}). 

\begin{figure}
	\centering
	{\PGFcommands{}
		\renewcommand{\ell}{|\vec{x}-\vec{y}|}
		\import{PGF/}{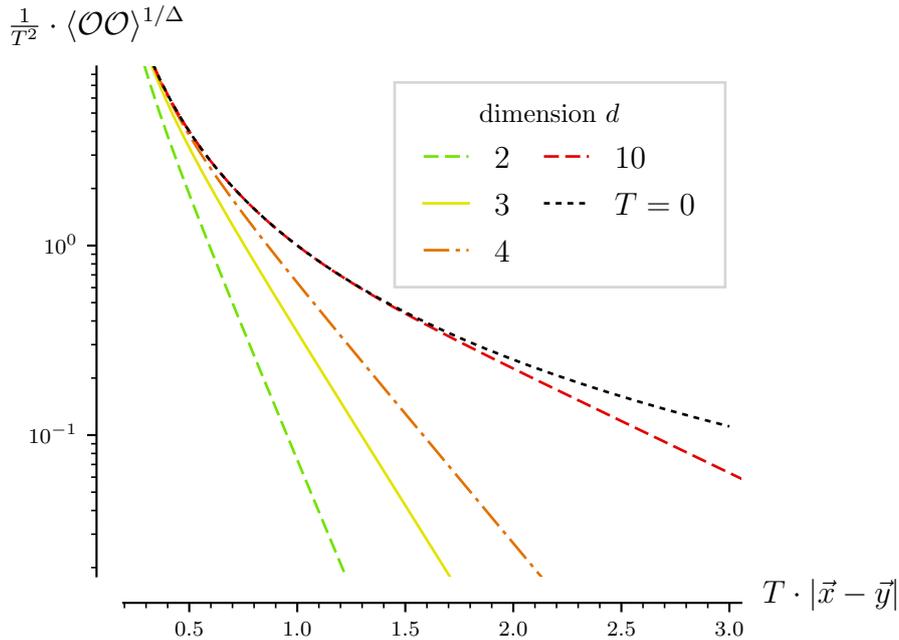}}
	\caption{Two-point function for different spacetime dimensions.
	}
	\label{Fig:2ptFct}
\end{figure}

Figure \ref{Fig:2ptFct} shows a plot of our analytical results  for the three lowest spacetime dimensions $d=2,3,4$, as well as in $d=10$ as an example of a high spacetime dimension. The analytical expressions for the three lowest spacetime dimensions are discussed in section \ref{sec:Results2pt}. Before moving on to these examples, let us have a look at characteristic behaviour of the expressions obtained.

For vanishing argument, generalised hypergeometric functions approach unity. Therefore, we can easily take the small-width limit (i.e.\ $
\zs/\zh \rightarrow 0$) and obtain
\begin{align}
\scl{1} =& 2\zs + \zs \cdot \mathcal{O}\zsh^{d}
\end{align}
for the distance $\ell$ and 
\begin{align}
\Area{1} =& 2 L \ln \left( \frac{2\zs}{\epsilon}\right)  +  \mathcal{O}\zsh^{d}, \nonumber\\
=& 2 L \ln \left( \frac{\ell}{\epsilon}\right)  +  \mathcal{O}\left(\ell \cdot T \right)^d
\end{align}
for the geodesic length. This leads to the characteristic power-law behaviour of the two-point function (c.f.\ \eqref{eqn:CFT2pt})
\begin{align}
\langle \mathcal{O}(t,\vec{x})  \mathcal{O}(t,\vec{y}) \rangle 
= \ell^{-2\Delta}\left[1 +  \mathcal{O}\left(\ell \cdot T \right)^d \right] \, ,\label{eqn:2ptT0}
\end{align}
where $\ell$ is the distance between $\vec{x}$ and $\vec{y}$.\footnote{This agrees with the result in terms of a power series, which we reviewed in the appendix (c.f.\ \eqref{eqn:SolutionAsPowerSeries} and \eqref{eqn:SolutionAsPowerSeriesn1}).} Figure \ref{Fig:2ptFct} shows how our result for finite temperature approaches the zero-temperature result (dotted black line) at small-width.

In contrast, the large-width limit (i.e.\ $\zs \rightarrow \zh$) is more involved. The behaviour of the hypergeometric functions at unit argument depends on their parameters (c.f. \eqref{eqn:HgfDivergences}). In the case considered here, the results diverge logarithmically. The minimal area and the distance are proportional in leading order and we obtain the expected exponential decay for the two-point function (c.f.\ Figure \ref{Fig:2ptFct}).  Let us take a closer look at the large-width limit in the following section.

\subsection[Large-width behaviour]{Large-width behaviour of the two-point function}

In the large-width limit (i.e.\ for $\zs \rightarrow \zh$), the series for the distance $\scl{1}$ and the geodesic length $\Area{1}$ are diverging. The reason is that we start with a power series (c.f.\ \eqref{eqn:SolutionAsPowerSeries}), which has a finite radius of convergence. 
 
In the following, we use properties of generalised hypergeometric functions to re-write the result for the geodesic length. The hypergeometric functions for the distance $\scl{1}$ and the geodesic length $\Area{1}$ differ by integer values. These kind of hypergeometric functions are referred to as {\it  associated}. There are  linear relationships between them, called {\it contiguous relations} (c.f.\ \eqref{eqn:ContiguousRel}). This allows us to write the geodesic length as (c.f.\ \eqref{eqn:SolutionHTn1})
\begin{subequations}
\begin{align}
 \Area{1}&= \frac{L\scl{1}}{\zs}-2L+ 2L\ln\left(\frac{2\zs}{\epsilon}\right)+\frac{3\sqrt{\pi}L}{16} \frac{\GammaFunc{d}}{\GammaFunc{\frac{2d+3}{2}}}\zsh^{\mathclap{~2d}} \nonumber \\
 &~~~\times \Hgf{d+3}{d+2}{ 1,\frac{5}{4},\frac{7}{4},a\tpt_{-\frac{1}{2}},\mydots,a\tpt_{d-\frac{3}{2}};\frac{3}{2},2,a\tpt_1,\mydots,a\tpt_{d};\zsh^{2d}}{\Delta m=2}\hspace{-1.5em} \label{eqn:AArea2ptGeneral}  \\
 &~~~ +\frac{\sqrt{\pi}L\GammaFunc{\frac{d}{2}}}{4\GammaFunc{\frac{d+3}{2}}} \zsh^{\mathclap{d}}\Hgf{d+2}{d+1}{\frac{3}{4},\frac{5}{4},a\tpt_{-\frac{1}{2}},\mydots,a\tpt_{d-\frac{3}{2}};\frac{3}{2},a\tpt_{1},\mydots,a\tpt_{d};\zsh^{2d} }{\Delta m=1}\nonumber
\end{align}
for general dimensions and
\begin{align}
 \Area{1} &= %2L\ln (\zs) \nonumber \\
   \frac{\sqrt{\pi}L\GammaFunc{\frac{d}{2}}}{4\GammaFunc{\frac{d+3}{2}}}\zsh^{\mathclap{d}} \Hgf{\frac{d}{2}+2}{\frac{d}{2}+1}{\frac{3}{2},1,2a\tpt_{-\frac{1}{2}},\mydots,2a\tpt_{\frac{d}{2}-\frac{3}{2}};2,2a\tpt_{1},\mydots,2a\tpt_{\frac{d}{2}};\zsh^{d} }{\Delta m=1}\nonumber \\
 &~~~ + \frac{L\scl{1}}{\zs}-2L+ 2L\ln\left(\frac{2\zs}{\epsilon}\right)\label{eqn:AArea2ptEven}
\end{align}\label{eqn:AArea2pt}%
\end{subequations}
for even dimensions. Remarkably, these generalised hypergeometric functions are finite when their argument approaches one,  and the divergent behaviour is entirely captured by the term containing the distance $\scl{1}$. The behaviour of the hypergeometric functions changes because we shift one of the denominator parameters by unity,
\begin{align}\chi\, a\tpt_{d/\chi} &= \chi\, a\tpt_0 + 1, & \chi = \begin{cases}
1 & d\text{ odd} \\
2 & d\text{ even}
\end{cases}.\end{align}
Due to the logarithmic divergence of the distance $\ell$, the turning point $\zs$ approaches the horizon exponentially fast. Hence, the leading-order contribution to the geodesic length at large-width is
\begin{align}
 \Area{1} &\approx -L\ln A_d+ 2L\ln\left(\frac{\zh}{\epsilon}\right) + \frac{L\scl{1}}{\zh},
\end{align}
where the constant $A_d$ is
\begin{align}
 \ln \left( 4 A_d\right) &= 2- \frac{\sqrt{\pi}\GammaFunc{\frac{d}{2}}}{4\GammaFunc{\frac{d}{2}+\frac{3}{2}}} \Hgf{d+2}{d+1}{\frac{3}{4},\frac{5}{4},a\tpt_{-\frac{1}{2}},\mydots,a\tpt_{d-\frac{3}{2}};\frac{3}{2},a\tpt_{0},\mydots,a\tpt_{d-1};1}{\Delta m=1} \nonumber \\[0.5em]
 & ~~~-\frac{3\sqrt{\pi}\GammaFunc{d}}{16\GammaFunc{d+\frac{3}{2}}} \Hgf{d+3}{d+2}{ 1,\frac{5}{4},\frac{7}{4},a\tpt_{-\frac{1}{2}},\mydots,a\tpt_{d-\frac{3}{2}};\frac{3}{2},2,a\tpt_1,\mydots,a\tpt_{d};1}{\Delta m=2},
 \end{align}
 which in even spacetime dimensions simplifies to
 \begin{align}
  \ln \left(  4 A_d\right)&= 2-\frac{\sqrt{\pi}}{4} \frac{\GammaFunc{\frac{d}{2}}}{\GammaFunc{\frac{d}{2}+\frac{3}{2}}} \Hgf{\frac{d}{2}+2}{\frac{d}{2}+1}{\frac{3}{2},1,2a\tpt_{-\frac{1}{2}},\mydots,2a\tpt_{\frac{d}{2}-\frac{3}{2}};2,2a\tpt_{1},\mydots,2a\tpt_{\frac{d}{2}};1}{\Delta m=1} \hspace{-2em}.
\end{align}
 In \cite{Fischler:2012ca}, the authors derived this subleading term as a converging power series.\footnote{The constant $A_d$ in their conventions is $(\mathcal{A}_{d,\Delta})^{1/\Delta}$. Our result is for $A_d$ can also be obtained by constructing generalised hypergeometric functions from their result.} 
 
 From our closed-form expressions given above, we obtain the large-width behaviour of the two-point function as
\begin{align}
 \langle \mathcal{O}(t,x)\mathcal{O}(t,y)\rangle \approx A_d^\Delta \left( \frac{4\pi T}{d}\right)^{2\Delta}\exp\left(-\frac{4\pi \Delta}{d}\cdot T|x-y|\right). \label{eqn:2ptHighT}
\end{align}
This displays an   exponential decay, as  expected for the two-point function in a field theory at finite temperature.

\subsection{Results}
\label{sec:Results2pt}

We already discussed how the results for specific spacetime dimension interpolate between the zero-temperature result in the small-width limit and the large-width behaviour (c.f.\ Figure \ref{Fig:2ptFct}). In the following, let us have a look at the explicit results at all temperatures  for the three lowest spacetime dimensions. For these examples, we use the notation
\begin{align}
\Hypergeometric{ p+1}{ p}{a}{b}{u}= \,_{p+1}F_p\left(\begin{matrix} a_1,\dots,a_{p+1} \\ b_1,\dots,b_p \end{matrix} ; u \right)
\end{align}
to avoid lengthy expressions.

\subsubsection{\texorpdfstring{$\AdS_3/\CFT_2$}{AdS3/CFT2} }
\label{sec:2pt2d}
Let us start our discussion with $d=2$. The result simplifies since we consider an even boundary spacetime dimension. The length of the interval in terms of the turning point $\zs$ (c.f.\ \eqref{eqn:Width2ptEven}) is\vspace*{-1em}
\begin{align}
|\vec{x} - \vec{y}| &= 2\zs \Hgf{2}{1} {1,\frac{1}{2};\frac{3}{2};\zsh^2}{} \nonumber\\
 &= 2\zh~ \text{artanh}\zsh.
\end{align}
The hypergeometric function simplifies to the inverse hyperbolic tangent\footnote{See \eqref{eqn:Artanh} in our list of known closed expressions of hypergeometric functions.}, such that the turning point is\vspace*{-1em}
\begin{align}
\zs = \zh \tanh \left( \frac{|\vec{x} - \vec{y}|}{2\zh}\right).
\end{align}
Let us empathise that this is a special case for the two-point function in two dimensions. In general, we only obtain the results in terms of the turning point without being able to write the turning point as a function of the width. In particular, there is no simplification for our  examples in higher dimensions.

According to \eqref{eqn:Area2ptEven}, the length of the bulk geodesic is
\begin{align}
\Area{1} &=2L \ln\left(\frac{2\zs}{\epsilon}\right) +  L\zsh^2 \Hgf{2}{1}{1, 1; 2;\zsh^2}{}\nonumber, \\
&= 2L \ln\left(\frac{2\zs}{\epsilon}\right) -L \ln\left(1-\zsh^2\right),
\end{align}
where we replaced to hypergeometric function by their known form \eqref{eqn:Ln}. Also, we may write the result in terms of the distance $\ell$ and the inverse temperature $\beta = T^{-1}$ (c.f.\ \eqref{eqn:Temperature}),
\begin{align}
\Area{1} &=2L \ln \left(\frac{\beta}{\pi\epsilon}\sinh\left(\frac{|\vec{x} - \vec{y}|\pi}{\beta} \right) \right).\label{eqn:2dResult}
\end{align}
The alternative form for the minimal area in \eqref{eqn:AArea2ptEven} simplifies to the same expression.

Combining this with the saddle-point approximation for the two-point function (c.f.\ \eqref{eqn:CFT2pt}) yields
\begin{align}
	\langle \mathcal{O}(t,\vec{x})  \mathcal{O}(t,\vec{y}) \rangle &=\lim \limits_{\epsilon \rightarrow 0}  \epsilon^{- 2 \Delta } \exp\left(-\Delta \cdot\frac{\Area{1}}{L}\right), \nonumber \\ &=~ \left(\frac{\beta }{\pi } \text{sinh}\left( \frac{\pi |\vec{x}-\vec{y}|}{\beta}\right)\right)^{-2\Delta},
	\end{align}
where we are able to express our result completely in field theory observables.
\subsubsection{\texorpdfstring{$\AdS_4/\CFT_3$}{AdS4/CFT3} }
The next higher dimension brings us to $d=3$. This time the spacetime dimension is odd and the two-point functions consists of two terms. The result for the distance can be found in \eqref{eqn:Width2ptGeneral} and for $d=3$ yields
\begin{align}
\scl{1} &= |\vec{x} - \vec{y}| \nonumber \\
&= 2 \zs \, \,_5F_4\left(\begin{matrix} \frac{1}{3},\frac{2}{3},1,\frac{1}{4},\frac{3}{4} \\ \frac{1}{2},\frac{5}{6},\frac{7}{6},\frac{1}{2} \end{matrix} ; {\zsh}^{6} \right)+\frac{3 \pi z_\star}{16}  {\zsh}^{3} \,_4F_3\left(\begin{matrix} \frac{5}{6},\frac{7}{6},\frac{3}{4},\frac{5}{4} \\ \frac{4}{3},\frac{5}{3},1 \end{matrix} ; {\zsh}^{6} \right) \, .
\end{align}
Unfortunately, we are not able to invert this expression. For the geodesic length, we obtain two expression: one where the hypergeometric functions diverge in the large-width limit (c.f.\ \eqref{eqn:Area2ptGeneral}) and one where the divergent part is contained in the width $\ell$ (c.f.\ \eqref{eqn:AArea2ptGeneral}). These two forms yield
\begin{subequations}
\begin{align}
	\langle \mathcal{O}(t,\vec{x})  \mathcal{O}(t,\vec{y}) \rangle &= \lim \limits_{\epsilon \rightarrow 0}  \epsilon^{- 2 \Delta } \exp\left(-\Delta \cdot\frac{\Area{1}}{L}\right),\nonumber \\
	&=\frac{1}{\left(2 \, {z_\star} \right)^{2\Delta}} \exp\left[ - \frac{2\Delta}{5} \, {\zsh}^{6} \,_6F_5\left(\begin{matrix} 1,\frac{4}{3},\frac{5}{3},\frac{5}{4},\frac{7}{4},1 \\ \frac{7}{6},\frac{3}{2},\frac{11}{6},\frac{3}{2},2 \end{matrix} ; {\zsh}^{6} \right) 
	\right. \nonumber \\
	&~~~-\left. \frac{\pi\Delta }{4} \,  {\zsh}^{3} \,_5F_4\left(\begin{matrix} \frac{1}{2},\frac{5}{6},\frac{7}{6},\frac{3}{4},\frac{5}{4} \\ \frac{2}{3},\frac{4}{3},1,\frac{3}{2} \end{matrix} ; {\zsh}^{6} \right)\right]
\end{align}
and equivalently
\begin{align}
	\langle \mathcal{O}(t,\vec{x})  \mathcal{O}(t,\vec{y}) \rangle &=\left(\frac{e}{2\zs }\right)^{2\Delta} e^{- \ell\Delta/\zs}~\exp\left[-\frac{\pi \Delta }{16 }  \zsh^3 \,_5F_4\left(\begin{matrix} \frac{1}{2},\frac{3}{4},\frac{5}{6},\frac{7}{6},\frac{5}{4} \\ 1,\frac{4}{3},\frac{3}{2},\frac{5}{3} \end{matrix} ;  \zsh^6\right) \nonumber\right. \\ &\left.~~~- \frac{2\Delta}{35 } \zsh^6 \,_6F_5\left(\begin{matrix} 1,1,\frac{5}{4},\frac{4}{3},\frac{5}{3},\frac{7}{4} \\ \frac{3}{2},\frac{3}{2},\frac{11}{6},2,\frac{13}{6} \end{matrix} ; \zsh^6 \right)\right]
\end{align}
\end{subequations}
for the two-point function (c.f.\ \eqref{eqn:CFT2pt}). The first form is most suited for the small-width behaviour (i.e.\ $\zs/\zh \rightarrow 0$, where $\ell \approx 2 \zs$); the expression in the exponential is subleading and we obtain the characteristic power-law behaviour. The second form is most suited for the large-width limit (i.e.\ $\zs \approx \zh$), where we see the characteristic exponential decay of the two-point function.

\subsubsection{\texorpdfstring{$\AdS_5/\CFT_4$}{AdS5/CFT4} }
Let us now turn to the result for four spacetime dimensions, as relevant to the dual of $\mathcal{N}=4 ~ SU(N)$ supersymmetric Yang-Mills theory. Since we have an even boundary dimension, the results simplify compared to the result in three dimensions. For the distance $\ell$ we obtain (c.f.\ \eqref{eqn:Width2ptEven})
\begin{align}
\scl{1} &= |\vec{x} - \vec{y}| \nonumber \\ &= 2 \, {z_\star} \,_3F_2\left(\begin{matrix} \frac{1}{2},\frac{1}{2},1 \\ \frac{3}{4},\frac{5}{4} \end{matrix} ; \zsh^4 \right).
\end{align}
We obtain two equivalent results for the geodesic length (c.f.\ \eqref{eqn:Area2ptEven} and \eqref{eqn:AArea2ptEven}). In terms of field-theory observables, this yields (c.f.\ \eqref{eqn:CFT2pt})
\begin{subequations}
\begin{align}
\langle \mathcal{O}(t,\vec{x})  \mathcal{O}(t,\vec{y}) \rangle &= \lim \limits_{\epsilon \rightarrow 0}  \epsilon^{- 2 \Delta } \exp\left(-\Delta \cdot\frac{\Area{1}}{L}\right),\nonumber \\
&=\frac{1}{(2\zs)^{2\Delta} }\exp\left[-\frac{2\Delta \,  {z_\star}^{4}}{3 \, {z_h}^{4}} \,_4F_3\left(\begin{matrix} 1,1,\frac{3}{2},\frac{3}{2} \\ \frac{5}{4},\frac{7}{4},2 \end{matrix} ; \frac{{z_\star}^{4}}{{z_h}^{4}} \right) \right],\\
&= \left(\frac{e}{2\zs }\right)^{2\Delta} e^{-\Delta \ell/\zs}~\exp\left[-\frac{2 \Delta }{15 }\zsh^4 \,_4F_3\left(\begin{matrix} 1,1,\frac{3}{2},\frac{3}{2} \\ \frac{7}{4},2,\frac{9}{4} \end{matrix} ; \zsh^4 \right)\right].
\end{align}
\end{subequations}The generalised hypergeometric functions in the first line diverge at unit argument, whereas the ones in the second line converge and the divergent behaviour is completely captured by the distance $\ell$. Therefore, the form in the second line is more suitable for analysing the large-width behaviour.

  \section{Spatial Wilson loop}
\label{sec:WL}
A further important non-local observable in gauge theories is the expectation value of Wilson loops (see \cite{Makeenko:2009dw} for a review). It is proportional to the phase factor associated to the parallel transport of a quark around a closed loop, $\mathcal{C}$
\begin{align}
 W (\mathcal{C}) &= \frac{1}{N}\Tr\left( \mathcal{P} \exp \oint_\mathcal{C} dx^\mu A_\mu\right),
\end{align}
where $\mathcal{P}$ is the path-ordering operator and $A_\mu$ is the gauge field. The expectation value of this gauge invariant quantity is an order parameter for confinement. According to \cite{Maldacena:1998im}, the holographic dual is
\begin{align}
 \langle W (\mathcal{C})  \rangle &= \exp\left(-S_{NG} \right), \label{eqn:CFTWL}
\end{align}
where $S_{NG}$ is the on-shell Nambu-Goto action of a string with boundary $\mathcal{C}$,
\begin{align}
 S_{NG} &= \frac{1}{2\pi \alpha'} \int d\tau d\sigma \sqrt{|\det g_{\mu \nu } \partial_\chi x^\mu \partial_\beta x^\nu |}, \nonumber \\
 &=  \frac{1}{2\pi \alpha'} \Area{2},
\end{align}
where $\alpha'$ is the square-root of the string length $l_s$. Geometrically, this is the area $\Area{2}$ of the two-dimensional extremal surface anchored at $\mathcal{C}$. 

\begin{figure}[b]
  \center
  \def\svgwidth{0.6\linewidth}
  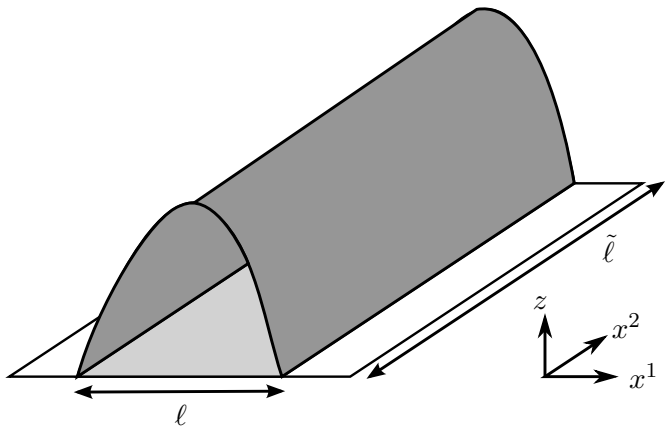
  \caption{Holographic calculation of the Wilson loop.}
  \label{fig:HWL}
\end{figure}

In this section we consider a Wilson loop of width $\ell$ in one and width $\tilde \ell \gg \ell$ in the other spatial direction (c.f.\  Figure \ref{fig:HWL}). 
Let us emphasise that this is a spatial and not the often considered temporal Wilson loop. However, we can double-Wick rotate our metric (i.e.\ $t \rightarrow i \tau,~x_1 \rightarrow i \hat t$~) and obtain
\begin{subequations}
	\begin{align}
	ds^2 &= \frac{L^2}{z^2} \left( b(z) d\tau^2 + \frac{dz^2}{b(z)} - d\hat{t}^2 +\sum\limits_{i=2}^{d-1} d x_i^2 \right), \\
	b(z) &= 1-\frac{z^d}{z_h^d}, 
	\end{align}\label{eqn:Solitonmetric}%
\end{subequations}
which is the AdS-soliton metric (c.f.\ \cite{Horowitz:1998ha,CasalderreySolana:2011us,Brandhuber:1998er}). This is a zero-temperature theory with one compact spatial direction. The geometry ends smoothly at $\zh$, where the compact $\tau$ direction shrinks to zero. The dual theory is confining (see also \cite{Sonnenschein:1999if}). $\zh$ can be identified with the inverse of the QCD scale,
\begin{align}
  \zh = \Lambda^{-1}.
\end{align}
In the IR limit the $\tau$ cycle shrinks to zero. The effective IR theory is a non-conformal $d-1$ dimensional pure gauge theory, since the fermions and the scalars acquire mass of order $\Lambda$.

For this configuration, the Wilson loop is temporal.  Since we consider the limit of infinite length $\tilde \ell \gg \ell$, it is related to the quark-antiquark potential $V_{q}$ by
\begin{align}
  V_{q} &= -\lim\limits_{\tilde \ell \rightarrow \infty} \frac{\ln \langle W \rangle}{\tilde \ell} =  -\lim\limits_{\tilde \ell \rightarrow \infty} \frac{S_{NG}}{\tilde \ell} \, .
\end{align}
This expression is UV-divergent due to the infinite quark mass and hence requires regularisation. For the choice of finite counterterms we follow the  arguments given for the  finite-temperature case in \cite{Ewerz:2016zsx}: The physics in the UV, i.e.\ in the $\ell \Lambda\ll 1$ limit, is not affected by the confinement scale $\Lambda$. We therefore regularise by just subtracting the $1/\epsilon$-term to avoid a $\Lambda$-dependent $\mathcal{O}(\ell)^0$ term,\footnote{The final results of this section will provide further justification for this choice of regularisation.} 
\begin{align}
V_q &= \frac{S_{NG}}{\tilde \ell}- \frac{L^2}{\pi \alpha' \epsilon}, \nonumber\\
&=\frac{\mathcal{A}}{2\pi \alpha '\tilde \ell}- \frac{L^2}{\pi \alpha' \epsilon}  \label{eqn:PotentialRegularisation} 
\end{align}
The subtracted term depends on the UV cut-off $\epsilon$, and on the AdS-radius $L$ and on $\alpha'$, but not on the width of the strip $\ell$.
As shown in \cite{Brandhuber:1998er}, the AdS-soliton geometry is confined, i.e.\ the potential scales as 
\begin{align}
V_q &= \underbrace{\text{linear~term}}_{\propto \ell}-2\kappa+\cdots
\end{align}
for large quark-antiquark distance $\ell\Lambda \gg 1$.\footnote{The quark-antiquark potential is derived from the temporal Wilson loop. This is the reason we consider AdS-Soliton instead of AdS-Schwarzschild. For a spatial Wilson loop, the potential derived in the analogous way is called pseudo-potential and connected to string tension and drag force (see e.g.\ \cite{Andreev:2017bvr,Andreev:2006eh}).}

In the following, the derive an analytical result for the quark-antiquark potential $V_q$. In particular, we consider the large-width limit (i.e.\ $\ell\Lambda \gg 1$) and derive an analytical expression for the subleading term $\kappa$.

\subsection[Analytical result]{Analytical result for the Wilson loop}
The holographic Wilson loop is calculated by determining the minimal area of the attached dimension two surface (i.e.\ it corresponds to $n=2$ in our notation). The calculation of the minimal area for general dimension $n$ is given in appendix \ref{sec:MinimalSurface}. Here we consider the case $n=2$ for general spacetime dimension $d$.   The results are of the form of a finite sum containing generalised hypergeometric functions.\footnote{For a review of these functions, see appendix \ref{sec:AppHGF}.} In the case that the greatest common denominator of the spacetime dimension $d$ and four is larger than one, this result simplifies further. To keep the result general, it is convenient to introduce the greatest common denominator $\chi$
by 
\begin{align}
 \chi &= \begin{cases}
            4 & \text{for $d$ divisible by four,} \\
            2 & \text{for $d$ even,} \\
            1 & \text{else}.
           \end{cases}
\end{align}

Let us introduce the parameters (c.f.\ \eqref{eqn:Parameters})
\begin{subequations}
\begin{align}
  a\wl_i =& \frac{\chi}{4d}\left( \Delta m d +1 + 4 i\right),\\
  b\wl_j =& \frac{\chi}{4}\left(\Delta m +j\right),
\end{align}
\end{subequations}
where $\Delta m$ is an integer. Applying the result for the width \eqref{eqn:SolutionAsHGFWidth} to the two-dimensional case yields
\begin{align}
	\scl{2} &= \frac{\sqrt{\pi}\zs}{2} \sum\limits_{\Delta m=0}^{\frac{4}{\chi}-1}\frac{1}{\Delta m!} \Pochhammer{\frac{1}{2}}{\Delta m} \zsh^{\Delta m d} \frac{\GammaFunc{\frac{d}{\chi} a\wl_{1/2}}}{\GammaFunc{\frac{d}{\chi} a\wl_1}}   \label{eqn:HGFWidthWL}\\
	&~~~\times\Hgf{\frac{4+d}{\chi}+1}{\frac{4+d}{\chi}} {1, a\wl_{\frac{1}{2}},\mydots,  a\wl_{\frac{d}{\chi}-\frac{1}{2}}, b\wl_{\frac{1}{2}},\mydots,  b\wl_{\frac{4}{\chi}-\frac{1}{2}}; a\wl_1,\mydots,  a\wl_{\frac{d}{\chi}},  b\wl_1,\mydots,   b\wl_{\frac{4}{\chi}};\zsh^{\frac{4d}{\chi}}}{}. \nonumber
\end{align}
The parameter $\zs$ is the turning point of the minimal surface and $\zh$ is the position of the horizon (c.f.\ \eqref{eqn:Temperature}).
In the same way, we use \eqref{eqn:SolutionAsHGFArea} for the minimal area and obtain the quark-antiquark potential \eqref{eqn:CFTWL},
\begin{align}
	V_q &=   \frac{\sqrt{\pi}L^2 }{4\pi \alpha'}\frac{1}{\zs} \sum\limits_{\Delta m=0}^{\frac{4}{\chi}-1} \frac{1}{\Delta m!} \Pochhammer{\frac{1}{2}}{\Delta m} \zsh^{\Delta m d} \frac{\GammaFunc{\frac{d}{\chi} a\wl_{-1/2}}}{\GammaFunc{\frac{d}{\chi} a\wl_0}}    \label{eqn:HGFAreaWL}\\
	&~~~\times\Hgf{\frac{4+d}{\chi}+1}{\frac{4+d}{\chi}}{1, a\wl_{-\frac{1}{2}},\mydots,  a\wl_{\frac{d}{\chi}-\frac{3}{2}},   b\wl_{\frac{1}{2}},\mydots,b\wl_{\frac{4}{\chi}-\frac{1}{2}}; a\wl_0,\mydots, a\wl_{\frac{d}{\chi}-1}, b\wl_1,\mydots,  b\wl_{\frac{4}{\chi}};\zsh^{\frac{4d}{\chi}}}{}. \nonumber
\end{align}
These sums contain four terms in general, which simplify to two or one term for dimension divisible by two or four, respectively.

The sums above are special, as they sum up to Meijer $G$-functions, which are reviewed in appendix \ref{sec:AppMGF}. For these, we introduce the parameters (c.f.\ \eqref{eqn:ParameterG})
\begin{subequations}
	\begin{align}
	\hat a\wl_i &= \frac{\chi}{d}i, \\
	\hat b\wl_j &= \frac{\chi}{4} \left(j + \frac{1}{d}\right).
	\end{align}
\end{subequations}
Using the results in \eqref{eqn:ResultAsMGF}, the result in Meijer $G$-functions is

{\renewcommand*{\arraystretch}{1.5}
	\begin{align}
	\scl{2} = \frac{2 \pi \zh}{\sqrt{4d}}~ G_{\frac{4+d}{\chi},\frac{4+d}{\chi}}^{\,\frac{4}{\chi},\frac{d}{\chi}}\!\left(\left.{\begin{matrix}\hat a\wl_{\frac{1}{2}},\dots ,\hat a\wl_{\frac{d}{\chi} -\frac{1}{2}}, \hat b\wl_{\frac{1}{2}},\dots,\hat b\wl_{4/\chi-\frac{1}{2}}\\\hat b\wl_{0},\dots,\hat b\wl_{\frac{4}{\chi}-1},\hat a\wl_{0},\dots ,\hat a\wl_{\frac{d}{\chi} -1}\end{matrix}}\;\right|\,\zsh^{\frac{4d}{\chi}}\right)\label{eqn:GWidthWL}
	\end{align}
	for the width of the strip and
	\begin{align}
	V_q &=  \frac{L^2}{\sqrt{2d} \alpha'} \frac{\zh}{\zs^2} ~ G_{\frac{4+d}{\chi},\frac{4+d}{\chi}}^{\,\frac{4}{\chi},\frac{d}{\chi}}\!\left(\left.{\begin{matrix}\hat a\wl_{\frac{3}{2}},\dots ,\hat a\wl_{\frac{d}{\chi} +\frac{1}{2}}, \hat b\wl_{\frac{1}{2}},\dots,\hat b\wl_{\frac{4}{\chi}-\frac{1}{2}}\\\hat b\wl_{0},\dots,\hat b\wl_{\frac{4}{\chi}-1},\hat a\wl_{1},\dots ,\hat a\wl_{\frac{d}{\chi} }\end{matrix}}\;\right|\,\zsh^{\frac{4d}{\chi}}\right)\label{eqn:GAreaWL}
	\end{align}
}%
	for the quark-antiquark potential. The representation in terms of the Meijer $G$-function has the advantage of being more compact, while the representation in terms of hypergeometric functions is more useful for explicit computations.

Let us turn to the small-width limit (i.e.\ $\zs \ll \zh$), which can easily be derived from the small-argument expansion of the hypergeometric functions. We obtain
\begin{align}
V_q &= - \frac{4\pi^2 L^2}{\alpha' \GammaFunc{\frac{1}{4}}^4} ~\frac{1}{\ell}\left[1+ \mathcal{O}(\ell \Lambda)^d \right]. \label{eqn:WLT0}
\end{align}
Following the arguments of \cite{Ewerz:2016zsx}, the UV physics should not depend on $\Lambda$, which fixes the regularisation. Subtracting an additional finite term would yield an unwanted $\Lambda$-dependence of the UV limit of the quark-antiquark potential. In the next section, we turn to  the opposite limit $\ell \Lambda \gg 1$.

\subsection[Large-width behaviour]{Large-width behaviour of the Wilson loop}

In contrast to \eqref{eqn:WLT0}, the large-width behaviour (i.e.\ $\zs \rightarrow \zh$) is far more involved since both the width and the potential diverge in this limit. Considering the general result \eqref{eqn:HGFAreaWL} in terms of a power series (as reviewed in appendix \ref{sec:MinimalSurfacePS}), it  can no longer be approximated by taking a finite number of terms: the divergence is due to the divergence of the power series and not captured by a finite number of terms. Our result allows for accurate results for arbitrary quark-antiquark distance (i.e.\ $\zs$ arbitrary close to $\zh$). For this, we use the results derived in appendix \ref{sec:LowHighT}. There, we use properties of hypergeometric functions or Meijer $G$-functions to split the result for the area into a finite part and into a part which diverges as the  width $\ell$  is taken to infinity. In particular, we show that the leading contribution is proportional to $\ell \Lambda^2$ and determine the subleading contribution.

We use \textit{contiguous relations} for the generalised hypergeometric functions \eqref{eqn:ContiguousRel} or equivalently \textit{recurrence relations} \eqref{eqn:ReccurenceMG} for the Meijer $G$-functions, which are reviewed in the appendix. Following the detailed calculation in section \ref{sec:LowHighT}, the quark-antiquark potential may be written  as sum of a  term divergent in the large-width limit (i.e.\ $\zs\rightarrow \zh$) and of a finite expression involving generalised hypergeometric functions, as derived in \eqref{eqn:SolutionHT},
\begin{subequations}
\begin{align}
 V_q &= \frac{L^2 }{2 \zs^2 \pi \alpha'}\cdot\scl{2}+  \frac{\sqrt{\pi}L^2 }{8 \pi \alpha'}~\frac{1}{\zs} \sum\limits_{\Delta m=0}^{\frac{4}{\chi}-1} \frac{1}{\Delta m!} \Pochhammer{\frac{1}{2}}{\Delta m} \zsh^{\Delta m d} \frac{\GammaFunc{\frac{d}{\chi} a\wl_{-1/2}}}{\GammaFunc{\frac{d}{\chi} a\wl_1}}  
 \label{eqn:HGFAAreaWL} \\
 &~~~\times \Hgf{\frac{d+4}{\chi}+1}{\frac{d+4}{\chi}
 	}{1,  a\wl_{-\frac{1}{2}},\mydots, a\wl_{\frac{d}{\chi}-\frac{3}{2}}, b\wl_\frac{1}{2},\mydots, b\wl_{\frac{4}{\chi}-\frac{1}{2}};  a\wl_1,\mydots, a\wl_{\frac{d}{\chi}}, b\wl_1,\mydots, b\wl_{\frac{4}{\chi}};\zsh^{\frac{4d}{\chi}}}{} . \nonumber
\end{align}
Isolating the divergent terms, this may be rewritten as
{\renewcommand*{\arraystretch}{1.5}
\begin{align}
 V_q &=   \frac{ L^2}{2\pi\alpha'\zs^2}\cdot \scl{2}+ \frac{\chi  L^2 }{\sqrt{d^3} 2 \pi \alpha'}~ \frac{ \zh}{\zs^2}   \label{eqn:GAAreaWL}\\
&~~~\times~ G_{\frac{4+d}{\chi},\frac{4+d}{\chi}}^{\,\frac{4}{\chi},\frac{d}{\chi}}\!\left(\left.{\begin{matrix}\hat a\wl_{\frac{3}{2}},\dots ,\hat a\wl_{\frac{d}{\chi} +\frac{1}{2}}, \hat b\wl_{\frac{1}{2}},\dots,\hat b\wl_{\frac{4}{\chi}-\frac{1}{2}}\\\hat b\wl_{0},\dots,\hat b\wl_{\frac{4}{\chi}-1},\hat a\wl_{0},\dots ,\hat a\wl_{\frac{d}{\chi}-1 }\end{matrix}}\;\right|\,\zsh^{\frac{4d}{\chi}}\right)\nonumber
\end{align}}\label{eqn:AreaWL}%
\end{subequations}
in terms of a Meijer $G$-function (see \eqref{eqn:SolutionHTG}).
The parameters for the hypergeometric functions and the Meijer $G$-function are shifted compared to \eqref{eqn:HGFAreaWL} and \eqref{eqn:GAreaWL}. This is the important difference to the previous result: the second term in the previous two expressions \eqref{eqn:AreaWL} is finite for $\zs \rightarrow \zh$, whereas all terms in \eqref{eqn:HGFAreaWL} and \eqref{eqn:GAreaWL} diverge.

Furthermore, corrections to the turning point decay exponentially in the large-width limit (as can be seen from the behaviour of the hypergeometric functions, c.f.\ \eqref{eqn:HgfDivergences}). The large-width behaviour of the quark-antiquark potential is therefore 
\begin{align}
 V_q &= \frac{L^2  \Lambda^2}{2\pi\alpha' } \cdot\scl{2} -2 \kappa, 
\end{align}
where the constant $\kappa$ is
\begin{subequations}
\begin{align}
\kappa &= -  \frac{\sqrt{\pi}L^2 }{16 \pi \alpha'}~\Lambda \sum\limits_{\Delta m=0}^{\frac{4}{\chi}-1} \frac{1}{\Delta m!} \Pochhammer{\frac{1}{2}}{\Delta m}\frac{\GammaFunc{\frac{d}{\chi} a\wl_{-1/2}}}{\GammaFunc{\frac{d}{\chi} a\wl_1}}  
 \\[0.5em]
&~~~\times \Hgf{\frac{d+4}{\chi}+1}{\frac{d+4}{\chi}
}{1,  a\wl_{-\frac{1}{2}},\mydots, a\wl_{\frac{d}{\chi}-\frac{3}{2}}, b\wl_\frac{1}{2},\mydots, b\wl_{\frac{4}{\chi}-\frac{1}{2}};  a\wl_1,\mydots, a\wl_{\frac{d}{\chi}}, b\wl_1,\mydots, b\wl_{\frac{4}{\chi}};1}{}, \nonumber \\[1em]
&=- \frac{\chi  L^2 }{\sqrt{d^3} 4 \pi \alpha'}~ \Lambda ~ G_{\frac{4+d}{\chi},\frac{4+d}{\chi}}^{\,\frac{4}{\chi},\frac{d}{\chi}}\!\left(\left.{\begin{matrix}\hat a\wl_{\frac{3}{2}},\dots ,\hat a\wl_{\frac{d}{\chi} +\frac{1}{2}}, \hat b\wl_{\frac{1}{2}},\dots,\hat b\wl_{\frac{4}{\chi}-\frac{1}{2}}\\\hat b\wl_{0},\dots,\hat b\wl_{\frac{4}{\chi}-1},\hat a\wl_{0},\dots ,\hat a\wl_{\frac{d}{\chi}-1 }\end{matrix}}\;\right|\,1\right).
\end{align}%
\end{subequations}
This term has the interpretation of a finite renormalisation of the quark mass within the confined meson: similarly to the subtracted term in \eqref{eqn:PotentialRegularisation}, it does not depend on the distance $\ell$ of the quark-antiquark pair.

The large-width behaviour of the expectation value of the Wilson loop is therefore (cf.\ \eqref{eqn:CFTWL} and \eqref{eqn:PotentialRegularisation})
\begin{align}
  \langle \mathcal{W} \rangle \approx  \exp\left(2\tilde \ell  \kappa -\frac{L^2}{\alpha'} \frac{\tilde \ell}{\pi \epsilon} \right) \exp\left(-\frac{L^2}{2\pi\alpha'} \cdot \tilde \ell \scl{2} \Lambda^2\right).
\end{align}
$\frac{L^2}{\alpha'}$ is related to the 't Hooft coupling of the field theory.

\subsection{Results}
\begin{figure}
	\centering
	{\PGFcommands
		\import{PGF/}{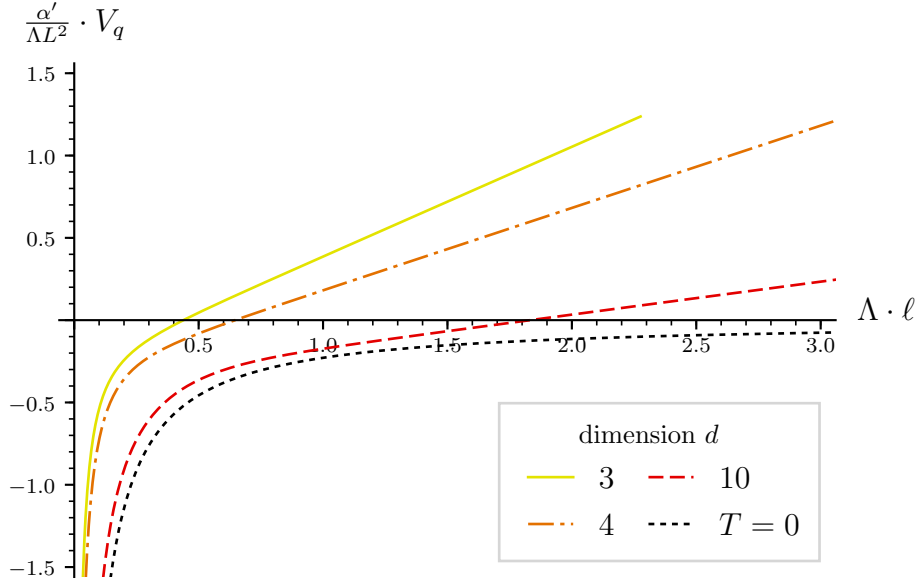}}
	\caption{Quark-antiquark potential.}
	\label{Fig:OnShellAction}
\end{figure}

Figure \ref{Fig:OnShellAction} shows the result for the potential for $d=3,4$ and $d=10$. Since we need two spatial dimensions for the spatial Wilson loop, we consider only theories with $d>2$.
In the following, let us have a look at the results for three and four dimensions. For these examples, we use the notation
\begin{align}
\Hypergeometric{ p+1}{ p}{a}{b}{u}= \,_{p+1}F_p\left(\begin{matrix} a_1,\dots,a_{p+1} \\ b_1,\dots,b_p \end{matrix} ; u \right)
\end{align}
to avoid lengthy expressions.
\subsubsection{\texorpdfstring{$\AdS_4/\CFT_3$}{AdS4/CFT3} }
\label{sec:WL3d}
For the Wilson loop expectation value, the result can be expressed in generalised hypergeometric functions or Meijer $G$-functions. We start with the result in terms of hypergeometric functions. Inserting $d=3$ into  \eqref{eqn:HGFWidthWL}, we obtain for the width of the strip
	\begin{align}
	\scl{2} &= \frac{\zs}{6} \, {\zsh}^{9} \,_7F_6\left(\begin{matrix} \frac{7}{8},1,\frac{9}{8},\frac{4}{3},\frac{11}{8},\frac{13}{8},\frac{5}{3} \\ \frac{7}{6},\frac{5}{4},\frac{3}{2},\frac{3}{2},\frac{7}{4},\frac{11}{6} \end{matrix} ; {\zsh}^{12} \right) \nonumber \\
	&~~~+ \frac{\zs \, \sqrt{\pi}   \Gamma\left(\frac{9}{4}\right)}{7\Gamma\left(\frac{3}{4}\right)} {\zsh}^{6}\,_6F_5\left(\begin{matrix} \frac{5}{8},\frac{7}{8},\frac{13}{12},\frac{9}{8},\frac{11}{8},\frac{17}{12} \\ \frac{11}{12},\frac{5}{4},\frac{5}{4},\frac{3}{2},\frac{19}{12} \end{matrix} ; {\zsh}^{12} \right) \nonumber \\
	&~~~ + \frac{\zs \, \pi}{8} {\zsh}^{3} \,_6F_5\left(\begin{matrix} \frac{3}{8},\frac{5}{8},\frac{5}{6},\frac{7}{8},\frac{9}{8},\frac{7}{6} \\ \frac{2}{3},\frac{3}{4},1,\frac{5}{4},\frac{4}{3} \end{matrix} ; {\zsh}^{12} \right) \nonumber \\ &~~~
	+ \frac{2\zs \, \sqrt{\pi}  \Gamma\left(\frac{3}{4}\right)}{\Gamma\left(\frac{1}{4}\right)} \,_6F_5\left(\begin{matrix} \frac{1}{8},\frac{3}{8},\frac{7}{12},\frac{5}{8},\frac{7}{8},\frac{11}{12} \\ \frac{5}{12},\frac{1}{2},\frac{3}{4},\frac{3}{4},\frac{13}{12} \end{matrix} ; {\zsh}^{12} \right).\label{eqn:Widthd3}
	\end{align}
The quark-antiquark potential is calculated from \eqref{eqn:HGFAreaWL} and for $d=3$ yields 
\begin{align}
V_q &=  \frac{L^2  5}{48\pi \alpha' \, {z_\star}} \, {\zsh}^{9} \,_7F_6\left(\begin{matrix} \frac{2}{3},1,\frac{4}{3},\frac{7}{8},\frac{9}{8},\frac{11}{8},\frac{13}{8} \\ \frac{5}{6},\frac{7}{6},\frac{3}{2},\frac{5}{4},\frac{3}{2},\frac{7}{4} \end{matrix} ; {\zsh}^{12} \right) \nonumber \\ &~~~
+ \frac{ \,L^2  \Gamma\left(\frac{9}{4}\right)}{10\sqrt{\pi} \alpha' \, {z_\star}\Gamma\left(\frac{3}{4}\right)}{\zsh}^{6} \,_6F_5\left(\begin{matrix} \frac{5}{12},\frac{13}{12},\frac{5}{8},\frac{7}{8},\frac{9}{8},\frac{11}{8} \\ \frac{7}{12},\frac{11}{12},\frac{5}{4},\frac{5}{4},\frac{3}{2} \end{matrix} ; {\zsh}^{12} \right)  \nonumber \\
&~~~  + \frac{L^2}{8  \alpha' \, {z_\star}} {\zsh}^{3} \,_6F_5\left(\begin{matrix} \frac{1}{6},\frac{5}{6},\frac{3}{8},\frac{5}{8},\frac{7}{8},\frac{9}{8} \\ \frac{1}{3},\frac{2}{3},\frac{3}{4},1,\frac{5}{4} \end{matrix} ; {\zsh}^{12} \right) \nonumber \\ &~~~ - \frac{ \,L^2    \Gamma\left(\frac{3}{4}\right)}{\sqrt{\pi} \alpha' \, {z_\star}\Gamma\left(\frac{1}{4}\right)}\,_6F_5\left(\begin{matrix} -\frac{1}{12},\frac{7}{12},\frac{1}{8},\frac{3}{8},\frac{5}{8},\frac{7}{8}\\ \frac{1}{12},\frac{5}{12},\frac{3}{4},\frac{1}{2},\frac{3}{4} \end{matrix} ; {\zsh}^{12} \right).\label{eqn:Potd3}
\end{align}
The hypergeometric functions in this expression all diverge when the turning point approaches the horizon $\zs \rightarrow \zh$. Equivalently, we may write the potential as (c.f.\ \eqref{eqn:HGFAAreaWL})
\begin{align}
V_q &=  \frac{\scl{2} L^2}{2\pi\alpha'\zs^2}+\frac{L^{2}}{16 \, \zs\alpha'}\zsh^3 \,_6F_5\left(\begin{matrix} \frac{1}{6},\frac{3}{8},\frac{5}{8},\frac{5}{6},\frac{7}{8},\frac{9}{8} \\ \frac{2}{3},\frac{3}{4},1,\frac{5}{4},\frac{4}{3} \end{matrix} ; \zsh^{12} \right)  \nonumber \\
&~~~+ \frac{\, L^{2}\Gamma\left(\frac{5}{4}\right)}{28 \, \sqrt{\pi} \zs\Gamma\left(\frac{3}{4}\right)\alpha'}\zsh^6 \,_6F_5\left(\begin{matrix} \frac{5}{12},\frac{5}{8},\frac{7}{8},\frac{13}{12},\frac{9}{8},\frac{11}{8} \\ \frac{11}{12},\frac{5}{4},\frac{5}{4},\frac{3}{2},\frac{19}{12} \end{matrix} ; \zsh^{12} \right)  \nonumber \\
&~~~+ \frac{L^{2} \Gamma\left(-\frac{1}{4}\right)}{2 \, \sqrt{\pi} {z_\star} \Gamma\left(\frac{1}{4}\right)\alpha'} \,_6F_5\left(\begin{matrix} -\frac{1}{12},\frac{1}{8},\frac{3}{8},\frac{7}{12},\frac{5}{8},\frac{7}{8} \\ \frac{5}{12},\frac{1}{2},\frac{3}{4},\frac{3}{4},\frac{13}{12} \end{matrix} ; \zsh^{12} \right)  \nonumber \\
&~~~ + \frac{L^{2}  }{48 \, \pi \zs\alpha'} \zsh^9 \,_7F_6\left(\begin{matrix} \frac{2}{3},\frac{7}{8},1,\frac{9}{8},\frac{4}{3},\frac{11}{8},\frac{13}{8} \\ \frac{7}{6},\frac{5}{4},\frac{3}{2},\frac{3}{2},\frac{7}{4},\frac{11}{6} \end{matrix} ; \zsh^{12} \right).\label{eqn:Potd3Alt}
\end{align}
Due to a unit shift of one of the parameters, the hypergeometric functions converge and the divergent behaviour of the potential is captured by the first term. This term yields the linear behaviour for large quark-antiquark distance and describes the confining behaviour. The remaining terms yield a constant subleading contribution due to the renormalisation of the quark mass within the meson.

Meijer $G$-functions may be used to simplify these results. This yields
\begin{align}
  \scl{2} &=  \frac{\pi \zh}{\sqrt{3}}~ G_{7,7}^{\,4,3}\!\left(\left.{\begin{matrix}\frac{1}{6},\frac{1}{2},\frac{5}{6}, \frac{5}{24},\frac{11}{24},\frac{17}{24},\frac{23}{24},\\\frac{1}{12},\frac{1}{3},\frac{7}{12},\frac{5}{6},0,\frac{1}{3},\frac{2}{3}\end{matrix}}\;\right|\,\zsh^{12}\right)
  \end{align}
  for the width (c.f.\ \eqref{eqn:GWidthWL}) and
  \begin{align}
  V_q  &=   \frac{L^2}{2\pi \alpha' \sqrt{3}} \frac{\zh}{\zs^2} ~ G_{7,7}^{\,4,3}\!\left(\left.{\begin{matrix}\frac{1}{2},\frac{5}{6},\frac{7}{6},\frac{5}{24},\frac{11}{24},\frac{17}{24},\frac{23}{24}\\\frac{1}{12},\frac{1}{3},\frac{7}{12},\frac{5}{6},\frac{1}{3}, \frac{2}{3}, 1\end{matrix}}\;\right|\,\zsh^{12}\right) \nonumber\\
  &= \frac{\scl{2} L^2}{2\pi\alpha'\zs^2} +  \frac{L^2}{12\pi \alpha' \sqrt{3}} \frac{\zh}{\zs^2} ~ G_{7,7}^{\,4,3}\!\left(\left.{\begin{matrix}\frac{1}{2},\frac{5}{6},\frac{7}{6},\frac{5}{24},\frac{11}{24},\frac{17}{24},\frac{23}{24}\\\frac{1}{12},\frac{1}{3},\frac{7}{12},\frac{5}{6},0,\frac{1}{3}, \frac{2}{3}\end{matrix}}\;\right|\,\zsh^{12}\right) 
\end{align}
for the quark-antiquark potential (c.f.\ \eqref{eqn:GAreaWL} and \eqref{eqn:GAAreaWL}). For the quark-antiquark potential we again obtain two forms: for the first one the Meijer $G$-function diverges for $\zs \rightarrow \zh$, whereas the one in the second form converges. This in particular allow to take the large-width limit and obtain
\begin{align}
V_q 
&= \frac{ L^2 \Lambda^2}{2\pi\alpha'}\cdot \scl{2} +  \frac{L^2 \Lambda}{12\pi \alpha' \sqrt{3}}  ~ G_{7,7}^{\,4,3}\!\left(\left.{\begin{matrix}\frac{1}{2},\frac{5}{6},\frac{7}{6},\frac{5}{24},\frac{11}{24},\frac{17}{24},\frac{23}{24}\\\frac{1}{12},\frac{1}{3},\frac{7}{12},\frac{5}{6},0,\frac{1}{3}, \frac{2}{3}\end{matrix}}\;\right|\,1\right) + \cdots, \label{eqn:WLd3Constant}
\end{align}
where the second term yields an $\ell$-independent constant. While the above derived expressions are involved, we note again that they may be given in closed form.

The Wilson loop expectation value is simply the exponential of the negative quark-antiquark potential
\begin{align}
\langle W (\mathcal{C})  \rangle &= \exp\left(-\tilde \ell\cdot V_q-\frac{L^2}{\alpha'} \frac{\tilde \ell}{\pi \epsilon} \right).
\end{align}
For the quark-antiquark potential we subtracted the $\epsilon$-term, which corresponds to the quark mass. Therefore, the width-independent constant in \eqref{eqn:WLd3Constant} is a renormalisation of the quark mass confined in the meson. In section \ref{sec:WLConstant} we look at the numerical value of this constant for $d=3$ and other spacetime dimensions.

\subsubsection{\texorpdfstring{$\AdS_5/\CFT_4$}{AdS5/CFT4} }
Let us move on to four dimensions. This is relevant in particular for $AdS_5\times S^5$, which is dual to $\mathcal{N}=4~ SU(N)$ SYM theory. This string theory embedding yields the AdS-CFT dictionary where the 't Hooft coupling $\lambda $ is \cite{Maldacena:1997re}
\begin{align}
  \lambda = \frac{1}{8\pi^2} \frac{L^4}{\alpha'^2}.
\end{align}

As shown above, the result for the Wilson loop expectation value simplifies in dimension divisible by four. For the width of the strip (c.f.\ \eqref{eqn:HGFWidthWL} and \eqref{eqn:GWidthWL}) we obtain
\begin{subequations}
	\begin{align}
	\scl{2} &= \frac{\sqrt{\pi} {z_\star} \Gamma\left(\frac{3}{4}\right)}{2 \, \Gamma\left(\frac{5}{4}\right)} \,_2F_1\left(\begin{matrix} \frac{3}{4},\frac{1}{2} \\ \frac{5}{4} \end{matrix} ; \zsh^4 \right), \\[0.5em]
	&= \frac{1}{2} \, \pi {z_h} ~ G_{2,2}^{\,1,1}\!\left(\left. {\begin{matrix}\frac{1}{2}, \frac{3}{4} \\\frac{1}{4},0\end{matrix}} \;\right|\,\zsh^{4}\right)
	\end{align}
\end{subequations}
and for the quark-antiquark potential (c.f.\ \eqref{eqn:HGFAreaWL} and \eqref{eqn:GAreaWL})
\begin{subequations}
	\begin{align}
	V_q &= \frac{L^2 \tilde \ell {z_h}^{2}\Gamma\left(-\frac{1}{4}\right)}{4\sqrt{\pi} \alpha' \, {z_\star} \Gamma\left(\frac{1}{4}\right)}  \,_2F_1\left(\begin{matrix} -\frac{1}{4},\frac{1}{2} \\ \frac{1}{4} \end{matrix} ; \zsh^4 \right) , \\
	&=\frac{L^2 \tilde \ell{z_h}}{4\alpha' \, {z_\star}^{2}} ~ G_{2,2}^{\,1,1}\!\left(\left. {\begin{matrix}\frac{3}{2}, \frac{3}{4} \\\frac{1}{4},1\end{matrix}} \;\right|\,\zsh^{4}\right) .
	\end{align}
\end{subequations}
Both of these expressions for the potential diverge for $\zs \rightarrow \zh$. Alternatively, it can be written as (c.f.\ \eqref{eqn:HGFAAreaWL} and \eqref{eqn:GAAreaWL}) 
\begin{subequations}
	\begin{align}
	V_q &=\frac{\scl{2} L^2}{2\pi\alpha'\zs^2}+ \frac{L^2  {z_h}^{2}\Gamma\left(-\frac{1}{4}\right)}{2\sqrt{\pi} \alpha' \, {z_\star} \Gamma\left(\frac{1}{4}\right)}  \,_2F_1\left(\begin{matrix} -\frac{1}{4},\frac{1}{2} \\ \frac{5}{4} \end{matrix} ; \zsh^4 \right) , \\
	&=\frac{\scl{2} L^2}{2\pi\alpha'\zs^2}+\frac{L^2 {z_h}}{8\alpha' \, {z_\star}^{2}} ~ G_{2,2}^{\,1,1}\!\left(\left. {\begin{matrix}\frac{3}{2}, \frac{3}{4} \\\frac{1}{4},0\end{matrix}} \;\right|\,\zsh^{4}\right) .
	\end{align}
\end{subequations}
Here, the second term of each expression is finite in the considered limit. Therefore, the large-width expansion of the potential is
	\begin{align}
	V_q &=\frac{ L^2}{2\pi\alpha'}~\Lambda^2\scl{2}-\Lambda~\frac{L^2}{\alpha'\pi} ,
	\end{align}
where we expressed the hypergeometric function at unit argument with the known result \eqref{eqn:2F1Unit}. The linear term is the confining potential and the width-independent term is due to the renormalisation of the quark mass. We now look at this constant more closely.

\begin{figure}
	\begin{minipage}{0.3\textwidth}
		\begin{center}
			\captionof{table}{Subleading term.}\vspace{1em}
			\label{Tab:WL}
			\begin{tabular}{|l|l|}
				\hline\rule{0pt}{1.5em}  $d$& $\frac{\alpha' }{L^2 \Lambda}\cdot\kappa$ \\[0.5em] \hline
				$3$ & $0.14010$ \\
				$4$ & $0.15915$ \\
				& $=(2 \pi)^{-1}$\\
				$5$ & $0.16854$ \\
				$6$ & $0.17401$ \\
				$7$ & $0.17753$ \\
				$8$ & $0.17996$ \\
				$9$ & $0.18172$ \\
				$10$ & $0.18305$ \\
				$11$ & $0.18407$ \\\hline
			\end{tabular}
		\end{center}
	\end{minipage} \hfill
	\begin{minipage}{0.675\textwidth}
		{\PGFcommands
			\input{PGF/WLConstant.pgf}}
		\captionof{figure}{Subleading term $\kappa$.}\label{Fig:WLConstant}
	\end{minipage}
\end{figure}
\subsubsection{Subleading term in large-width limit}
\label{sec:WLConstant}
Lastly, let us have a look at the large-width behaviour. The quark-antiquark potential has the characteristic form
\begin{align}
V_q &= \underbrace{\text{linear~term}}_{\propto \ell}-2\kappa+\cdots.
\end{align}
The leading term is the linear term expected for confinement and causes the area-term in the Wilson loop expectation value. The subleading term is independent of the width $\ell$ and has therefore the same structure as the term subtracted due do the quark mass in \eqref{eqn:PotentialRegularisation}. It corresponds to a finite renormalisation of the quark mass within the confined meson. Table \ref{Tab:WL} shows the results for this constant for different spacetime dimensions. Figure \ref{Fig:WLConstant} shows these results graphically.

  \section{Entanglement entropy}
\label{sec:HEE}

The  entanglement entropy $S_{EE}$ measures entanglement for a bipartite state. The entanglement entropy for a degrees of freedom in $B$ is the von Neumann entropy of the reduced density matrix $\rho_B$
\begin{align}
S_{EE}(B) =& -\Tr_B \rho_B \ln \rho_B.\label{eqn:EE}
\end{align} 
This entropy is related to the degrees of freedom and can be used as an order parameter for quantum phase transitions. However, its calculation is a complicated quantum calculation and there barely exist exact results for higher-dimensional (i.e.\ with spacetime dimension $d>2$) field theories \cite{band2013quantum,Calabrese:2004eu,Calabrese:2005zw,Calabrese:2009qy,2017JHEP...03..089C}.

\begin{figure}[b]
	\center
	\def\svgwidth{0.45\linewidth}
	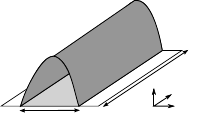
	\caption{Holographic minimal surface calculation of the entanglement entropy. \\ $B$ is the entangling region and $C$ is its complement.}
	\label{fig:HEE}
\end{figure}

In this section, we use the AdS/CFT correspondence to calculate the entanglement entropy at finite temperature. Previously, a closed form was only known for $d=2$ and not for higher-dimensional cases. The considered region is a strip $B$ with width $\ell$. To calculate the holographic entanglement entropy, we follow Ryu's and Takayanagi's conjecture \cite{Ryu:2006bv,Ryu:2006ef,Nishioka:2009un,Rangamani:2016dms}
\begin{subequations}
\begin{align}
  S_{EE}(B) &=  \frac{\Area{d-1}}{4G_N}, 
\end{align}
\label{eqn:HEE}% 
\end{subequations}
where $\Area{d-1}$ is the bulk minimal hypersurface in a constant time-slice anchored on the boundary of $B$. $G_N$ is the $(d+1)$-dimensional Newton's constant. Figure \ref{fig:HEE} shows this construction. Therefore, the calculation of the entanglement entropy reduces to the calculation of a minimal hypersurface. For regularisation, we introduce a bulk cut-off $\epsilon$. We keep the divergent terms explicitly, instead of removing them e.g.\ by introducing counterterms \cite{Taylor:2016aoi,Taylor:2016kic,Taylor:2017zzo}.

Whereas the analytical result for zero temperature is known, this is not the case for finite temperature. We are only aware of numerical results (e.g.\ \cite{Faraggi:2007fu}) and results in terms of an infinite series \cite{Fischler:2012ca}. In the following, we present our new analytical result.

\subsection[Analytical result]{Analytical result for the entanglement entropy}

Following Ryu's and Takayanagi's proposal \eqref{eqn:HEE}, the relevant surface for the entanglement entropy is a hypersurface in a constant time slice, i.e.\ it is  co-dimension one surface. For $d=2$, the considered surface is a geodesic, which is calculated in section \ref{sec:2pt} and reproduces the know result from \cite{Ryu:2006bv,Ryu:2006ef}.  We look at this case in section \ref{sec:CFT2} and disregard it here.

The result can be written in terms of power series \cite{Fischler:2012ca}. By rearranging this series, we simply the result to a finite sum containing generalised hypergeometric functions. Furthermore, we show that this finite sum is a special case of the Meijer $G$-function.  Since the calculation of the minimal surface is similar for different surface dimension, the detailed calculation is placed in appendix \ref{sec:MinimalSurface} and we summarise these results for the considered case in the following. It is convenient to introduce a parameter $\chi$, with
\begin{align}
 \chi = \begin{cases}
           2 & \text{for even dimension $d$,} \\
           1 & \text{else,}
          \end{cases}
\end{align}
for simplification. It captures whether the spacetime dimension is even or odd. This definition allows to write the result general, but at the same time in the most simplified form.

First, let us present the result in terms of generalised hypergeometric functions. Besides their argument, these functions depend on a large number of parameter (see section \ref{sec:AppHGF} for a short review of hypergeometric functions). Let us define (c.f.\ \eqref{eqn:ParameterEE} for the parameters for a surface of general dimension)
\begin{subequations}
	\begin{align} 
	a_i\EE =& \frac{\chi}{2(d-1)d}~\Big( \Delta m d +1 + 2(d-1) i\Big),\\
	b_j\EE =& \frac{\chi}{2(d-1)}~\Big(\Delta m +j\Big),
	\end{align}\label{eqn:ParameterEE}%
\end{subequations}
where $\Delta m$ is a parameter.
We use the general result from \eqref{eqn:SolutionAsHGF} and apply it to the case of a $d-1$-dimensional surface, which yields
\begin{subequations}
	\begin{align}
	&\ell = \frac{\sqrt{\pi}\zs}{d-1} \sum\limits_{\Delta m=0}^{\frac{2(d-1)}{\chi}-1}\frac{1}{\Delta m!} \Pochhammer{\frac{1}{2}}{\Delta m} \zsh^{\Delta m d} \frac{\GammaFunc{\frac{d}{\chi} a\EE_{1/2}}}{\GammaFunc{\frac{d}{\chi} a\EE_1}}   \label{eqn:WidthEntanglementEntropy}\\
	&~~~\times\Hgf{\frac{3d-2}{\chi}+1}{\frac{3d-2}{\chi}} {1, a\EE_{\frac{1}{2}},\mydots,  a\EE_{\frac{d}{\chi}-\frac{1}{2}}, b\EE_{\frac{1}{2}},\mydots,  b\EE_{\frac{2(d-1)}{\chi}-\frac{1}{2}}; a\EE_1,\mydots,  a\EE_{\frac{d}{\chi}},  b\EE_1,\mydots,   b\EE_{\frac{2(d-1)}{\chi}};\zshs^{\frac{2(d-1)d}{\chi}}}{}\nonumber
	\end{align}
	for the width of the strip and
	\begin{align}
	&S_{EE} = \frac{L^{d-1}\left(\tilde \ell/\epsilon\right)^{d-2}}{2(d-2)  G_N}   +\frac{\sqrt{\pi}L^{d-1} }{4(d-1)G_N}\frac{\tilde \ell^{d-2}}{\zs^{d-2}} \sum\limits_{\Delta m=0}^{\frac{2(d-1)}{\chi}-1} \frac{\Pochhammer{1/2}{\Delta m}}{\Delta m!}  \frac{\GammaFunc{\frac{d}{\chi} a\EE_{-1/2}}}{\GammaFunc{\frac{d}{\chi} a\EE_0}}  \zsh^{~~\mathclap{\Delta md}}\hspace*{-1em} \label{eqn:EntanglementEntropy}     \\
	&\times\hspace{-0.5em}\Hgf{\frac{3d-2}{\chi}+1}{\frac{3d-2}{\chi}}{1, a\EE_{-\frac{1}{2}},\mydots,  a\EE_{\frac{d}{\chi}-\frac{3}{2}},   b\EE_{\frac{1}{2}},\mydots,b\EE_{\frac{2(d-1)}{\chi}-\frac{1}{2}}; a\EE_0,\mydots, a\EE_{\frac{d}{\chi}-1}, b\EE_1,\mydots,  b\EE_{\frac{2(d-1)}{\chi}};\zshs^{\frac{2(d-1)d}{\chi}}}{} \nonumber
	\end{align}\label{eqn:HEEHGF}%
	for the holographic entanglement entropy. $\zs$ is the turning point of the hypersurface and $\zh$ is the position of the horizon, which is proportional to the inverse temperature (c.f.\ \eqref{eqn:Temperature}). The result for $\chi=1$ is valid for any spacetime dimension. For even $d$, the simplified version for $\chi=2$ can be used. This means the sum contains $2(d-1)$ terms, which can be simplified to $d-1$ terms for even dimension. The factor $L^{d-1}/G_N$ is proportional to the number of degrees of freedom, i.e.\ to $N^2$ for adjoint degrees of freedom of a gauge group $SU(N)$. The specific relationship can be obtained from a top-down approach or from matching the conformal anomaly.
\end{subequations}

The above sum of hypergeometric functions can be summed up to a Meijer $G$-function (see review in section \ref{sec:AppMGF}). Using the result in \eqref{eqn:ResultAsMGF}\footnote{For our detailed calculation see section \ref{sec:MinimalSurfaceMGF}.}, the width can be written as
\begin{subequations}
	{\renewcommand*{\arraystretch}{1.5}
	\begin{align}
	\ell = \frac{2 \pi \zh}{\sqrt{2{(d-1)}d}}~ G_{\frac{3d-2}{\chi},\frac{3d-2}{\chi}}^{\,\frac{{2(d-1)}}{\chi},\frac{d}{\chi}}\!\left(\left.{\begin{matrix}\hat a\EE_{\frac{1}{2}},\dots ,\hat a\EE_{\frac{d}{\chi} -\frac{1}{2}}, \hat b\EE_{\frac{1}{2}},\dots,\hat b\EE_{\frac{2(d-1)}{\chi}-\frac{1}{2}}\\\vspace{0.5em}\hat b\EE_{0},\dots,\hat b\EE_{\frac{2{(d-1)}}{\chi}-1},\hat a\EE_{0},\dots ,\hat a\EE_{\frac{d}{\chi} -1}\end{matrix}}\;\right|\,\zsh^{\frac{2{(d-1)}d}{\chi}}\right) \label{eqn:WidthEntanglementEntropyG}
	\end{align}
	and the holographic entanglement entropy as
	\begin{align}
	S_{EE} &= \frac{L^{d-1}}{2(d-2)G_N} \left(\frac{\tilde \ell}{\epsilon}\right)^{d-2}+ \frac{ \pi L^{d-1}}{\sqrt{2^3({d-1})d}\cdot G_N} \frac{\tilde \ell^{d-2}\zh}{\zs^{d-1}}   \label{eqn:EntanglementEntropyG} \\
	&~~~\times~ G_{\frac{3d-2}{\chi},\frac{3d-2}{\chi}}^{\,\frac{2({d-1})}{\chi},\frac{d}{\chi}}\!\left(\left.{\begin{matrix}\hat a\EE_{\frac{3}{2}},\dots ,\hat a\EE_{\frac{d}{\chi} +\frac{1}{2}}, \hat b\EE_{\frac{1}{2}},\dots,\hat b\EE_{\frac{2({d-1})}{\chi}-\frac{1}{2}}\\\hat b\EE_{0},\dots,\hat b\EE_{\frac{2({d-1})}{\chi}-1},\hat a\EE_{1},\dots ,\hat a\EE_{\frac{d}{\chi} }\end{matrix}}\;\right|\,\zsh^{\frac{2({d-1})d}{\chi}}\right)\nonumber.
	\end{align}%
}\label{eqn:HEEMGF}%
\end{subequations}
Following our derivation in section \ref{sec:MinimalSurfaceMGF}, the new parameters are (c.f.\ \eqref{eqn:ParameterG})
\begin{subequations}
	\begin{align}
	\hat a\EE_i &= \frac{\chi}{d}i, \\
	\hat b\EE_j &= \frac{\chi}{2(d-1)} \left(j + \frac{1}{d}\right).
	\end{align}\label{eqn:ParameterEEG}%
\end{subequations}

Since the Meijer $G$-function inherits its properties from the hypergeometric functions, the different forms of the result yield the same properties for the entanglement entropy and the width. The result as Meijer $G$-function only contains one term.

\subsection[Large-width behaviour]{Large-width behaviour of the entanglement entropy}

In the small-width limit, the leading contribution agrees with the the zero-temperature result \cite{Ryu:2006ef}, whereas the subleading term is determined by entanglement thermodynamics (c.f.\ \eqref{eqn:IntroET} and \eqref{eqn:EnergyDensity}),
\begin{align}
S_{EE} =& \frac{L^{d-1}}{2(d-2)G_N} \left(\frac{\tilde \ell}{\epsilon}\right)^{d-2} - \frac{L^{d-1}}{4(d-2) G_N } \left(\frac{2\sqrt{\pi} \GammaFunc{\frac{d}{2(d-1)}}}{\GammaFunc{\frac{1}{2(d-1)}}} \right)^{d-1}\cdot  \left( \frac{\tilde \ell}{\ell} \right)^{d-2} \nonumber\\[0.5em]
& + \langle T_{tt} \rangle~ \ell \tilde \ell^{d-2}\cdot \underbrace{T_{\text{ent}}^{-1}}_{\frac{2\pi}{d+1} \ell},
\label{eqn:EET0}
\end{align}
where $T_{\text{ent}}$ is the entanglement temperature \cite{Bhattacharya:2012mi}. Let us note that the leading order correction is positive and increases the entanglement entropy. This result can be easily derived from the result in terms of a power series \eqref{eqn:SolutionAsPowerSeries}.

In contrast, the large-width limit is more involved. In the following, we use properties of generalised hypergeometric functions and Meijer $G$-functions to obtain a large-width expansion. In this limit, the width of the strip and the minimal area are divergent and proportional in leading order. We derive a closed form for the subleading contribution. 

Following section \ref{sec:LowHighT}, the result for the entanglement entropy may be written as
\begin{subequations}
\begin{align}
&S_{EE} = \frac{L^{d-1}}{2(d-2)G_N} \left(\frac{\tilde \ell}{\epsilon}\right)^{~~\mathclap{d-2}} + \frac{L^{d-1} \tilde \ell^{d-2}}{4 \zs^{d-1}G_N } \ell \\ & + \frac{\sqrt{\pi}L^{d-1} }{8(d-1)G_N}\frac{\tilde \ell^{d-2}}{\zs^{d-2}} \sum\limits_{\Delta m=0}^{\frac{2(d-1)}{\chi}-1} \frac{1}{\Delta m!} \Pochhammer{\frac{1}{2}}{\Delta m} \zsh^{\Delta m d} \frac{\GammaFunc{\frac{d}{\chi} a\EE_{-\frac{1}{2}}}}{\GammaFunc{\frac{d}{\chi} a\EE_1}}   \nonumber\\
&\times\Hgf{\frac{3d-2}{\chi}+1}{\frac{3d-2}{\chi}}{1,  a\EE_{-\frac{1}{2}},\mydots,  a\EE_{\frac{d}{\chi}-\frac{3}{2}},   b\EE_{\frac{1}{2}},\mydots,   b\EE_{\frac{2(d-1)}{\chi}-\frac{1}{2}}; a\EE_1,\mydots, a\EE_{\frac{d}{\chi}}, b\EE_1,\mydots,   b\EE_{\frac{2(d-1)}{\chi}};\zshs^{\frac{2(d-1)d}{\chi}}}{} \nonumber
\end{align}
or in terms of Meijer $G$-functions 
\begin{align}
S_{EE} &= \frac{L^{d-1}}{2(d-2)G_N} \left(\frac{\tilde \ell}{\epsilon}\right)^{d-2}+  \frac{\tilde \ell^{d-2}\ell L^{d-1}}{4\zs^{d-1}G_N}+ \frac{\chi \pi L^{d-1} }{\sqrt{8(d-1)d^3}\cdot G_N} \frac{\tilde \ell^{d-2}\zh}{\zs^{d-1}}   \\
&~~~\times~ G_{\frac{3d-2}{\chi},\frac{3d-2}{\chi}}^{\,\frac{2(d-1)}{\chi},\frac{d}{\chi}}\!\left(\left.{\begin{matrix}\hat a_{\frac{3}{2}},\dots ,\hat a_{\frac{d}{\chi} +\frac{1}{2}}, \hat b_{\frac{1}{2}},\dots,\hat b_{\frac{2(d-1)}{\chi}-\frac{1}{2}}\\\hat b_{0},\dots,\hat b_{\frac{2(d-1)}{\chi}-1},\hat a_{0},\dots ,\hat a_{\frac{d}{\chi}-1 }\end{matrix}}\;\right|\,\zsh^{\frac{2(d-1)d}{\chi}}\right).\nonumber
\end{align}\label{eqn:HEEHighT}%
\end{subequations}
This result is equivalent to the previous one, but more suitable for the large-width limit. In this limit, the turning point approaches the horizon, i.e.\ $\zs\rightarrow \zh$. The width of the strip and the entanglement entropy diverge in this limit. Examining the previous form \eqref{eqn:HEE}, all hypergeometric functions are divergent in this limit (c.f.\ \eqref{eqn:HgfDivergences}). In this alternative form \eqref{eqn:HEEHighT} however, the divergent behaviour of the entanglement entropy is captured in the second term, whereas the remaining part remains finite. The improved form is compatible with the notation used in \cite{Gushterov:2017vnr}, where the authors introduced a function $C$ to express the entanglement entropy as
\begin{align}
  S_{EE} = \frac{L^{d-1} \tilde \ell^{d-2}}{2(d-2) G_N\epsilon^{d-2}} + \frac{L^{d-1}}{4 G_N }\frac{\tilde \ell^{d-2}}{\zs^{d-2}}\cdot \frac{\ell}{\zs}+ \frac{L^{d-1} }{2 G_N }\frac{\tilde \ell^{d-2}}{\zs^{d-2}} C(\zs).\label{eqn:HEEHighT2}
\end{align}
Comparing this to our alternative result derived above \eqref{eqn:HEEHighT},
the function $C$ may be written as 
\begin{subequations}
\begin{align}
  &C(\zs) = \frac{\sqrt{\pi} }{4(d-1)} \sum\limits_{\Delta m=0}^{\frac{2(d-1)}{\chi}-1} \frac{1}{\Delta m!} \Pochhammer{\frac{1}{2}}{\Delta m} \zsh^{\Delta m d} \frac{\GammaFunc{\frac{d}{\chi} a\EE_{-\frac{1}{2}}}}{\GammaFunc{\frac{d}{\chi} a\EE_1}}   \\
  &~~\times\hspace{-0.5em}\Hgf{\frac{3d-2}{\chi}+1}{\frac{3d-2}{\chi}}{1,  a\EE_{-\frac{1}{2}},\mydots,  a\EE_{\frac{d}{\chi}-\frac{3}{2}},   b\EE_{\frac{1}{2}},\mydots,   b\EE_{\frac{2(d-1)}{\chi}-\frac{1}{2}}; a\EE_1,\mydots, a\EE_{\frac{d}{\chi}}, b\EE_1,\mydots,   b\EE_{\frac{2(d-1)}{\chi}};\zshs^{\frac{2(d-1)d}{\chi}}}{} \nonumber, \\[0.5em]
  &= \frac{\chi \pi  }{\sqrt{2(d-1)d^3}} \frac{\zh}{\zs} ~ G_{\frac{3d-2}{\chi},\frac{3d-2}{\chi}}^{\,\frac{2(d-1)}{\chi},\frac{d}{\chi}}\!\left(\left.{\begin{matrix}\hat a_{\frac{3}{2}},\dots ,\hat a_{\frac{d}{\chi} +\frac{1}{2}}, \hat b_{\frac{1}{2}},\dots,\hat b_{\frac{2(d-1)}{\chi}-\frac{1}{2}}\\\hat b_{0},\dots,\hat b_{\frac{2(d-1)}{\chi}-1},\hat a_{0},\dots ,\hat a_{\frac{d}{\chi}-1 }\end{matrix}}\;\right|\,\zsh^{\frac{2(d-1)d}{\chi}}\right)
\end{align}\label{eqn:HEEC}%
\end{subequations}
and only depends on the ratio $\zs/\zh$.

Now, it is only a small step to obtain the large-width expansion. Expanding the turning point for large width only leads to exponentially decaying subleading contributions (c.f.\  \eqref{eqn:HgfDivergences}). The leading order terms for $\zs \rightarrow \zh$ are consequently
\begin{align}
  S_{EE} = \frac{L^{d-1} \tilde \ell^{d-2}}{2(d-2) G_N \epsilon^{d-2}} + \frac{L^{d-1}}{4 G_N }\frac{\tilde \ell^{d-2}}{\zh^{d-2}}\cdot \frac{\ell}{\zh}+ \frac{L^{d-1} }{2 G_N }\frac{\tilde \ell^{d-2}}{\zh^{d-2}} C(\zh),
\end{align}
where $C(\zh)$ is a temperature-independent constant. The second term is the thermal entropy and scales with the volume of the strip. The last term however is again an area-law and hence has the same behaviour as the UV-divergent term. 

These results can also be applied to the mutual information between two strips $A$ and $B$
\begin{align}
  \mathcal{I}(A,B) = S_{EE}(A) + S_{EE}(B)-S_{EE}(A \cup B).
\end{align}
In the limit of two large parallel strips with small separation, the previously derived subleading term is relevant as discussed in \cite{Fischler:2012uv}.

\subsection{Results}
\label{sec:ResultsEE}
Let us have a look at specific spacetime dimensions. We again consider $d=2,3,4$. Figure \ref{Fig:EE} shows the analytical result for the entanglement entropy \eqref{eqn:HEEHGF}. In particular, it shows how the analytical result interpolates between the zero-temperature behaviour, shown as dotted line, and the extensive large-width behaviour.

In the following, we use the notation
\begin{align}
\Hypergeometric{ p+1}{ p}{a}{b}{u}= \,_{p+1}F_p\left(\begin{matrix} a_1,\dots,a_{p+1} \\ b_1,\dots,b_p \end{matrix} ; u \right)
\end{align}
to avoid lengthy expressions.
\begin{figure}
	\begin{subfigure}{0.495\linewidth}
    	{\PGFcommands
    		\hspace{-3em}
    		\import{PGF/}{EE2.pgf}}
    	\caption{$d=2$}
    \end{subfigure}
	\begin{subfigure}{0.495\linewidth}
		{\PGFcommands
			\import{PGF/}{EE3.pgf}}
		\caption{$d=3$}
	\end{subfigure}\\
	\begin{subfigure}{0.495\linewidth}
		{\PGFcommands
			\hspace{-3em}
			\import{PGF/}{EE4.pgf}}
		\caption{$d=4$}
	\end{subfigure}
	\begin{subfigure}{0.495\linewidth}
		{\PGFcommands
			\import{PGF/}{EE10.pgf}}
		\caption{$d=10$}
	\end{subfigure}
	\caption{Entanglement entropy for different dimensions. \protect\\
		We subtracted the cut-off term, which is proportional to $\ln(\epsilon/\zh)$ for $d=2$. The dotted black line is the zero-temperature result.}
	\label{Fig:EE}
\end{figure}

\subsubsection{\texorpdfstring{$\AdS_3/\CFT_2$}{AdS3/CFT2} }
\label{sec:CFT2}
The holographic entanglement entropy in in $AdS_3$ is related to the length of the geodesic. Therefore, the entanglement entropy is related to the saddle-point approximations of the two-point function
\begin{subequations}
	\begin{align}
	\langle \mathcal{O}(t,\vec{x})  \mathcal{O}(t,\vec{y}) \rangle &=\lim \limits_{\epsilon \rightarrow 0}  \epsilon^{- 2 \Delta } \exp\left(-\Delta \cdot\frac{\Area{1}}{L}\right),\\
	S_{EE} &=  \frac{\Area{1}}{4G_N}.
	\end{align}
\end{subequations}
We already obtained the result for the two-point function in \ref{sec:2pt2d}. Using \eqref{eqn:2dResult} we have
\begin{align}
S_{EE} &= \frac{c}{3} \ln \left(\frac{\beta}{\pi\epsilon}\sinh\left(\frac{|\vec{x} - \vec{y}|\pi}{\beta} \right) \right),
\end{align}
where we used the central charge $c$ is given by \cite{Brown:1986nw}
\begin{align}
  c = \frac{3L}{2 G_N}.
\end{align}
Therefore, in $d=2$ our general result \eqref{eqn:HEEHGF} simplifies to the known holographic result \cite{Ryu:2006bv,Ryu:2006ef}, which agrees with field-theory calculations \cite{Calabrese:2004eu,Calabrese:2005zw}.

\subsubsection{\texorpdfstring{$\AdS_4/\CFT_3$}{AdS4/CFT3} }
The holographic entanglement entropy in $AdS_4$ is related to the area of a two-dimensional surface. Therefore, the entanglement entropy is related to the quark-antiquark potential of the double-Wick rotated solution, 
\begin{subequations}
	\begin{align}
	V_q &= \frac{\mathcal{A}}{2\pi \alpha '\tilde \ell}- \frac{L^2}{\pi \alpha' \epsilon},\\
	S_{EE} &=  \frac{\Area{2}}{4G_N}.
	\end{align}
\end{subequations}
We already considered the potential in section \ref{sec:WL3d}. The width of the entangling region is the width of the Wilson loop (c.f.\, \eqref{eqn:Widthd3}). For the potential, we obtained a result \eqref{eqn:Potd3} containing generalised hypergeometric functions which diverge in the large-width limit, as well as an equivalent expression \eqref{eqn:Potd3Alt} with converging hypergeometric functions.

\subsubsection{\texorpdfstring{$\AdS_5/\CFT_4$}{AdS5/CFT4} }
Let us turn to $d=4$. The most famous example is SUGRA on $AdS_5\times S^5$. which is dual to $\mathcal{N}=4$ SYM theory \cite{Maldacena:1997re}. The AdS-CFT dictionary yields
\begin{align}
  \frac{L^3}{4 G_N} =& \frac{N^2}{2\pi}, \nonumber\\
  =& \frac{2}{\pi} \cdot a
\end{align}
hence the entanglement entropy is proportional to the central charge $a = N^2/4$ of the field theory. The second line is also valid if we take a general five-dimensional Einstein manifold instead of $S^5$.

In this case, the relevant surface for the entanglement entropy is three-dimensional and not related to the other considered observables. We obtain
\begin{align}
\scl{3} &= \frac{\sqrt{\pi} z_\star  \Gamma\left(\frac{4}{3}\right)}{6 \,  \Gamma\left(\frac{11}{6}\right)} \zsh^4\,_4F_3\left(\begin{matrix} \frac{1}{2},\frac{5}{6},\frac{7}{6},\frac{7}{6} \\ \frac{11}{12},\frac{4}{3},\frac{17}{12} \end{matrix} ; \zsh^{12} \right) + \frac{\sqrt{\pi} {z_\star} \Gamma\left(\frac{2}{3}\right)}{3 \, \Gamma\left(\frac{7}{6}\right)} \,_4F_3\left(\begin{matrix} \frac{1}{6},\frac{1}{2},\frac{5}{6},\frac{5}{6} \\ \frac{7}{12},\frac{2}{3},\frac{13}{12} \end{matrix} ; \zsh^{12} \right)\nonumber\\ & ~~~ + \frac{z_\star}{6 \, }\zsh^8\,_5F_4\left(\begin{matrix} \frac{5}{6},1,\frac{7}{6},\frac{3}{2},\frac{3}{2} \\ \frac{5}{4},\frac{4}{3},\frac{5}{3},\frac{7}{4} \end{matrix} ; \zsh^{12} \right)
\end{align}
for the width of the strip and
\begin{align}
S_{EE} &= \frac{\sqrt{\pi} L^3 \tilde \ell^2 \Gamma\left(\frac{1}{3}\right)}{24 G_N \, {z_\star}^{2}  \Gamma\left(\frac{5}{6}\right)}\zsh^4 \,_4F_3\left(\begin{matrix} \frac{1}{6},\frac{1}{2},\frac{5}{6},\frac{7}{6} \\ \frac{5}{12},\frac{11}{12},\frac{4}{3} \end{matrix} ; \zsh^{12} \right) \nonumber \\ 
& ~~~ + \frac{\sqrt{\pi} L^3 {\tilde \ell^2}  \Gamma\left(-\frac{1}{3}\right)}{12 G_N \, {z_\star}^{2} \Gamma\left(\frac{1}{6}\right)}\,_4F_3\left(\begin{matrix} -\frac{1}{6},\frac{1}{6},\frac{1}{2},\frac{5}{6} \\ \frac{1}{12},\frac{7}{12},\frac{2}{3} \end{matrix} ; \zsh^{12} \right)  \nonumber \\
& ~~~+ \frac{L^3 {\tilde \ell^2}}{16 G_N \, \zs^2} \zh^8 \,_5F_4\left(\begin{matrix} \frac{1}{2},\frac{5}{6},1,\frac{7}{6},\frac{3}{2} \\ \frac{3}{4},\frac{5}{4},\frac{4}{3},\frac{5}{3} \end{matrix} ; \zsh^{12} \right)+ \frac{L^3 {\tilde \ell}^2}{4G_N{\epsilon}^{2}}
\end{align}
for the entanglement entropy (c.f.\ \eqref{eqn:HEEHGF}). Each of the appearing hypergeometric functions diverges logarithmically in the large-width limit (i.e.\ the turning point approaches the horizon, $\zs \rightarrow \zh$). Therefore, it is convenient to transform the hypergeometric functions to obtain (c.f.\ \eqref{eqn:HEEHighT})
\begin{align}
S_{EE} &= \frac{\sqrt{\pi} L^{3} \tilde \ell^{2} \Gamma\left(\frac{1}{3}\right)}{48 G_N \, \zs^{2} \Gamma\left(\frac{11}{6}\right)}\zsh^4 \,_4F_3\left(\begin{matrix} \frac{1}{6},\frac{1}{2},\frac{5}{6},\frac{7}{6} \\ \frac{11}{12},\frac{4}{3},\frac{17}{12} \end{matrix} ; \zsh^{12} \right) \nonumber \\
	&~~~ + \frac{\sqrt{\pi} L^{3} \tilde \ell^{2} \Gamma\left(-\frac{1}{3}\right)}{24 G_N \, {z_\star}^{2} \Gamma\left(\frac{7}{6}\right)} \,_4F_3\left(\begin{matrix} -\frac{1}{6},\frac{1}{6},\frac{1}{2},\frac{5}{6} \\ \frac{7}{12},\frac{2}{3},\frac{13}{12} \end{matrix} ; \zsh^{12} \right) \nonumber \\
		&~~~ + \frac{L^{3} \tilde \ell^{2} }{48 G_N \, \zs^2}\zsh^8 \,_5F_4\left(\begin{matrix} \frac{1}{2},\frac{5}{6},1,\frac{7}{6},\frac{3}{2} \\ \frac{5}{4},\frac{4}{3},\frac{5}{3},\frac{7}{4} \end{matrix} ; \zsh^{12} \right) + \frac{L^{3} \tilde \ell^{2}}{4 G_N {\epsilon}^{2}} + \frac{L^3 \tilde \ell^2 \ell}{\zs^3 4 G_N}.
\end{align}
This result splits of the divergent term as a term proportional to $\ell/\zs^3$, whereas the appearing hypergeometric functions are finite at unit argument. 

These results can be expressed in terms of Meijer $G$-functions, where we obtain
\begin{align}
  \scl{3} &= \frac{\pi {z_h}}{\sqrt{6}} \, ~ G^{3,2}_{\,5,5}\!\left(\left. {\begin{matrix}\frac{1}{4}, \frac{3}{4}, \frac{1}{4}, \frac{7}{12}, \frac{11}{12} \\\frac{1}{12}, \frac{5}{12}, \frac{3}{4}, 0, \frac{1}{2}\end{matrix}} \;\right|\,\zsh^{12}\right)
\end{align}
for the width and
\begin{subequations}
\begin{align}
S_{EE} & = \frac{\pi L^3 \tilde\ell^2\zh}{4 \sqrt{6} G_N \, {z_\star}^{3}}~ G_{5,5}^{\,3,2}\!\left(\left. {\begin{matrix}\frac{3}{4}, \frac{5}{4}, \frac{1}{4}, \frac{7}{12}, \frac{11}{12} \\ \frac{1}{12}, \frac{5}{12}, \frac{3}{4}, 0, \frac{1}{2}\end{matrix}} \;\right|\,\zsh^{12}\right)+ \frac{L^3 {\tilde \ell^2} }{4 G_N{\epsilon}^{2}},\\
 & = \frac{\pi L^3 \tilde\ell^2\zh}{16 \sqrt{6} G_N \, {z_\star}^{3}}~ G_{5,5}^{\,3,2}\!\left(\left. {\begin{matrix}\frac{3}{4}, \frac{5}{4}, \frac{1}{4}, \frac{7}{12}, \frac{11}{12} \\ \frac{1}{12}, \frac{5}{12}, \frac{3}{4},  \frac{1}{2},0\end{matrix}} \;\right|\,\zsh^{12}\right)+ \frac{L^3 {\tilde \ell^2} }{4 G_N{\epsilon}^{2}} + \frac{L^3 \tilde \ell^2 \ell}{\zs^3 4 G_N}
\end{align}
\end{subequations}
for the entanglement entropy (c.f.\ \eqref{eqn:HEEMGF}). The Meijer $G$-function in the first line diverges in the large-width limit, whereas the one in the second line converges and the divergent behaviour is captured by the $\ell/\zs^3$ term.

  \subsection{Entanglement density}
\label{sec:HED}

Our results have a particularly useful application to the {\it  entanglement density}, considered in \cite{Gushterov:2017vnr}. For a strip entangling region, the entanglement density\footnote{This is not the entanglement density defined as variation of the entanglement entropy, as defined in \cite{Nozaki:2013wia,Bhattacharya:2014vja}.} as defined in \eqref{eq:sigma} is given by 
	\begin{align}
		\sigma =& \frac{1}{\tilde \ell^{d-2}\ell}\left[S_{EE}(\ell)-S_{EE}(\ell)|_{T=0}  \right]  \label{eqn:ED}
	\end{align}
where $\ell$ is the width of the strip and $\tilde \ell$ is its length, see Figure~\ref{fig:HEE}. 
The transverse volume $\tilde \ell^{d-2}$ appears as an overall proportionality constant in the entanglement entropy and drops out in the entanglement density. The zero-temperature result for the entanglement entropy is subtracted before dividing by the volume of the entangling region, which makes the density UV-finite. 
As discussed in the introduction, in the large-width limit, the expected asymptotic behaviour is
\begin{align}
  \sigma = s\left[1- \frac{1}{\ell \cdot T }  \Delta \hat \alpha + \cdots \right]. \label{eqn:DoF}
\end{align}
where $ \Delta \hat \alpha  $ is a dimensionless number obtained from $\Delta \alpha$ in \eqref{eqn:IntroAlpha} using that $s \propto T^{d-1}$ and $\Delta \alpha \propto T^{d-2}$,
\begin{align}
  \Delta \hat \alpha = \frac{T}{s}\Delta \alpha.
\end{align}
It may be expressed in terms of our previously defined function $C(\zs)$ (c.f.\ \eqref{eqn:HEEC} and \eqref{eqn:HEEHighT2}),
\begin{align}
  \Delta \hat \alpha = -2 C(\zh).
\end{align}

We use our analytical results 
\eqref{eqn:HEEHGF} and \eqref{eqn:HEEMGF} for the entanglement entropy
to calculate the entanglement density. Figure \ref{Fig:HED} shows a plot of our analytical result. We observe a linear behaviour at small $\ell \cdot T$   with a positive slope. This is consistent with 
the first law of entanglement thermodynamics \cite{Blanco:2013joa,Bhattacharya:2012mi}  given in \eqref{eqn:IntroET} which 
implies 
\begin{align}
  \sigma \propto \langle T_{tt} \rangle \ell \,  ,
\end{align}
i.e.~a linear behaviour in the width $\ell$ at small $\ell$ for constant temperature.  In contrast, the entanglement density is expected to approach a constant in the large-width limit: The entanglement entropy becomes extensive (i.e.\ it scales with  volume term) and the entanglement density approaches the thermal entropy density. The results displayed in figure \ref{Fig:HED} indeed show this expected behaviour. Moreover, figure \ref{Fig:HED} shows how our analytical result interpolates between the small and large width regimes. 
\begin{figure}[htbp]
	\begin{minipage}{0.675\textwidth}
		{\PGFcommands{}
			\import{PGF/}{HED.pgf}}
		\captionof{figure}{Holographic entanglement density $\sigma$.}
		\label{Fig:HED}
	\end{minipage}\hfill
	\begin{minipage}{0.3\textwidth}
		\begin{center}
			\captionof{table}{Derivative of $\sigma$.}\label{Tab:ED}\vspace{1.5em}
			\begin{tabular}{|l|l|l|}\hline
				\rule{0pt}{1.5em} $d$  & $ C(z_h)$ & $\frac{\partial\sigma}{\partial \ell}$ \\[0.5em] \hline
				\rule{0pt}{1.15em}$3$ & $-0.880$ & \multirow{ 4}{*}{$>0$} \\
				$4$ & $-0.333$& \\
				$5$ & $-0.142$ &\\
				$6$ & $-0.0444$ &\\\hline
				\rule{0pt}{1.15em}$7$ & $0.0148$&\multirow{5}{*}{$<0$}  \\
				$8$ & $0.0545$& \\
				$9$ & $0.0829$ &\\
				$10$ & $0.104$ &\\
				$11$ & $0.121$ &\\\hline
			\end{tabular}
		\end{center}
	\end{minipage} 
\end{figure} 
In particular, for field-theory spacetime dimension $d>6$, we observe a non-monotonic
behaviour and a global maximum in the entanglement density. This global maximum appears at a finite value of $\ell T$. To investigate this further, we look at the derivative of the entanglement density. As argued above, for small $\ell \cdot T$ we have 
\begin{align}
  \frac{\partial \sigma}{ \partial (\ell\cdot T)} = const. >0 
\end{align}
due to the first law of entanglement thermodynamics. For the large-width limit, the extensive term drops out and we obtain
\begin{align}
\frac{\partial \sigma}{ \partial (\ell\cdot T)} = - s C(\zh)  \frac{d}{2\pi (T\ell)^2}.
\end{align}
This implies that the derivative vanishes asymptotically for  $\ell \cdot T \rightarrow \infty$.  However,  whether the entanglement density approaches the thermal entropy from above or below depends on the sign of $C(\zh)$, which was introduced in \eqref{eqn:HEEC} as the next-to-leading-order term of the entanglement entropy in this limit. Consequently, a global maximum appears for positive $C(\zh)$. 
Table \ref{Tab:ED} displays the values of  $C(\zh)$ and the sign of the first derivative for large $\ell \cdot T$  obtained from our analytical result \eqref{eqn:HEEHGF} and \eqref{eqn:HEEMGF}. These analytical results show perfect agreement with the numerical results in \cite{Gushterov:2017vnr}. 

We see from Figure~\ref{Fig:HED} that the entanglement density approaches the thermal entropy density from below for field-theory dimensions $d \leq 6$. However, for $d \geq 7$ field-theory spacetime dimensions,  it approaches the thermal entropy density from above and the first derivative for large $\ell \cdot T$  changes sign accordingly. Consequently, $\Delta \hat \alpha$ becomes negative for large $\ell \cdot T$. This corresponds to a violation of the area theorem. 
As substantiated further by the consideration of entanglement negativity in the following section, we assume that we may identify $\Delta \hat \alpha$  with a measure of the number degrees of freedom. Our results thus imply that there appears to be a larger number of degrees of freedom at low energies. The large-dimension limit of the Schwarzschild geometry may yield a clue for the origin of the additional degrees of freedom: As was shown in \cite{Emparan:2013xia,Emparan:2013moa}, for very large dimensions the near-horizon geometry approaches a two-dimensional black hole of string theory. The new degrees of freedom in the IR may be due to the additional conformal symmetry. Our results indicate that this behaviour sets in above a critical number of dimensions.

  \subsection{Entanglement negativity}
\label{sec:HEN}
For a pure state, the entanglement entropy is a measure for the entanglement in a bi-partite system. However, it is not a good measure for a mixed state, such as the thermal state we consider. This is due to the fact that  it also contains contributions from classical correlations. We already saw this above, where the entanglement entropy became extensive at large width. For a finite-temperature state, the {\it entanglement negativity} has been suggested  as a measure for entanglement \cite{Calabrese:2012ew,Calabrese:2012nk,Calabrese:2014yza,2002PhRvA..65c2314V,2014NJPh...16l3020E}.

\subsubsection{Review}

Let us review why entanglement negativity is a measure of entanglement at finite temperature.
We begin by considering Pere's criterion for separability \cite{Peres:1996dw}, which considers the eigenvalues of the partial transpose $\rho^{T_B}$ of the density matrix. This is defined such that
\begin{align}
  \langle e_i^B e_j^C | \rho^{T_B}| e^B_k e^C_l \rangle = \langle e_k^B e_j^C | \rho| e^B_i e^C_l \rangle,
\end{align}
where $e_i^B$ and $e_j^C$ are the basis vector of $B$ and its complement $C$, respectively. The partial transpose has unit trace and can have non-negative eigenvalues. The sum of negative eigenvalues is
\begin{align}
  \mathcal{N} =& \frac{1}{2} \left(||\rho^{T_B}||-1 \right),
\end{align}
where the trace-norm $||A|| = \sqrt{A^\dag A}$ is the sum of the absolute values of eigenvalues. Pere's criterion states that the state is not separable if the partial transpose has negative eigenvalues. Consequently, the entanglement negativity\footnote{Sometimes, $\mathcal{N}$ is called negativity and $\EN$ logarithmic negativity.}
\begin{align}
  \EN(B) = \ln ||\rho^{T_B}||\label{eqn:EN}
\end{align}
is a measure for entanglement, as it measures how `negative' the eigenvalues are. We use $\EN{}$ since it is an additive quantity in contrast to $\mathcal{N}$ \cite{Calabrese:2012ew,Calabrese:2012nk,Calabrese:2014yza,2002PhRvA..65c2314V}.

Similarly to the entanglement entropy, it is difficult to calculate the entanglement negativity in  field theory, in particular in higher dimensions. It is thus of special interest to consider its gravity dual. Furthermore, it is also a candidate for a generalised $c$-function \cite{Banerjee:2015coc,Liu:2012eea} as a measure for quantum entanglement at different energy scales. Here we follow the proposal of Chaturvedi, Malvimat and Sengupta \cite{Chaturvedi:2016rcn,Chaturvedi:2016rft}. Their starting point is the CFT result for $d=2$, where the entanglement negativity of an interval with width $\ell$ can be written as
\begin{align}
\EN(B) = \frac{3}{2} \left(S_{EE}(B) -S^{th}(B) \right) + f\left( e^{-2\pi \ell T}\right)
\end{align}
at finite temperature, where $S^{th}$ is the thermal entropy and $f$ is a non-universal function, which depends on the entire particle content. In holography, $f$ only yields a subleading contribution in the large-$N$ limit and can be neglected. The proposal generalises this to general spacetime dimension $d$, i.e.\ it reads
\begin{align}
\EN(B) = \frac{3}{2} \left(S_{EE}(B) -S^{th}(B) \right).\label{eqn:HEN}
\end{align}

\begin{figure}
	\center
	\def\svgwidth{0.7\linewidth}
	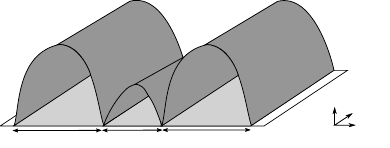
	\caption{Construction for entanglement negativity.}
	\label{fig:HEN}
\end{figure}
This also has an interesting relation to mutual information
\begin{align}
  \mathcal{I}(B,C) = S_{EE}(B) + S_{EE}(C)-S_{EE}(B\cup C).
\end{align}
 When considering a strip (i.e.\ $x^1 \in [-\ell/2,\ell/2]$), the complement $C$ may be split in $C_1$ and $C_2$ with $x^1<-\ell/2$ and $x^1>\ell/2$ respectively. This construction is shown in Figure \ref{fig:HEN}. The holographic entanglement entropy can be expressed in terms of mutual information as
\begin{align}
\EN(B) &= \frac{3}{4}\left(\mathcal{I}(B,C_1)+\mathcal{I}(B,C_2) \right). 
\end{align}

\subsubsection{Results}
In our case, the proposal holographic dual of entanglement negativity \eqref{eqn:HEN} reads
\begin{align}
\EN(B) = \frac{3}{2} \left(S_{EE}(B) -\frac{L^{d-1}}{4G_N} \frac{\tilde \ell^{d-2}\ell}{\zh^{d-1}} \right), \label{eq:neuEN}
\end{align}
which allows to use our previously derived results for the holographic entanglement entropy. The width $\ell$ is the same as for the entanglement entropy, as stated in \eqref{eqn:WidthEntanglementEntropy} in terms of hypergeometric functions and in \eqref{eqn:WidthEntanglementEntropyG} in terms of a Meijer $G$-function. Let us quickly remind you that we obtained two qualitatively different forms of the result for the holographic entanglement entropy, which can be characterised by their large-width behaviour: In the first result \eqref{eqn:HEE} each term diverges in this limit, whereas the divergent behaviour of the second result \eqref{eqn:HEEHighT} is captured by the term involving $\ell/\zs^{d-1}$.  We expressed the remaining part in terms of a function $C$, following the notation in \cite{Gushterov:2017vnr}. We derived the result for $C$ in \eqref{eqn:HEEC}. Using these results for evaluating \eqref{eq:neuEN}, we obtain
	\begin{align}
	&\varepsilon = \frac{3L^{d-1}}{4(d-2)G_N} \left(\frac{\tilde \ell}{\epsilon}\right)^{~~\mathclap{d-2}} + \frac{3L^{d-1} \tilde \ell^{d-2}\ell}{8 \zs^{d-1}G_N }\left(1-\zsh^{d-1} \right)  + \frac{3L}{4G_N} \frac{\tilde \ell^{d-2}}{\zs^{d-2}} C(\zs).
	\end{align}
We recall that $\zs$ is the turning point of the minimal surface in the radial direction.

The small-width limit is similar as for the entanglement entropy, which we considered in \eqref{eqn:EET0}. This yields
\begin{align}
\varepsilon = & \frac{3L^{d-1}}{4(d-2)G_N} \left(\frac{\tilde \ell}{\epsilon}\right)^{d-2} - \frac{3L^{d-1}}{8 (d-2) G_N } \left(\frac{2\sqrt{\pi} \GammaFunc{\frac{d}{2(d-1)}}}{\GammaFunc{\frac{1}{2(d-1)}}} \right)^{d-1}\cdot  \left( \frac{\tilde \ell}{\ell} \right)^{d-2} \nonumber \\
&-\frac{3L^{d-1}}{8G_N} \frac{\tilde \ell^{d-2}\ell}{\zh^{d-1}}
\end{align}
at zero temperature. The leading order correction to the zero-temperature result does not arise due to the correction to the entanglement entropy, but due to the subtracted thermal entropy. This one is negative and causes a decrease of the entanglement negativity. Therefore, a finite temperature decreases short-range entanglement independent of the spacetime dimension.

For the large-width behaviour, the minimal area of the bulk surface is extensive, producing the thermal volume term for the entanglement entropy. Since this volume term is subtracted for the entanglement negativity, the entanglement entropy approaches a finite value in this limit,
\begin{align}
 &\varepsilon =\frac{3 L^{d-1} }{4(d-2)G_N}\cdot \frac{\tilde \ell^{d-2}}{\epsilon^{d-2}}  + \frac{3 L^{d-1}\tilde \ell^{d-2}}{4 G_N \zh^{d-2}}C(\zh),
\end{align}
with $C(\zh)$ is a temperature-independent constant given by \eqref{eqn:HEEC}. The leading contribution is thus an area term and not an extensive volume term. At zero-temperature (i.e.\ $\zh \rightarrow \infty$), the second term vanishes.

Table \ref{Tab:ED} shows the result for $C(\zh)$: it is negative for $d\leq6 $ and positive for $d>6$. This sign change was already discussed for the entanglement density: a negative value for $C(\zh)$ causes a local maximum for the entanglement density and a violation of the area theorem, which hints at new IR degrees of freedom. 
The sign change implies that  the entanglement density has a local maximum at a finite value of $\ell \cdot T$. 
\begin{figure}[h]\centering
	{\PGFcommands{}
		\import{PGF/}{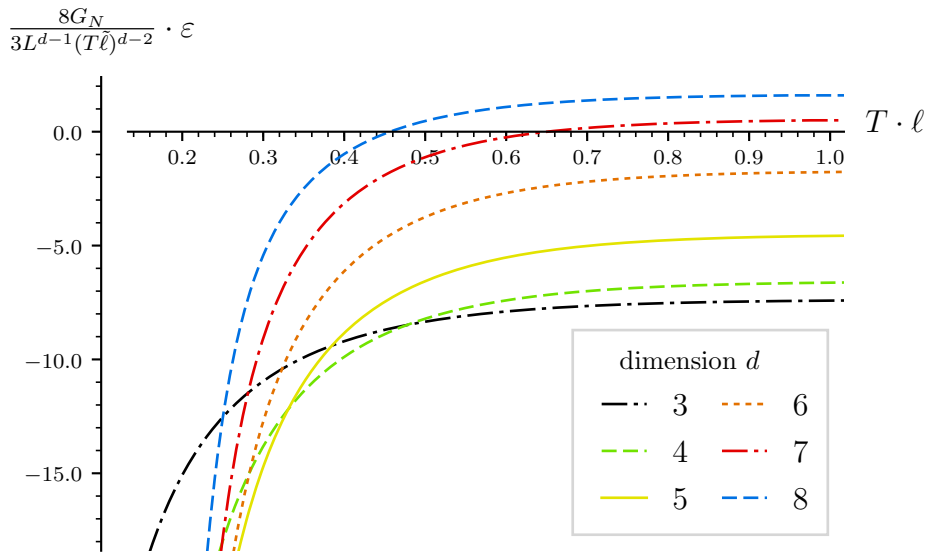}}
	\caption{Cut-off independent part of the entanglement negativity. \protect\\
		For plotting, we subtracted the $\mathcal{O}(\epsilon)^{2-d}$ term. For $d=2$, we subtracted $2L \ln(\epsilon/\zh)$. }
	\label{fig:EN2}
\end{figure} 

For the entanglement negativity, we see how $C(\zh)$ causes a temperature-dependent shift of the asymptotic value for $\ell T \rightarrow \infty$.  Figure \ref{fig:EN2} shows the entanglement negativity for several spacetime dimensions $d$. The plot clearly shows that the entanglement negativity asymptotically approaches a finite value  for $d>6$.

Figure \ref{fig:EN1} shows the result for specific spacetime dimensions in comparison to the zero-temperature result. We already considered $d=4$ in our examples section and additionally look at $d=10$ as an example for $d>6$.  For $d=4$, we have $C(\zh) < 0 $. The entanglement negativity at finite temperature is always smaller than the entanglement negativity at zero-temperature. In this case, the temperature decreases entanglement. This looks different for $d=10$, where we have $C(\zh) > 0 $. For a fixed temperature, there exists a critical temperature $\ell_{crit}$, at which the zero-temperature result agrees to the finite temperature result. The entanglement negativity is smaller than the zero-temperature result for $\ell < \ell_{crit}$, but larger for $\ell>\ell_{crit}$. A finite temperature decreases short-range entanglement, but increases long-range entanglement. These two regimes are shown in Figure \ref{fig:EN110}.
\begin{figure}
	\begin{subfigure}{0.49\linewidth}
		\hspace*{-4em}
		{\PGFcommands{}
			\import{PGF/}{EN4.pgf}}
		\caption{$d=4$}
		\label{fig:EN14}
	\end{subfigure}
	\begin{subfigure}{0.49\linewidth}
		{\PGFcommands{}
			\import{PGF/}{EN10.pgf}}
		\caption{$d=10$}
		\label{fig:EN110}
	\end{subfigure}
	\caption{Entanglement negativity for different dimensions. \protect\\
		We subtracted the cut-off term. The dotted black line is the zero-temperature result.}
	\label{fig:EN1}
\end{figure}

To summarise, $C(\zh)$ is a dimensionless constant, which depends solely on the spacetime-dimension. For $d<7$, it is negative and the entanglement negativity is always smaller compared to the zero-temperature result. For $ d \geq 7$, the constant is positive and we observe a cross-over: at small widths the entanglement negativity is smaller than the zero-temperature result, whereas it is larger at large widths. The value of the constant is shown in Table \ref{Tab:ED}. The appearance of this cross-over only depends on the sign of $C(\zh)$. The cross-over appears if and only if the entanglement density discussed in \ref{sec:HED} has a local maximum at a finite width. This shows an interesting connection between entanglement density and entanglement negativity.

  \section{Comparison to numerical results}
\label{sec:Numerics}
In section \ref{sec:CFT2}, we considered the entanglement entropy in $d=2$ and saw how our result in generalised hypergeometric functions \eqref{eqn:HEEHGF} simplifies to the known holographic result derived by \cite{Ryu:2006bv,Ryu:2006ef}. 

\begin{figure}
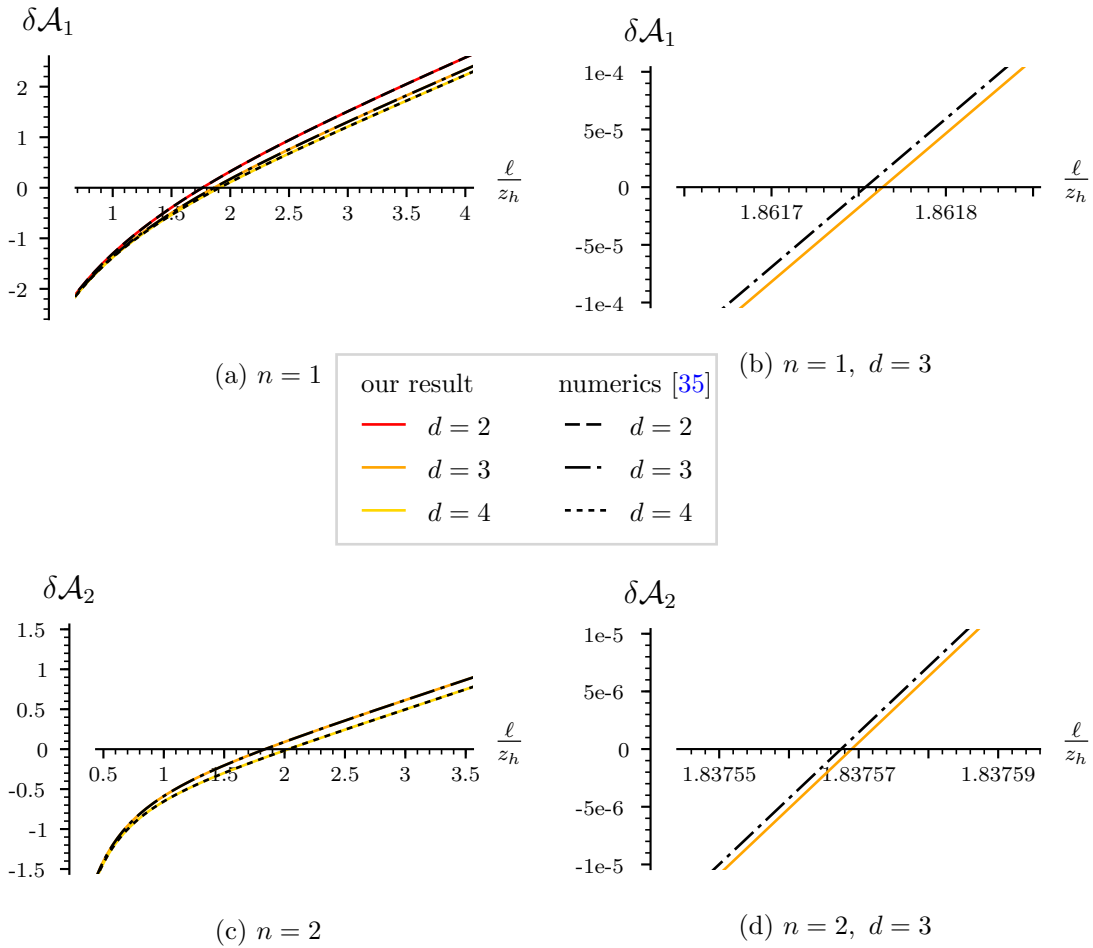

	\begin{subfigure}{0.485\linewidth}
		{\PGFcommands
			\import{PGF/}{Numerics2pt.pgf}}
		\caption{$n=1$}
		\label{Fig:Numericsn1}
	\end{subfigure} 
	\begin{subfigure}{0.485\linewidth}
		{\PGFcommands
			\import{PGF/}{Numerics2ptZoom.pgf}}
		\caption{$n=1,~d=3$}
		\label{Fig:Numericsn1Zoom}
	\end{subfigure}\\[-0.4in]
	\begin{center}
		\begin{subfigure}{0.2\linewidth}
			{\PGFcommands
				\hspace*{-1.45in} % Position good?
				%% Creator: Matplotlib, PGF backend
%%
%% To include the figure in your LaTeX document, write
%%   \input{<filename>.pgf}
%%
%% Make sure the required packages are loaded in your preamble
%%   \usepackage{pgf}
%%
%% Figures using additional raster images can only be included by \input if
%% they are in the same directory as the main LaTeX file. For loading figures
%% from other directories you can use the `import` package
%%   \usepackage{import}
%% and then include the figures with
%%   \import{<path to file>}{<filename>.pgf}
%%
%% Matplotlib used the following preamble
%%   \usepackage{fontspec}
%%   \setmainfont{DejaVu Serif}
%%   \setsansfont{DejaVu Sans}
%%   \setmonofont{DejaVu Sans Mono}
%%
\begingroup%
\makeatletter%
\begin{pgfpicture}%
\pgfpathrectangle{\pgfpointorigin}{\pgfqpoint{2.85in}{1.25in}}%
\pgfusepath{use as bounding box, clip}%
\begin{pgfscope}%
\pgfsetbuttcap%
\pgfsetmiterjoin%
\definecolor{currentfill}{rgb}{1.000000,1.000000,1.000000}%
\pgfsetfillcolor{currentfill}%
\pgfsetfillopacity{0.800000}%
\pgfsetlinewidth{1.003750pt}%
\definecolor{currentstroke}{rgb}{0.800000,0.800000,0.800000}%
\pgfsetstrokecolor{currentstroke}%
\pgfsetstrokeopacity{0.800000}%
\pgfsetdash{}{0pt}%
\pgfpathmoveto{\pgfqpoint{0.858659in}{0.211389in}}%
\pgfpathlineto{\pgfqpoint{2.827388in}{0.211389in}}%
\pgfpathlineto{\pgfqpoint{2.827388in}{1.201864in}}%
\pgfpathlineto{\pgfqpoint{0.858659in}{1.201864in}}%
\pgfpathclose%
\pgfusepath{stroke,fill}%
\end{pgfscope}%
\begin{pgfscope}%
\pgftext[x=1.262308in,y=1.004674in,left,base]{\sffamily\fontsize{10.000000}{12.000000}\selectfont \hspace{-2em}our result \hspace{2.5em}    numerics  \cite{Balasubramanian:2011ur}}%
\end{pgfscope}%
\begin{pgfscope}%
\pgfsetrectcap%
\pgfsetroundjoin%
\pgfsetlinewidth{1.003750pt}%
\definecolor{currentstroke}{rgb}{1.000000,0.000000,0.000000}%
\pgfsetstrokecolor{currentstroke}%
\pgfsetdash{}{0pt}%
\pgfpathmoveto{\pgfqpoint{0.996159in}{0.836792in}}%
\pgfpathlineto{\pgfqpoint{1.210048in}{0.836792in}}%
\pgfusepath{stroke}%
\end{pgfscope}%
\begin{pgfscope}%
\pgftext[x=1.332270in,y=0.783320in,left,base]{\sffamily\fontsize{10.000000}{12.000000}\selectfont \(\displaystyle d=2\)}%
\end{pgfscope}%
\begin{pgfscope}%
\pgfsetrectcap%
\pgfsetroundjoin%
\pgfsetlinewidth{1.003750pt}%
\definecolor{currentstroke}{rgb}{1.000000,0.647059,0.000000}%
\pgfsetstrokecolor{currentstroke}%
\pgfsetdash{}{0pt}%
\pgfpathmoveto{\pgfqpoint{0.996159in}{0.612549in}}%
\pgfpathlineto{\pgfqpoint{1.210048in}{0.612549in}}%
\pgfusepath{stroke}%
\end{pgfscope}%
\begin{pgfscope}%
\pgftext[x=1.332270in,y=0.559077in,left,base]{\sffamily\fontsize{10.000000}{12.000000}\selectfont \(\displaystyle d=3\)}%
\end{pgfscope}%
\begin{pgfscope}%
\pgfsetrectcap%
\pgfsetroundjoin%
\pgfsetlinewidth{1.003750pt}%
\definecolor{currentstroke}{rgb}{1.000000,0.843137,0.000000}%
\pgfsetstrokecolor{currentstroke}%
\pgfsetdash{}{0pt}%
\pgfpathmoveto{\pgfqpoint{0.996159in}{0.388307in}}%
\pgfpathlineto{\pgfqpoint{1.210048in}{0.388307in}}%
\pgfusepath{stroke}%
\end{pgfscope}%
\begin{pgfscope}%
\pgftext[x=1.332270in,y=0.334834in,left,base]{\sffamily\fontsize{10.000000}{12.000000}\selectfont \(\displaystyle d=4\)}%
\end{pgfscope}%
\begin{pgfscope}%
\pgfsetbuttcap%
\pgfsetroundjoin%
\pgfsetlinewidth{1.003750pt}%
\definecolor{currentstroke}{rgb}{0.000000,0.000000,0.000000}%
\pgfsetstrokecolor{currentstroke}%
\pgfsetdash{{5.600000pt}{2.400000pt}}{0.000000pt}%
\pgfpathmoveto{\pgfqpoint{2.041635in}{0.836792in}}%
\pgfpathlineto{\pgfqpoint{2.255524in}{0.836792in}}%
\pgfusepath{stroke}%
\end{pgfscope}%
\begin{pgfscope}%
\pgftext[x=2.377746in,y=0.783320in,left,base]{\sffamily\fontsize{10.000000}{12.000000}\selectfont \(\displaystyle d=2\)}%
\end{pgfscope}%
\begin{pgfscope}%
\pgfsetbuttcap%
\pgfsetroundjoin%
\pgfsetlinewidth{1.003750pt}%
\definecolor{currentstroke}{rgb}{0.000000,0.000000,0.000000}%
\pgfsetstrokecolor{currentstroke}%
\pgfsetdash{{9.600000pt}{2.400000pt}{1.600000pt}{2.400000pt}}{0.000000pt}%
\pgfpathmoveto{\pgfqpoint{2.041635in}{0.612549in}}%
\pgfpathlineto{\pgfqpoint{2.255524in}{0.612549in}}%
\pgfusepath{stroke}%
\end{pgfscope}%
\begin{pgfscope}%
\pgftext[x=2.377746in,y=0.559077in,left,base]{\sffamily\fontsize{10.000000}{12.000000}\selectfont \(\displaystyle d=3\)}%
\end{pgfscope}%
\begin{pgfscope}%
\pgfsetbuttcap%
\pgfsetroundjoin%
\pgfsetlinewidth{1.003750pt}%
\definecolor{currentstroke}{rgb}{0.000000,0.000000,0.000000}%
\pgfsetstrokecolor{currentstroke}%
\pgfsetdash{{2.200000pt}{2.200000pt}}{0.000000pt}%
\pgfpathmoveto{\pgfqpoint{2.041635in}{0.388307in}}%
\pgfpathlineto{\pgfqpoint{2.255524in}{0.388307in}}%
\pgfusepath{stroke}%
\end{pgfscope}%
\begin{pgfscope}%
\pgftext[x=2.377746in,y=0.334834in,left,base]{\sffamily\fontsize{10.000000}{12.000000}\selectfont \(\displaystyle d=4\)}%
\end{pgfscope}%
\end{pgfpicture}%
\makeatother%
\endgroup%
} % legend.pgf contains the legend from the NumericsWL plot
		\end{subfigure} \vspace*{-2em}
	\end{center}
	\begin{subfigure}{0.485\linewidth}
		{\PGFcommands
			\import{PGF/}{NumericsWL.pgf}}
		\caption{$n=2$}
		\label{Fig:Numericsn2}
	\end{subfigure}  
	\begin{subfigure}{0.485\linewidth}
		{\PGFcommands
			\import{PGF/}{NumericsWLZoom.pgf}}
		\caption{$n=2,~d=3$}
		\label{Fig:Numericsn2Zoom}
	\end{subfigure} 
	\caption{Comparison to numerical results of the minimal $n$-dimensional area. \\
		We set  the AdS-radius $L=1$. The results for the two-point function are shifted by $\ln 2$ and the results for the Wilson loop are divided by twice the length $2 \tilde \ell$ to match the normalisation used for the numerical results. }
	\label{Fig:Numerics}
\end{figure}

In the following, let us compare our analytical results to numerical results. In \cite{Balasubramanian:2011ur}, the authors considered non-local observables in a thermalisation scenario of an infalling shell. In particular, they examined numerically how non-local observables in this time-dependent geometry approach the thermal-equilibrium result. In this context, they first determined the minimal areas associated to two-point function for $d=2,3,4$ and to the Wilson loop for $d=3,4$ in the static case, i.e.\ for our setup planar AdS-Schwarzschild. They did not consider the entanglement entropy for a strip explicitly, however for $d=2,3$ it is related to the two-point function and the Wilson loop, respectively. Consequently, these cases also cover the entanglement entropy for $d=2,3$ and offer a reference for the result for the here considered observables for low spacetime dimension $d$.
Figure \ref{Fig:Numerics} compares our analytical result to the numerical result from \cite{Balasubramanian:2011ur}. To see the overall agreement, let us have a look at subfigures \ref{Fig:Numericsn1} and \ref{Fig:Numericsn2}: we see no deviation neither in the small- nor in the large-width limit. To have an estimate of the deviation between our analytical and the numerical result, subfigures \ref{Fig:Numericsn1Zoom} and \ref{Fig:Numericsn2Zoom} zoom into the result for $d=3$. We see that the deviation is negligibly small.

Further confidence may be gained by comparing our results for the entanglement density in Figure \ref{Fig:HED} to the numerical result of \cite{Gushterov:2017vnr}.

  \section{Conclusions and outlook}

For general dimensions, we have obtained closed form analytical expressions for physical
quantities holographically dual to the area of minimal surfaces of
varying codimension, i.e. for two-point functions, the Wilson loop and
the entanglement entropy. Our expressions coincide with previous
numerical results to great accuracy.

Our results allow in particular for a consistent expansion in the
regime where the relevant length scale (such as the size of the
entangling region for the example of the entanglement entropy)  is
large, such that the dual surface probes the deep interior of the
bulk. This corresponds to low energies in the field theory. In
particular, we were able to extract physical information from the
first subleading term in this expansion: For the Wilson loop in the
AdS soliton background, it
corresponds to a finite mass renormalisation. For the entanglement
density defined in \eqref{eqn:ED} for a strip entangling region in the AdS
Schwarzschild background, the subleading term corresponds to an area
term. While this term satisfies an area theorem for RG flows, here we find that for field theories in
dimension $d \geq 7$, the area theorem is violated when comparing
zero- and finite temperature at fixed entangling region.

We refer to the extensive recent discussion of this issue in
\cite{Gushterov:2017vnr}, where not only the AdS-Schwarzschild background, but also the
Reissner-Nordstr\"om solution with finite charge as well as further examples were considered in a
numerical approach. It was argued that the violation of the area law
in these geometries is tied to an inhomogeneous scaling of time and
spatial directions. For the AdS-Schwarzschild case, this appears in
the limit of very large dimensions, i.e.\ $d \rightarrow \infty$. 
\cite{Emparan:2013moa,Emparan:2013xia}
Here however, we observe a change in behaviour at an intermediate
value of $d = 7$ on the field-theory side. One possibility is that new
degrees of freedom are generated in the IR. This remains an open
question worth to be studied. 

A further possibility is that in phase transitions occur which restore
area law behaviour. For instance, when the boundary direction in which
the strip width expands is compactified on a circle,  for large strip
width $\ell$ there is a transition in the entanglement entropy when
$\ell$ is increased: Beyond a critical $\ell$, the minimal area is
given by the surface over the complement of $\ell$ plus the black hole
entropy \cite{Hubeny:2013gta}. It will be interesting to investigate
if the area law violation is absent when this behaviour is taken into
account. This should be possible using the formulae given in the
present paper.

We expect that the analytical expressions given here
will be useful for investigating many further issues in holography. Since the method described in the appendix only depends on the form of the power series, it may also yield analytic results for other geometries.\footnote{Possible candidates include \cite{Kusuki:2017jxh,Kundu:2016dyk}.}

\bigskip

{\bf Acknowledgements}\\[1em]
We are grateful to Nikola Gushterov, Andy O'Bannon and Ronnie Rodgers
for discussions and for sharing their numerical results in \cite{Gushterov:2017vnr}  on entanglement
density prior to publication. Moreover, we are grateful to Martin Ammon, to Andreas Karch, to Haye Hinrichsen, to Mario Flory and to Oleg Andreev for discussions. Furthermore, we thank Ben Craps and Wieland Staessens for providing the numerics of \cite{Balasubramanian:2011ur} for comparison.

\bigskip

  \appendix
  \newpage
\section{Generalised hypergeometric functions and Meijer \texorpdfstring{$G$}{G}-Function}
\label{sec:AppFunctions}
The results for the extremal surfaces can be expressed in terms of generalised hypergeometric functions and Meijer $G$-functions. In the following, we review these functions and their properties. \cite{luke1969special,beals2013meijer,slater1966generalized,solomonovich2014table}
\subsection{Generalised hypergeometric functions}
\label{sec:AppHGF}
A generalised hypergeometric function is the power series
\begin{subequations}
\begin{align}
  \Hypergeometric{ p}{ q}{a}{b}{z}
  &= \sum\limits_{n=0}^\infty \frac{1}{n!} \frac{\Pochhammer{a_1}{n}\cdots \Pochhammer{a_p}{n}}{\Pochhammer{b_1}{n} \cdots \Pochhammer{b_q}{n}}
  % \frac{\prod\limits_{i=1}^p \Pochhammer{a_i}{n}}{\prod\limits_{i=1}^q \Pochhammer{b_i}{n}}
  ~ z^n,\label{eqn:HGFDef}\\
  &= \sum\limits_{n=0}^\infty c_n, \label{eqn:HGFCoeff}
\end{align}
\end{subequations}
where $\Pochhammer{a}{n}$ is the (rising) Pochhammer symbol
\begin{align}
 \Pochhammer{a}{n} = \begin{cases}
                      1 & \text{if } n=0, \\
                      a \cdot (a+1) \cdot \dots \cdot (a+n-1) & \text{if }n \in \mathbb{N}.
                     \end{cases}\label{eqn:Pochhammer}
\end{align}
The parameters $a_i$ and $b_i$ are the numerator and denominator parameters respectively, whereas $z$ is the variable or argument of the hypergeometric function. Another common notation is
\begin{align}
  \, _{p}F_{q}\left(\left.  \begin{matrix}
  a_1,\dots,a_p\\
  b_1,\dots,b_q
  \end{matrix}\right|z\right),
\end{align}
which we will use occasionally to avoid lengthy expressions. In this work, we construct hypergeometric functions from a known power series, i.e.\ for known $c_n$ (c.f.\ \eqref{eqn:HGFCoeff}) normalised such that $c_0=1$. This can be done by calculating the ratio between successive coefficients
\begin{align}
 \frac{c_{n+1}}{c_n} = z \cdot \frac{\prod_{m=1}^p \left(a_m+n\right)}{\prod_{m=1}^q \left(b_m+n\right)} \frac{1}{n+1} \label{eqn:AppHyGeoCon}.
\end{align}

For a power series, the radius of convergence is important. A generalised hypergeometric function converges absolutely 
\begin{itemize}
\item for all values of $|z|$ if $p \leq q$,
\item for $|z|<1$ if $p=q+1$,
\item for $|z|=1$ if $p=q+1$ under the condition that
\begin{align}
 \Psi = \sum_{i=1}^{p} b_i-\sum_{i=1}^{p+1} a_i>0. \label{eqn:ConvergenceCondition}
\end{align}
\end{itemize}
Let us look closer at the case $p=q+1$. For $_2F_1$, the result at unit argument is known in the case that it is finite
\begin{align}
{}_2F_1 (a,b;c;1)&= \frac{\Gamma(c)\Gamma(c-a-b)}{\Gamma(c-a)\Gamma(c-b)}, \qquad   \text{for }\Re(c)>\Re(a+b) \label{eqn:2F1Unit}.
\end{align}
Unfortunately, this is not the case for general $_{p+1}F_p$. However, we can examine the divergent behaviour for $\Psi \leq 0$. 
\begin{subequations}
	\begin{align}
	\Hypergeometric{ p+1}{ p}{a}{b}{z} &= - \frac{\prod_{i=1}^{p} \GammaFunc{b_i}}{\prod_{i=1}^{p+1} \GammaFunc{a_i}}\cdot \ln(1-z)&\text{ for } \Psi = 0,  \\[0.5em]
	 \Hypergeometric{ p+1}{ p}{a}{b}{z} &= \GammaFunc{-\Psi} \frac{\prod_{i=1}^{p} \GammaFunc{b_i}}{\prod_{i=1}^{p+1} \GammaFunc{a_i}}\cdot (1-z)^{\Psi}&\text{ for } \Re( \Psi) < 0.
	\end{align}
	\label{eqn:HgfDivergences}%
\end{subequations}

Let us finish this section with possible simplifications. From the series representation \eqref{eqn:HGFDef} we notice the trivial one: coinciding numerator and denominator parameter cancel each other
\begin{align}
\, _{p+1}F_{q+1}\left(a_1,\dots,a_p,a_{p+1};b_1,\dots,b_q,a_{p+1} ;z\right)&=\, _pF_q\left(a_1,\dots,a_p;b_1,\dots,b_q ;z\right). \label{eqn:GHFkuerzen}
\end{align}
Another interesting simplification is possible if two hypergeometric functions are associated or contiguous to each other, what means that their parameters differ by integer values. One can find a linear relationship between them, so called contiguous relations. One simple case is
\begin{align}
&&  a_1 \cdot \, _pF_q\left(a_1+1,a_2,\dots,a_p;b_1,\dots,b_q ;z\right)& \nonumber\\
&& -(b_1-1)\cdot \, _pF_q\left(a_1,\dots,a_p;b_1-1,b_2,\dots,b_q ;z\right)& \nonumber\\
&& +(b_1-a_1-1)\cdot \, _pF_q\left(a_1,\dots,a_p;b_1,\dots,b_q ;z\right)&=0. \label{}
\label{eqn:ContiguousRel}
\end{align}
Finally, for some parameters a closed form for the hypergeometric function or the value at unit argument is known. In particular, we will use
\begin{subequations}
\begin{align}
  \Hypergeometric{1}{0}{a}{}{z}&=(1-z)^{-a} \label{eqn:AppSqrtHGF},\\
  \Hypergeometric{2}{1}{1,\frac{1}{2}}{\frac{3}{2}}{z}&= \frac{1}{\sqrt{z}}\operatorname{artanh}\left(\sqrt{z}\right),\label{eqn:Artanh}\\
  \Hypergeometric{2}{1}{1,1}{2}{z}&=-\frac{1}{z} \log(1-z),\label{eqn:Ln} \\
  \Hypergeometric{3}{2}{1,1,\frac{3}{2}}{2,\frac{5}{2}}{z} &=-\frac{6 \tanh ^{-1}\left(\sqrt{z}\right)}{z^{3/2}}+\frac{6}{z}-\frac{3 \log (1-z)}{z}, \label{eqn:HgfA2}\\
  \Hypergeometric{3}{2}{1,1,\frac{3}{2}}{2,2}{1} &= 4\ln2,  \label{eqn:App3F21} \\
  \Hypergeometric{3}{2}{1,1,\frac{3}{2}}{2,\frac{5}{2}}{1} &=3 (2 - \ln 4). \label{eqn:HgfA23}
\end{align}
\end{subequations}

\subsection{Meijer \texorpdfstring{$G$}{G}-Function}
\label{sec:AppMGF}

The sum of generalised hypergeometric functions
\begin{align}
	&G_{p,q}^{\,m,n}\!\left(\left.{\begin{matrix}a_{1},\mydots ,a_{p}\\b_{1},\mydots ,b_{q}\end{matrix}}\;\right|\,z\right)=\sum _{h=1}^{m}{ ~\frac {\prod\limits_{j=1}^{n}\Gamma (1+b_{h}-a_{j})~\prod\limits _{\mathclap{j=1,~j\neq h}}^{m}\Gamma (b_{j}-b_{h})}{\prod\limits_{\mathclap{j=m+1}}^{q}\Gamma (1+b_{h}-b_{j})~\prod\limits_{\mathclap{j=n+1}}^{p}\Gamma (a_{j}-b_{h})}} z^{b_{h}}  \\
	&~~~\times \;_{p}F_{q-1}\!\left(1+b_{h}-a_{1},\mydots,1+b_{h}-a_{p} ; \underbrace{1+b_{h}-b_{1} ,\mydots,1+b_{h}-b_{q}}_{\text{without }b_h};(-1)^{p-m-n}\;z\right) \nonumber \label{eqn:MeijerGHGF}
\end{align}
is a Meijer $G$-function, where for $m \leq q,~n \leq p$. In general, this function is defined as
\begin{align}
	G_{p,q}^{\,m,n}\!\left(\left.{\begin{matrix}a_{1},\mydots ,a_{p}\\b_{1},\mydots ,b_{q}\end{matrix}}\;\right|\,z\right)={\frac {1}{2\pi i}}\int _{L}{\frac {\prod\limits _{j=1}^{m}\Gamma (b_{j}-s)\prod\limits _{j=1}^{n}\Gamma (1-a_{j}+s)}{\prod\limits _{j=m+1}^{q}\Gamma (1-b_{j}+s)\prod\limits _{j=n+1}^{p}\Gamma (a_{j}-s)}}\,z^{s}\,ds,
\end{align}
which is well defined for
\begin{subequations}
\begin{align}
0\leq m &\leq q,~  0 \leq n \leq p&, \\
a_k - b_j &\notin \mathbb{N} &\forall k=1,\dots,n\text{ and }j=1,\dots,m, \\
z \neq 0.
\end{align}
\end{subequations}
The path of integration $L$ is chosen in such a way that it splits the poles of $\Gamma (b_{j}-s)$ from the ones of $\Gamma (1-a_{j}+s)$. 

The advantage of writing this sum of hypergeometric function as a Meijer $G$-function is that the Meijer $G$-function inherits its properties from the hypergeometric functions. Therefore, we have similar properties but only have to consider one term instead of $m$. One example are the convergence for unit argument. Meijer $G$-function converge for
\begin{align}
  \nu =\sum _{j=1}^{q}b_{j}-\sum _{j=1}^{p}a_{j} < -1 \label{eqn:ConvergenceMG},
\end{align}
which is the analogue of \eqref{eqn:ConvergenceCondition}. Another more involved example are the recurrence relations
\begin{align}
  (a_{1}-b_{q}-1)\;G_{p,q}^{\,m,n}\!\left(\left.{\begin{matrix}a_1,\mydots,a_p \\b_1,\mydots,b_q \end{matrix}}\;\right|\,z\right)&=G_{p,q}^{\,m,n}\!\left(\left.{\begin{matrix}a_{1},\mydots ,a_{p}\\b_{1},\mydots ,b_{q-1},b_{q}+1\end{matrix}}\;\right|\,z\right)  \label{eqn:ReccurenceMG}\\
  &~~~ -G_{p,q}^{\,m,n}\!\left(\left.{\begin{matrix}a_{1}-1,\mydots ,a_{p-1},a_{p}\\b_{1},\mydots ,b_{q}\end{matrix}}\;\right|\,z\right),\quad n<p,\;m<q,\nonumber
\end{align}
which are the analogue of the contiguous relations \eqref{eqn:ContiguousRel} of the hypergeometric functions.

\section{Calculation of minimal area of bulk surfaces}
\label{sec:MinimalSurface}
We consider an $n$-dimensional strip on the boundary, as shown in Figure \ref{Fig:MinimalSurface}.\footnote{We look at a constant time-slice.} We calculate the minimal area anchored on it. In the following, we first review the calculation in form of a power series, as done in \cite{Fischler:2012ca} for $n=1,2,d-1$. Afterwards, we simplify this result in terms of generalised hypergeometric functions and later Meijer $G$-functions. The area for $n=1$ has some subtleties and is considered separately in section \ref{sec:MinimalSurfacen1}.
\begin{figure}
  \center
  \def\svgwidth{0.45\linewidth}
  \input{Inkscape/MinimalSurface.pdf_tex}
\caption{Boundary region and associated bulk surface.\\
 The strip has the width $\ell$ in direction $x^1$ and extends infinitely in the directions $x^i$ with  $i=2,\cdots,n$. For regularisation, we take the finite width $\tilde \ell \gg \ell$. The transverse directions (i.e.\ $x^j$ with $j=n+1,\cdots,d-1$) are not shown.
 }
\label{Fig:MinimalSurface}
  \label{Fig:BulkMinimalSurface}
\end{figure}

The strip has the width $\ell$ in one spatial direction and the length $\tilde \ell$ in $n-1$ spatial directions in the limit $\tilde \ell \gg \ell$ (c.f. Figure \ref{Fig:MinimalSurface}). For planar AdS-Schwarzschild (c.f. \eqref{eqn:metric}), the area of a surface parametrised by $x^1(z) = x(z)$ is
\begin{align}
 \Area{} %&= 2 \int\limits_{\epsilon}^{\zs}dz\int dx_2 \cdots dx_n ~ \sqrt{ g_{\text{ind}} }, \nonumber \\
 &=  2 L^n \tilde \ell^{n-1} \int\limits_{\epsilon}^{\zs}dz~ z^{-n} \sqrt{\frac{1}{b(z)} + x'(z)^2}.
\end{align}
Instead of integrating all the way down to $z=0$, we introduce the bulk cut-off $\epsilon$, the regularize the area. For the minimal surface, the quantity
\begin{align}
 \frac{x'(t)}{z^n} \frac{1}{\sqrt{\frac{1}{b(z)} + x'(z)^2}} = \frac{1}{\zs^n}
\end{align}
is conserved.\footnote{The right-hand side is obtained by considering the turning point $\zs$ of the minimal surface, where $x'$ diverges.} Consequently, the embedding for an extremal surface with turning point $\zs$ is described by
\begin{align}
 x'(t) = \pm \left(\frac{z}{\zs}\right)^{n} \frac{1}{\sqrt{b}} \frac{1}{\sqrt{1-\left(z/\zs\right)^{2n}}}
\end{align}
and the area of the extremal surface is\footnote{The blackening factor is $b(z) = 1-(z/\zh)^d$.}
\begin{subequations}
\begin{align}
 \Area{} &= 2L^n \tilde \ell^{n-1} \int\limits_{\epsilon}^{\zs} dz~ z^{-n} \frac{1}{\sqrt{b(z)}} \frac{1}{\sqrt{1-\left(z/\zs\right)^{2n}}}.
\end{align} 
Furthermore, the width $\ell$ of the strip is
\begin{align}
 \ell &= 2\int_0^\ell dz ~x'(z), \nonumber \\
 &=  2 \int\limits_{0}^{\zs} dz~ \left(\frac{z}{\zs}\right)^{\mathclap{n}} \frac{1}{\sqrt{b(z)}} \frac{1}{\sqrt{1-\left(z/\zs\right)^{2n}}}.
\end{align}\label{eqn:SolutionAsIntegral}%
\end{subequations}
This approach is the general procedure to calculate the minimal area in cases where we have a conserved quantity. For a strip and a general metric this is discussed in more generality in \cite{Bilson:2010ff}.

\subsection{Minimal area and width as power series}
\label{sec:MinimalSurfacePS}
In this section, we are making the first step towards solving these integrals. The square-roots in the integrals \eqref{eqn:SolutionAsIntegral} are a special case of hypergeometric functions and can be written as power series (see \eqref{eqn:AppSqrtHGF}). Since we have the hierarchy $z\leq\zs<\zh$, these series are absolutely convergent. This allows piecewise integration, yielding
\begin{subequations}
\begin{align}
 \Area{} &= \frac{2L^{n} }{n-1} \frac{\tilde \ell^{n-1}}{\epsilon^{n-1}} +  2L\frac{\tilde \ell^{n-1}}{\zs^{n-1}} \sum\limits_{m_1,m_2=0}^{\infty}\frac{\Pochhammer{\frac{1}{2}}{m_1}\Pochhammer{\frac{1}{2}}{m_2}}{m_1!~m_2!}   \frac{(\zs/\zh)^{m_1 d}}{m_1 d+ 2n m_2 -n+1}, \nonumber\\[0.5em]
   &= \frac{2L^{n} }{n-1} \frac{\tilde \ell^{n-1}}{\epsilon^{n-1}} +  \frac{\sqrt{\pi} L^n }{n}\frac{\tilde \ell^{n-1}}{\zs^{n-1}} \sum\limits_{m=0}^{\infty}\frac{1}{m!} \Pochhammer{\frac{1}{2}}{m} \zsh^{md}  \frac{\GammaFunc{\frac{1}{2n}\left( md-n+1\right)}}{\GammaFunc{\frac{1}{2n}\left( md+1\right)}}
\end{align}
for the minimal area and
\begin{align}
 \ell &= 2\zs \sum\limits_{m_1,m_2=0}^{\infty}\frac{\Pochhammer{\frac{1}{2}}{m_1}\Pochhammer{\frac{1}{2}}{m_2}}{m_1!~m_2!}   \frac{(\zs/\zh)^{m_1 d}}{m_1 d+ 2n m_2 +n+1}, \nonumber \\[0.5em]
   &= \frac{\zs \sqrt{\pi}}{n} \sum\limits_{m=0}^{\infty}\frac{1}{m!}\Pochhammer{\frac{1}{2}}{m}   \zsh^{m d} \frac{\GammaFunc{\frac{1}{2n}\left(md+n+1 \right)}}{\GammaFunc{\frac{1}{2n}\left(md+2n+1 \right)}}
\end{align}%
\label{eqn:SolutionAsPowerSeries}%
\end{subequations}
for the width.\footnote{At this point, it is obvious that the area for $n=1$ has to be considered separately. The cut-off divergence is not determined by a power law, but by a logarithmic divergence. } \footnote{The sum over $m_2$ is a hypergeometric function $_{2}F_1$ evaluated at unit argument. Its value can be expressed in terms of Gamma functions, see \eqref{eqn:2F1Unit}.} In the special cases $d=1,2,d-1$,this agrees with the results from \cite{Fischler:2012ca}.

\subsection{Minimal area and width in terms of hypergeometric functions}
\label{sec:MinimalSurfaceHGF}
We simplify these results \eqref{eqn:SolutionAsPowerSeries} by constructing generalised hypergeometric functions. Due to the non-integer factor $d/2n$ in the Gamma functions, the ratio between successive coefficients is not a rational function of the index of summation $m$. Our trick is to rearrange the sum by defining $m$ as
\begin{align}
 m &= \Delta m + \frac{2n}{\chi} \delta m,\\
 \delta m &= 0,\dots,\infty.
\end{align}
The range of $\Delta m$ is $\Delta m = 0,\mydots, \frac{2n}{\chi}-1$ and $\chi$ is the greatest common denominator of $d$ and $2n$. This redefinition allows to construct generalised hypergeometric functions with respect to $\delta m$. We rearrange the series by first keeping $\delta m$ fixed and performing the sum over $\delta m$, but this doesn't change the result since the series is absolutely converging.

Using the construction procedure \eqref{eqn:AppHyGeoCon} results in
\begin{subequations}
\begin{align}
 \ell &= \frac{\sqrt{\pi}\zs}{n} \sum\limits_{\Delta m=0}^{\frac{2n}{\chi}-1}\frac{1}{\Delta m!} \Pochhammer{\frac{1}{2}}{\Delta m} \zsh^{\Delta m d} \frac{\GammaFunc{\frac{d}{\chi} a_{1/2}}}{\GammaFunc{\frac{d}{\chi} a_1}}   \label{eqn:SolutionAsHGFWidth}\\[0.5em]
 &~~~\times\Hgf{\frac{2n+d}{\chi}+1}{\frac{2n+d}{\chi}} {1, a_{\frac{1}{2}},\mydots,  a_{\frac{d}{\chi}-\frac{1}{2}}, b_{\frac{1}{2}},\mydots,  b_{\frac{2n}{\chi}-\frac{1}{2}}; a_1,\mydots,  a_{\frac{d}{\chi}},  b_1,\mydots,   b_{\frac{2n}{\chi}};\zsh^{\frac{2nd}{\chi}}}{}\nonumber
\end{align}
for the width of the strip and
\begin{align}
 \Area{} &= \frac{2L^n}{n-1} \left(\frac{\tilde \ell}{\epsilon}\right)^{n-1} +  \frac{\sqrt{\pi}L^n }{n}\frac{\tilde \ell^{n-1}}{\zs^{n-1}} \sum\limits_{\Delta m=0}^{\frac{2n}{\chi}-1} \frac{1}{\Delta m!} \Pochhammer{\frac{1}{2}}{\Delta m} \zsh^{\Delta m d} \frac{\GammaFunc{\frac{d}{\chi} a_{-1/2}}}{\GammaFunc{\frac{d}{\chi} a_0}}    \label{eqn:SolutionAsHGFArea}\\[0.5em]
 &~~~\times\Hgf{\frac{2n+d}{\chi}+1}{\frac{2n+d}{\chi}}{1, a_{-\frac{1}{2}},\mydots,  a_{\frac{d}{\chi}-\frac{3}{2}},   b_{\frac{1}{2}},\mydots,b_{\frac{2n}{\chi}-\frac{1}{2}}; a_0,\mydots, a_{\frac{d}{\chi}-1}, b_1,\mydots,  b_{\frac{2n}{\chi}};\zsh^{\frac{2nd}{\chi}}}{} \nonumber
 \end{align}
 for the minimal area of the surface.
\label{eqn:SolutionAsHGF}%
\end{subequations}
The parameters in the hypergeometric functions are
\begin{subequations}
\begin{align} 
  a_i =& \frac{\chi}{2nd}\left( \Delta m d +1 + 2n i\right),\\[0.5em]
  b_j =& \frac{\chi}{2n}\left(\Delta m +j\right).
\end{align}\label{eqn:Parameters}%
\end{subequations}

Let us emphasise the simplification of this result compared to the power series in \eqref{eqn:SolutionAsPowerSeries}: this equation has a finite number of terms and is no longer an infinite sum. 

Looking at the properties of hypergeometric functions, we notice that while each of these terms converges for $\zs < \zh$, they diverge logarithmically for $\zs \rightarrow \zh$ (c.f.\ chapter \ref{sec:AppHGF}). This implies $\Area{} \propto \ell$ in this limit. We come back to this divergence later. First, let us write this result in terms of Meijer $G$-functions.

\subsection{Result in terms of Meijer \texorpdfstring{$G$}{G}-Function}
\label{sec:MinimalSurfaceMGF}
In the last section, we simplified the result to a sum over hypergeometric functions (c.f.\ \eqref{eqn:SolutionAsHGF}). Comparing this with the Meijer $G$-function \eqref{eqn:MeijerGHGF} yields
\begin{subequations}
\begin{align}
\ell = \frac{2 \pi \zh}{\sqrt{2nd}}~ G_{\frac{2n+d}{\chi},\frac{2n+d}{\chi}}^{\,\frac{2n}{\chi},\frac{d}{\chi}}\!\left(\left.{\begin{matrix}\hat a_{1/2},\dots ,\hat a_{d/\chi -1/2}, \hat b_{1/2},\dots,\hat b_{2n/\chi-1/2}\\\hat b_{0},\dots,\hat b_{2n/\chi-1},\hat a_{0},\dots ,\hat a_{d/\chi -1}\end{matrix}}\;\right|\,\zsh^{\frac{2nd}{\chi}}\right)
\end{align}
for the width of the strip and
\begin{align}
 \Area{} &= \frac{2L^n}{n-1} \left(\frac{\tilde \ell}{\epsilon}\right)^{n-1}+ \frac{2 \pi L^n}{\sqrt{2nd}} \frac{\tilde \ell^{n-1}\zh}{\zs^n}   \\[0.5em]
 &~~~\times~ G_{\frac{2n+d}{\chi},\frac{2n+d}{\chi}}^{\,\frac{2n}{\chi},\frac{d}{\chi}}\!\left(\left.{\begin{matrix}\hat a_{3/2},\dots ,\hat a_{d/\chi +1/2}, \hat b_{1/2},\dots,\hat b_{2n/\chi-1/2}\\\hat b_{0},\dots,\hat b_{2n/\chi-1},\hat a_{1},\dots ,\hat a_{d/\chi }\end{matrix}}\;\right|\,\zsh^{\frac{2nd}{\chi}}\right)\nonumber
\end{align}\label{eqn:ResultAsMGF}%
\end{subequations}
for the area of the minimal surface. The new parameters are
\begin{subequations}
	\begin{align}
	\hat a_i &= \frac{\chi}{d}i, \\[0.5em]
	\hat b_j &= \frac{\chi}{2n} \left(j + \frac{1}{d}\right).
	\end{align}\label{eqn:ParameterG}%
\end{subequations}
Comparing this with criterion \eqref{eqn:ConvergenceMG}, we see that both of this quantities diverge in the limit $\zs \rightarrow \zh$, which matches our earlier observation. In the following, we derive an alternative form of the result and take a closer look at the large-width limit $\zs \rightarrow \zh$.

\subsection{Large-width behaviour}
\label{sec:LowHighT}
The large-width limit corresponds to $\zs \rightarrow \zh$, while $\zh$ and hence the temperature is kept fixed.\footnote{The reason why we cannot simply take the dimensionless ratio $\ell \cdot T$ to infinity is that our strip has two length-scales: $\ell$ and $\tilde \ell$.}
Each term in the previous results for the minimal area \eqref{eqn:SolutionAsHGF} and \eqref{eqn:ResultAsMGF} diverges in this limit. In the following, we rewrite the area to split of the divergent part.

Comparing the result in terms of hypergeometric functions for width and minimal area (c.f.\ \eqref{eqn:SolutionAsHGF}), one notices that they are associated. Consequently, we rewrite the result using the contiguous relation from \eqref{eqn:ContiguousRel}, yielding
\begin{align}
\Area{} &= \frac{2L^n}{n-1} \left(\frac{\tilde \ell}{\epsilon}\right)^{~~\mathclap{n-1}} + \frac{L^n \tilde \ell^{n-1}}{\zs^n} \ell+  \frac{\sqrt{\pi}L^n }{2n}\frac{\tilde \ell^{n-1}}{\zs^{n-1}} \sum\limits_{\Delta m=0}^{\frac{2n}{\chi}-1} \frac{1}{\Delta m!} \Pochhammer{\frac{1}{2}}{\Delta m} \zsh^{\Delta m d} \frac{\GammaFunc{\frac{d}{\chi} a_{-1/2}}}{\GammaFunc{\frac{d}{\chi} a_1}}   \nonumber\\[0.5em]
&\times\Hgf{\frac{2n+d}{\chi}+1}{\frac{2n+d}{\chi}}{1,  a_{-\frac{1}{2}},\mydots,  a_{\frac{d}{\chi}-\frac{3}{2}},   b_{\frac{1}{2}},\mydots,   b_{\frac{2n}{\chi}-\frac{1}{2}}; a_1,\mydots, a_{\frac{d}{\chi}}, b_1,\mydots,   b_{\frac{2n}{\chi}};\zsh^{\frac{2nd}{\chi}}}{}\label{eqn:SolutionHT}.
\end{align}
This shifts one of the parameters in the hypergeometric functions by unit, which causes them to converge at unit argument. Analogously, we use the recurrence relations \eqref{eqn:ReccurenceMG} for Meijer $G$-functions, which results in
\begin{align}
\Area{} &= \frac{2L^n}{n-1} \left(\frac{\tilde \ell}{\epsilon}\right)^{n-1}+  \frac{\tilde \ell^{n-1}\ell L^n}{\zs^n}+ \frac{\chi \pi L^n }{\sqrt{2nd^3}} \frac{\tilde \ell^{n-1}\zh}{\zs^n}   \\[0.5em]
&~~~\times~ G_{\frac{2n+d}{\chi},\frac{2n+d}{\chi}}^{\,\frac{2n}{\chi},\frac{d}{\chi}}\!\left(\left.{\begin{matrix}\hat a_{3/2},\dots ,\hat a_{d/\chi +1/2}, \hat b_{1/2},\dots,\hat b_{2n/\chi-1/2}\\\hat b_{0},\dots,\hat b_{2n/\chi-1},\hat a_{0},\dots ,\hat a_{d/\chi-1\label{eqn:SolutionHTG} }\end{matrix}}\;\right|\,\zsh^{\frac{2nd}{\chi}}\right).\nonumber
\end{align}

In both expressions, the second term is the one which diverges in the limit $\zs \rightarrow \zh$ and yields the thermal entropy for the region. The third term however is finite in this limit, as can be seen from \eqref{eqn:ConvergenceMG} or \eqref{eqn:ConvergenceCondition}. The width $\ell$ diverges logarithmically (c.f.\  \eqref{eqn:HgfDivergences})
\begin{align}
 \ell \propto \ln \left( 1-\frac{\zs}{\zh} \right).
\end{align} Therefore, the large-width behaviour of the area is
\begin{subequations}
\begin{align}
\Area{} &= \frac{2L^n}{n-1} \left(\frac{\tilde \ell}{\epsilon}\right)^{n-1}+  \frac{  L^n}{\zh^n}\tilde\ell^{n-1}\ell+ \frac{\chi \pi L^n }{\sqrt{2nd^3}} \frac{\tilde \ell^{n-1}}{\zh^{n-1}}   \nonumber\\[0.5em]
&~~~\times~ G_{\frac{2n+d}{\chi},\frac{2n+d}{\chi}}^{\,\frac{2n}{\chi},\frac{d}{\chi}}\!\left(\left.{\begin{matrix}\hat a_{3/2},\dots ,\hat a_{d/\chi +1/2}, \hat b_{1/2},\dots,\hat b_{2n/\chi-1/2}\\\hat b_{0},\dots,\hat b_{2n/\chi-1},\hat a_{0},\dots ,\hat a_{d/\chi-1 }\end{matrix}}\;\right|\,1\right)\nonumber \\[0.5em]
&~~~+ \text{subleading terms}
\end{align}
or expressed in hypergeometric functions
\begin{align}
\Area{} &= \frac{2L^n}{n-1} \left(\frac{\tilde \ell}{\epsilon}\right)^{n-1} + \frac{L^n }{\zh^n}\tilde \ell^{n-1} \ell+  \frac{\sqrt{\pi}L^n }{2n}\frac{\tilde \ell^{n-1}}{\zh^{n-1}} \sum\limits_{\Delta m=0}^{\frac{2n}{\chi}-1} \frac{1}{\Delta m!} \Pochhammer{\frac{1}{2}}{\Delta m} \frac{\GammaFunc{\frac{d}{\chi} a_{1/2}}}{\GammaFunc{\frac{d}{\chi} a_1}}    \nonumber\\[0.5em]
&~~~\times\Hgf{\frac{2n+d}{\chi}+1}{\frac{2n+d}{\chi}}{1, a_{\frac{1}{2}},\mydots, a_{\frac{d}{\chi}-\frac{3}{2}},  b_{\frac{1}{2}},\mydots,  b_{\frac{2n}{\chi}-\frac{1}{2}};  a_1,\mydots, a_{\frac{d}{\chi}},  b_1,\mydots,  b_{\frac{2n}{\chi}};1}{}\nonumber \\[0.5em]
&~~~+ \text{subleading terms}.
\end{align}
\end{subequations}
In both expressions, the second term has a volume scaling and hence the leading contribution in the large-width limit. The remaining terms are an area scaling, as they scale with the boundary area of the strip.

This agrees with the observations in \cite{Fischler:2012ca}. However, the authors here derived the subleading term as a convergent infinite series. Instead of using the properties of hypergeometric functions and Meijer $G$-functions, the same subleading term can be derived by constructing hypergeometric functions out of this series. This section therefore nicely shows how we can use known properties of hypergeometric functions or Meijer $G$-functions to learn more about our result.

\subsection{Special case: geodesic length}
\label{sec:MinimalSurfacen1}
\begin{figure}
	\center
	\def\svgwidth{0.45\linewidth}
	\input{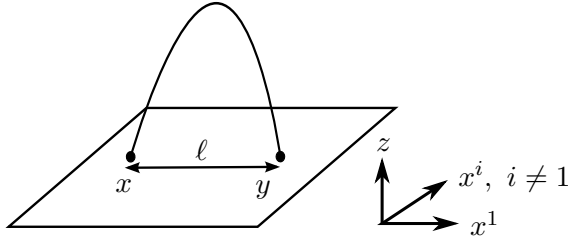}
	\caption{Width and associated bulk geodesic. }
	\label{Fig:MinimalSurfacen1}
\end{figure}
In this section, we turn to the calculation of the geodesic length between two points on the boundary, which we excluded earlier. Following the notation above, this is the case $n=1$. This case is sketched in Figure \ref{Fig:MinimalSurfacen1}. The integral representation in \eqref{eqn:SolutionAsIntegral} is still valid, yielding
\begin{align}
 \Area{1} &= 2L  \int\limits_{\epsilon}^{\zs} dz~ z^{-1} \frac{1}{\sqrt{b(z)}} \frac{1}{\sqrt{1-\left(z/\zs\right)^{2}}}.
\end{align}

The result for the width $\ell$ is already presented in \eqref{eqn:SolutionAsHGFWidth}. However, when calculating the geodesic length, we have to be careful. Writing the square-roots as power series (c.f.\ \eqref{eqn:AppSqrtHGF}) yields
\begin{align}
 \Area{1} &=2L\sum\limits_{\substack{m_1,m_2=0}}^{\infty}\frac{\Pochhammer{\frac{1}{2}}{m_1}\Pochhammer{\frac{1}{2}}{m_2}}{m_1!~m_2!}   \begin{cases}
 \ln (\zs/\epsilon) & \text{for }m_1=m_2=0 \\[0.5em]
 \dfrac{(\zs/\zh)^{m_1 d}}{m_1 d+ 2 m_2 } & \text{else} 
 \end{cases},  \nonumber \\[1em]
 &=2L \ln\left(\frac{\zs}{\epsilon}\right) + 2L\sum\limits_{\substack{m_1=1}}^{\infty}\sum\limits_{\substack{m_2=0}}^{\infty}\frac{\Pochhammer{\frac{1}{2}}{m_1}\Pochhammer{\frac{1}{2}}{m_2}}{m_1!~m_2!}   \frac{(\zs/\zh)^{m_1 d}}{m_1 d+ 2 m_2 } 
% &~~~+ 2L\sum\limits_{\substack{m=1}}^{\infty}\frac{\Pochhammer{\frac{1}{2}}{m}}{m!} \frac{1}{m}  \left(  \frac{(\zs/\zh)^{m d}}{m d }+\frac{1}{2 m }   \right) . 
+ 2L\sum\limits_{\substack{m_2=1}}^{\infty}\frac{\Pochhammer{\frac{1}{2}}{m_2}}{m_2!} \frac{1}{2m_2}  .
\end{align}
Simplifying this, the geodesic length can be written as a power series
 \begin{align}
  \Area{1} &= 2L \ln\left(\frac{2\zs}{\epsilon}\right) + \sqrt{\pi}L\sum\limits_{m=1}^\infty \frac{1}{m!} \Pochhammer{\frac{1}{2}}{m} \frac{\GammaFunc{\frac{md}{2}}}{\GammaFunc{\frac{1}{2}(md+1)}} \zsh^{dm}.\label{eqn:SolutionAsPowerSeriesn1}
 \end{align}
 It is worthwhile to compare this to the result for $n\neq 1$ in \eqref{eqn:SolutionAsPowerSeries} to notice that only the range of the sum is shifted. Therefore, the previous result can be used for shifted $\Delta m$, yielding
  \begin{align}
 \Area{1} &=2L \ln\left(\frac{2\zs}{\epsilon}\right) +  \sqrt{\pi}L \sum\limits_{\Delta m=1}^{\frac{2}{\chi}} \frac{1}{\Delta m!} \Pochhammer{\frac{1}{2}}{\Delta m} \zsh^{\Delta m d} \frac{\GammaFunc{\frac{d}{\chi} a_{-1/2}}}{\GammaFunc{\frac{d}{\chi} a_0}}   \label{eqn:SolutionAsHGFArean1}\\
 &~~~\times\Hgf{\frac{2+d}{\chi}+1}{\frac{2+d}{\chi}}{1, a_{-\frac{1}{2}},\mydots,  a_{\frac{d}{\chi}-\frac{3}{2}},   b_{\frac{1}{2}},\mydots,b_{\frac{2}{\chi}-\frac{1}{2}}; a_0,\mydots,a_{\frac{d}{\chi}-1}, b_1,\mydots, b_{\frac{2}{\chi}};\zsh^{\frac{2d}{\chi}}}{}\nonumber
\end{align}
for the geodesic length, where $\chi$ is the greatest common denominator of $d$ and $2$. The parameters are the ones introduced in \eqref{eqn:Parameters}. The previous simplification to Meijer $G$-functions cannot done as before, since due to the shift of $\Delta m$ the numerator parameter 1 is not always cancelled by a denominator parameter. However, this result is already simple enough on its own as it consists of maximal two terms.

Let us turn to the large-width limit. First, it is necessary to perform the same shift of $\Delta m$ for $\scl{1}$, yielding
\begin{align}
 \scl{1} &= 2 \zs+ \sqrt{\pi}\zs \sum\limits_{\Delta m=1}^{\frac{2}{\chi}}\frac{1}{\Delta m!} \Pochhammer{\frac{1}{2}}{\Delta m} \zsh^{\Delta m d} \frac{\GammaFunc{\frac{d}{\chi} a_{1/2}}}{\GammaFunc{\frac{d}{\chi} a_1}}   \nonumber \\
 &~~~\times\Hgf{\frac{2+d}{\chi}+1}{\frac{2+d}{\chi}} {1,    a_{\frac{1}{2}},\mydots,  a_{\frac{d}{\chi}-\frac{1}{2}},b_{\frac{1}{2}},\mydots, b_{\frac{2}{\chi}-\frac{1}{2}}; a_1,\mydots,a_{\frac{d}{\chi}}, b_1,\mydots, b_{\frac{2}{\chi}};\zsh^{\frac{2d}{\chi}}}{}.\label{eqn:SolutionAsHGFWidthn1Shift}
\end{align}
The minimal area can be transformed in the same way as in section \ref{sec:LowHighT}. Therefore, considering the result in \eqref{eqn:SolutionHT}, the result for $n=1$ is
 \begin{align}
 \Area{1} &=2L \ln\left(\frac{2\zs}{\epsilon}\right) + \frac{L}{\zs} \scl{1} -2L+  \frac{\sqrt{\pi}L}{2} \sum\limits_{\Delta m=1}^{\frac{2}{\chi}} \frac{1}{\Delta m!} \Pochhammer{\frac{1}{2}}{\Delta m} \zsh^{\Delta m d} \frac{\GammaFunc{\frac{d}{\chi} a_{-1/2}}}{\GammaFunc{\frac{d}{\chi} a_1}}   \nonumber\\[0.5em]
 &~~~\times\Hgf{\frac{2+d}{\chi}+1}{\frac{2+d}{\chi}}{1,    a_{-\frac{1}{2}},\mydots,  a_{\frac{d}{\chi}-\frac{3}{2}}, b_{\frac{1}{2}},\mydots,b_{\frac{2}{\chi}-\frac{1}{2}};    a_1,\mydots,  a_{\frac{d}{\chi}},   b_1,\mydots,   b_{\frac{2}{\chi}};\zsh^{\frac{2d}{\chi}}}{}.
\end{align}
The constant third term is due to the shift in \eqref{eqn:SolutionAsHGFWidthn1Shift}. Calculating the large-width limit yields
\begin{align}
 \Area{} &=2L \ln\left(\frac{2\zh}{\epsilon}\right)-2L + \frac{L}{\zh} \ell+  \frac{\sqrt{\pi}L}{2} \sum\limits_{\Delta m=1}^{\frac{2}{\chi}} \frac{1}{\Delta m!} \Pochhammer{\frac{1}{2}}{\Delta m}  \frac{\GammaFunc{\frac{d}{\chi} a_{-1/2}}}{\GammaFunc{\frac{d}{\chi} a_1}}   \nonumber\\[0.5em]
 &~~~\times\Hgf{\frac{2+d}{\chi}+1}{\frac{2+d}{\chi}}{1,  a_{-\frac{1}{2}},\mydots, a_{\frac{d}{\chi}-\frac{3}{2}},  b_{\frac{1}{2}},\mydots,  b_{\frac{2}{\chi}-\frac{1}{2}}; a_1,\mydots,a_{\frac{d}{\chi}},b_1,\mydots, b_{\frac{2}{\chi}};1}{}\nonumber \\[0.5em]
 &~~~+ \text{subleading terms}. \label{eqn:SolutionHTn1}
\end{align}

Therefore, we were able to obtain an analytical result for the minimal area and the width of the strip for the geodesic (i.e.\ the strip reduces to an interval).

  \nocite{*}
  \bibliography{main,books}
  \bibliographystyle{JHEP}

\end{document}